\documentclass[aps,pre,twocolumn]{revtex4-2}
\usepackage{amssymb}
\usepackage{amsmath,bm}
\usepackage{graphicx,color}
\usepackage{epsfig}
\usepackage{undertilde}
\usepackage{multirow}
\usepackage[cal=cm]{mathalpha}
\usepackage{accents}
\DeclareMathAccent{\wtilde}{\mathord}{largesymbols}{"65}
\usepackage{stackengine}
\usepackage{ulem}
\usepackage[mathscr]{euscript}
\usepackage{float}

\stackMath
\newcommand\tenq[2][1]{%
	\def\useanchorwidth{T}%
	\ifnum#1>1%
	\stackunder[0pt]{\tenq[\numexpr#1-1\relax]{#2}}{\scriptscriptstyle\sim}%
	\else%
	\stackunder[1pt]{#2}{\scriptscriptstyle\sim}%
	\fi%
}

\renewcommand{\vec}[1]{{\underline{#1}}}

\newcommand{\B}[1]{{\bm{#1}}}

\begin{document}
\title{The Outer Eshelby Problem in Amorphous Solids with Odd Dipole Screening}
	
\author{Avanish Kumar$^{1}$, Jin Shang$^{2}$ and Itamar Procaccia$^{1,2,3}$}
\affiliation{$^1$ Sino-Europe Complex Science Center, School of Mathematics, North University of China, Shanxi, Taiyuan 030051, China. \\$^2$Hangzhou International Innovation Institute, Beihang University, Hangzhou, China, 311115 \\$^3$Department of Chemical Physics, the Weizmann Institute of Science, Rehovot 7610001,}

\begin{abstract}
The so-called ``2-dimensional Eshelby problem'' was defined by JD Eshelby, where a circle is cut out of an elastic sheet, deformed into an ellipse, and forced back into the material. This creates a quadrupolar response which turned out to be very relevant for describing plastic deformations in amorphous solids. Moreover, the resulting strain field has been used to describe how elasto-plastic models respond to external strains. It is thus important to examine carefully the Eshelby problem in amorphous solids, especially when anomalous elasticity is at play. Here we solve the Eshelby problem in amorphous solids with both even and odd dipole screening and a complete analytical result is obtained for the displacement and strain fields. The analytical results are crucial for understanding the anomalous responses in displacement fields measured directly from experiments and numerical simulations. The computation of strain fields clearly demonstrates the deviations from their classical counterparts for perfectly elastic solids. The decomposition of the displacement field shows the excitation of higher-order Fourier modes. The presence of higher excitations other than the imposed Eshelby geometry therefore suggests the possible nonlinear mode coupling that can influence and modify the anomalous responses. 
\end{abstract}
\maketitle
\section{Introduction}
In the 1950's John Douglas Eshelby has defined and solved a seemingly abstract problem, where one extracts a circular piece of an infinite elastic sheet, deforms it into an ellipse and forces it back into the material \cite{57Esh}. Eshelby considered both the inner response inside the inclusion and the outer elastic response which is quadrupolar in nature, offering an analytic solution of the outer displacement field and the stress and strain responses. Eshelby's work stimulated a considerable amount of research in the material science community, focused on the mechanical response of a variety of elastic solid materials \cite{99ML,04ML}. In recent decades it turned out that the outer solution of this seemingly artificial procedure has intimate relation to the spontaneous plastic response of amorphous solids under external stress or strain \cite{12DHP,13DHP}. In contradistinction to electrostatics where charges (monopoles) cannot be created locally, but dipoles can, giving rise to dielectrics, 2-dimensional elasticity theory protects both monopoles {\bf and} dipoles, making a quadrupolar response the ``cheapest" way to release stress locally. This somewhat fortunate coincidence has given rise to considerable amount of modeling of the mechanical response of amorphous solids, known sometime as `mesoscopic models' or `elasto-plastic models' where Eshelby's outer strain and stress solutions were taken verbatim as central elements of the models \cite{18NFMB}. Quite recently it was pointed out that there is a fly in the ointment. When amorphous solids suffer from extended plastic responses, sometime known as `avalanches', classical elasticity no longer suffices to provide the correct description of events, and anomalous elasticity becomes relevant \cite{21LMMPRS,22BMP,22KMPS,22MMPRSZ,23MMPR}. 
	
Anomalous elasticity comes about in 2-dimensional amorphous solids when plastic responses are not isolated quasi-localized Eshelby-like events. Then one needs to take into account a quadrupolar {\bf field}, $\B Q (\B r)$ rather than isolated quadrupoles. It was shown that the quadrupolar field by itself re-normalizes the elastic moduli, but does not change the form of elasticity theory. But when the 2-dimensional quadrupolar field is non-uniform, its divergence is a dipolar field $\B P$, 
\begin{equation}
P^\beta \equiv \partial_\alpha Q^{\alpha\beta} \ . 
\end{equation}
The presence of the dipolar field is a game changer. It introduces screening parameters and the breaking of translational, rotational and chiral symmetries. It requires the construction of a new theory that is referred to as ``anomalous elasticity", and in particular it changes dramatically the outer solution of the Eshelby problem \cite{24HKPR}. 
	
The analytic form of the displacement field associated with the outer solution of the Eshelby problem in amorphous solids, was calculated so far only in a simplified version of anomalous elasticity, where only the breaking of translational symmetry was taken into account \cite{24HKPR}.  The aim of this paper is to present what appears to be the first full analytic solution of the Eshelby problem in the context of anomalous elasticity where also rotational and chiral symmetries are broken. Moreover, the present paper presents not only the displacement field, but also the stress and strain field of the outer Eshelby problem, which are the necessary inputs for any elasto-plastic model when avalanches of plastic events are present. The solution presented below will be computed as the response of an amorphous solid contained within a circle of outer radius $r_{\rm out}$ and an inner circle of radius $r_{\rm in}$ which turns into an ellipse. The solution of the stress and strain field are nevertheless useful for a variety of different geometries. 
	
In truth, until recently we thought that a full analytic solution would be too hard to get - indeed the reader may find the present paper somewhat technical and not easy to follow. We therefore made a particular effort to present the solution in some detail and with careful pedagogical outline. The next section introduces the equation for the displacement field for a system in radial geometry, which needs to be solved for any boundary conditions. In such a radial geometry it is natural to consider the radial and tangential components of the displacement field, denoted $d_r(r,\theta)$ and $d_\theta(r,\theta)$ respectively. We will find that the resulting equations are algebraically complex and therefore in Sect.~\ref{Helmholtz} we will introduce a Helmholtz decomposition to help overcome this complexity. This same section will present a gauge transformation that eventually allows a Fourier representation of the problem that results in determining the Helmholtz functions in terms of Bessel functions with coefficients that depend on the boundary conditions. Section \ref{radtan} presents the explicit solution of the displacement field that results from the Eshelby inner boundary conditions.  This is one of the central results of the present paper.
The displacement field associated with the outer Eshelby problem needs fitting the boundary conditions of the Eshelby problem, and this is accomplished in Sect.~\ref{boundary}. The next section, Sec.~\ref{strain}, focuses on the important calculation of the strain and stress fields that can be computed explicitly once the displacement field is determined. Section \ref{numerics} presents the detailed numerical computations and comparisons between theory and simulations. The paper is summarized with some additional comments in Sec.~\ref{summary}

\section{The Eshelby Problem with odd screening} \label{eshelby}
The main results of the theory of anomalous elasticity is the equation for the displacement field that results when a system in mechanical equilibrium is subjected to a small stress or strain. Here ``small" means that the theory remains linear in the displacement field, (and for the nonlinear extension that is necessary when the system undergoes a large response, the reader is referred to \cite{26KP}). Plastic deformations are taken into account, and it was shown and explained that quadrupolar and dipolar fields bring about screening that modifies the classical
La\'me-Navier equation for the displacement field. Explicitly, the equation that needs tobe solved includes a tensor of screening parameters $\tilde{\Gamma}$, i.e. 
\begin{equation}\label{L1}
	\mu \Delta \vec{d} + (\lambda +\mu)\nabla (\nabla \cdot \vec{d}) +\tilde{\Gamma} \vec{d} =0 \ ,
\end{equation}
where 
\begin{equation}
	\tilde{\Gamma}= \begin{bmatrix}
		\tilde{\kappa}_{e}^{2} & -\tilde{\kappa}_{o}^{2}  \\
		\tilde{\kappa}_{o}^{2} & \tilde{\kappa}_{e}^{2} 
	\end{bmatrix}.
\label{defgam}
\end{equation}
In previous publications it was explained that the `even' screening parameter $\kappa_e$ results from translational symmetry breaking, whereas the `odd' screening parameter is due to chiral and rotational symmetry breaking \cite{24FHKKP,25CSWDM,25FJPP}. In circular geometry it is natural to decompose the displacement field according to e:
\begin{equation}\label{L2}
	\B d = d_{r}(r,\theta)\hat{r} + d_{\theta}(r,\theta)\hat{\theta}.
\end{equation}

Now equation (\ref{L1}) can be decomposed into following coupled differential equations:
\begin{widetext}
\begin{equation}\label{L3}
	\frac{(\lambda +2\mu)}{r^2}\left[ r^2d_{r}^{\prime\prime} + rd_{r}^{\prime}-d_{r}   \right] -\frac{(\lambda +3\mu)}{r^2}\frac{\partial d_{\theta}}{\partial \theta} + \frac{\mu}{r^2}\frac{\partial^2 d_{r}}{\partial \theta^2} + \frac{(\lambda +\mu)}{r} \frac{\partial^2 d_{\theta}}{\partial r\partial \theta} +\tilde{\kappa}_{e}^{2}d_{r} -\tilde{\kappa}_{o}^{2}d_{\theta} =0 ,
\end{equation}
\begin{equation}\label{L4}
	\frac{\mu}{r^2}\left[ r^2d_{\theta}^{\prime\prime} + rd_{\theta}^{\prime} + \frac{\partial^2 d_{\theta}}{\partial\theta^2}-d_{\theta}   \right] + \frac{(\mu +\lambda)}{r^2}\left[ \frac{\partial^2 d_{\theta}}{\partial\theta^2} + r\frac{\partial^2 d_{r}}{\partial r\partial \theta} \right] + \frac{(\lambda +3\mu)}{r^2}\frac{\partial d_{r}}{\partial \theta} + \tilde{\kappa}_{o}^{2}d_{r} + \tilde{\kappa}_{e}^{2}d_{\theta} =0 .
\end{equation}
It is natural to seek a solution in the form 
\begin{align}\label{S2}
	d_{r}(r,\theta) = a_{r}(r)\cos(2\theta) + A_r(r)\sin(2\theta), \nonumber \\
	d_{\theta}(r,\theta) = b_{\theta}(r)\sin(2\theta) + B_\theta(r)\cos(2\theta) .
\end{align}
Substituting and rearranging the equations to be solved are
\begin{align}\label{newodes}
r^2a_{r}^{\prime\prime} + ra_{r}^{\prime} + \frac{[\kappa_{e}^2r^2 -\tilde{\lambda} -6]}{(\tilde{\lambda} +2)}a_{r} -\frac{\kappa_{o}^2r^2}{(\tilde{\lambda}+2)}B_\theta+ 2r\frac{(\tilde{\lambda
	} +1)}{(\tilde{\lambda}+2)}b_{\theta}^{\prime} - 2\frac{(\tilde{\lambda
	} +3)}{(\tilde{\lambda}+2)}b_{\theta} &=0 \nonumber \\ \nonumber \\
 r^2A_r^{\prime\prime} + rA_r^{\prime} + \frac{[\kappa_{e}^2r^2 -\tilde{\lambda} -6]}{(\tilde{\lambda} +2)}A_r -\frac{\kappa_{o}^2r^2}{(\tilde{\lambda}+2)}b_{\theta} - 2r\frac{(\tilde{\lambda} +1)}{(\tilde{\lambda}+2)}B_\theta^{\prime} + 2\frac{(\tilde{\lambda
	} +3)}{(\tilde{\lambda}+2)}B_\theta  & =0  \nonumber \\ \nonumber \\
r^2B_\theta^{\prime\prime} + rB_\theta^{\prime} +[\kappa_{e}^2r^2 -4\tilde{\lambda} -9]B_\theta + \kappa_{o}^{2}r^2a_{r} + 2r(\tilde{\lambda}+1)A_r^{\prime} + 2(\tilde{\lambda}+3)A_r &=0 \nonumber \\ \nonumber \\
r^2b_{\theta}^{\prime\prime} + rb_{\theta}^{\prime} +[\kappa_{e}^2r^2 -4\tilde{\lambda} -9]b_{\theta} + \kappa_{o}^{2}r^2A_r - 2r(\tilde{\lambda}+1)a_{r}^{\prime} - 2(\tilde{\lambda}+3)a_{r} &=0 \ ,
\end{align}
\end{widetext}
where  $\tilde{\lambda} = \lambda/\mu$; $\kappa_e^2 =\tilde{\kappa}^2_e/\mu$, and $\kappa_o^2 =\tilde{\kappa}^2_o/\mu$. 
The four coupled equations for $a_r$, $b_{\theta}$, $A_r$, and $B_\theta$ are algebraically complex. Given this complexity we seek next the elegant approach of Helmholtz decomposition that avoids this complexity. 
 \begin{figure}
 	\centering
 	\includegraphics[width=0.30\textwidth]{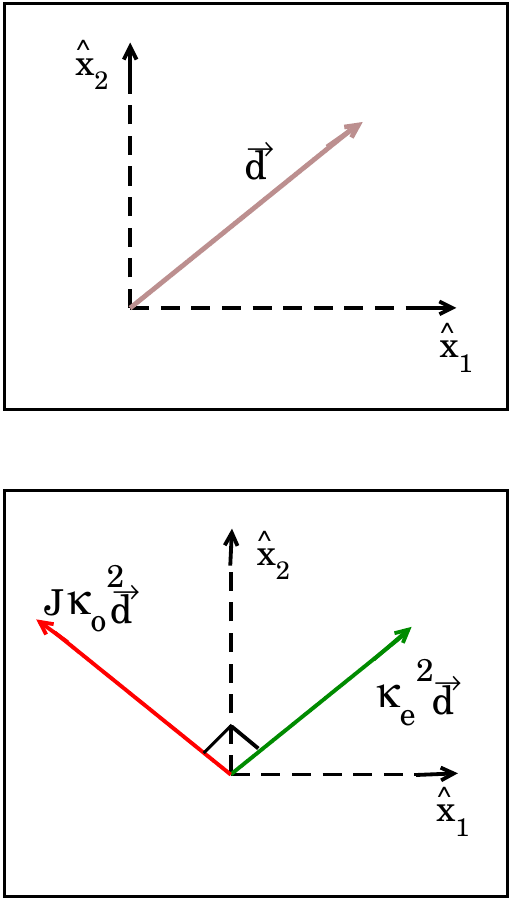}
 	\caption{Top panel: Representation of a two-dimensional displacement vector $\bf d$. Bottom panel: Effect of $\kappa_{e}^2$ and $J\kappa_{o}^2$ on $\bf d$. $\kappa_{e}^2$ make longitudinal transformation whereas $J\kappa_{o}^2$ rotates the $\bf d$ by $90^{\circ}$ making it transverse.}
 	\label{fig:LT}
 \end{figure}
\section{Displacement Field in terms of 2-dimensional Helmholtz decomposition} \label{Helmholtz}
Let us start again with the normalized screening matrix $\Gamma$ and apply it onto the displacement vector $\bf d$. The matrix $\Gamma$ can be thought of superposition of two matrices each acting on $\bf d$ in the following manner:
\begin{align}\label{IJ}
\Gamma {\bf d}&=
\begin{bmatrix}
		\kappa_{e}^{2} & -\kappa_{o}^{2}  \\
		\kappa_{o}^{2} & \kappa_{e}^{2}  
\end{bmatrix} {\bf d} \nonumber
\\
&= \begin{bmatrix}
	\kappa_{e}^{2} & 0  \\
	0 & \kappa_{e}^{2}  
\end{bmatrix} {\bf d} ~+~ 
\begin{bmatrix}
	0 & -\kappa_{o}^{2}  \\
	\kappa_{o}^{2} & 0  
\end{bmatrix}{\bf d} \nonumber
\\
& = \kappa_{e}^{2} \begin{bmatrix}
		1 & 0  \\
		0 & 1  
\end{bmatrix} {\bf d} ~+~ 
\kappa_{o}^{2} \begin{bmatrix}
		0 & - 1 \\
		1 & 0  
\end{bmatrix} {\bf d} \nonumber
\\
& = \kappa_{e}^{2} \mathbb{I} {\bf d}   ~+~ 
	\kappa_{o}^{2} J {\bf d} \nonumber
\\
& = \kappa_{e}^{2} {\bf d}   ~+~  J \kappa_{o}^{2} {\bf d}, 
\end{align}
where $\mathbb{I}$ is the identity matrix and the operator $J$ represents the matrix which rotates the $\bf d$ by $90^{\circ}$. Thus it is very clear that even screening affects the displacement field longitudinally and the odd screening has a transverse effect as shown in a schematic diagram in Fig.\ref{fig:LT}.

\subsection{Helmholtz transformation}
The above discussion shows that the screening matrix has both longitudinal and transverse transformations on the displacement vector $\bf d$; the two transformations being orthogonal to each other. This also suggests that the displacement field $\bf d$ can be written as superposition of two vectors which are orthogonal. Thus to solve for $\bf d$ we use the Helmholtz decomposition which breaks a vector field into its longitudinal and transverse components \cite{58Hel}.

Under appropriate boundary conditions a vector field in \textit{two dimensions} and hence the displacement field $\bf{d}$ can be written in terms of \textit{two scalar potentials $\phi$ and $\psi$} in the following form:
\begin{equation}
	\B d = \B \nabla \phi - \B J \B \nabla \psi \ .
	\label{Helm}
\end{equation}
where
\begin{equation}
\B 	J = \begin{pmatrix}
		0 & -1 \\
		1 & 0
	\end{pmatrix}, ~~~\text{and},~~~ J^2 = -\mathbb{I}.
\end{equation}
The first scalar potential $\phi$ is curl free and represents the longitudinal part, whereas the second scalar potential $\psi$ describes the transverse (divergence free) part.
With the help of these two scalar potentials, the components of $\bf d$ in polar coordinates are given by,
\begin{eqnarray}
 d_r  &= ~ \left( \frac{\partial \phi}{\partial r} + \frac{1}{r} \frac{\partial \psi}{\partial \theta} \right)\hat{r} \nonumber \\ 
 d_{\theta}  &=  ~ \left(\frac{1}{ r} \frac{\partial \phi}{\partial \theta} - \frac{\partial  \psi}{\partial r}  \right)\hat{\theta} \label{E0B}
\end{eqnarray}
Next, using the same definition of $\B J$, Eq.~(\ref{L1}) is rewritten in the form
\begin{equation}
	\nabla^2 \B d + (\tilde{\lambda} + 1)\nabla(\nabla \cdot \B d) + \kappa_e^{2} \B d + J \kappa_o^2 \B d = 0,
\end{equation}
and then using Eq.~(\ref{Helm})
\begin{widetext}
\begin{equation}
\nabla^2 \left(\B \nabla\phi -J\nabla \psi   \right) +(\tilde{\lambda}+1) \nabla \left(\B \nabla^2 \phi\right) + \kappa_e^{2} \left(\B  \nabla\phi -J\nabla \psi   \right) + J \kappa_o^2 \left( \B \nabla\phi -J\B \nabla \psi   \right) = 0
\end{equation}

Simplifying and Rearranging all the terms, we get
\begin{equation}
\B	\nabla \left[  (\tilde{\lambda} + 2)\nabla^2\phi + \kappa_e^2 \phi + \kappa_o^2\psi \right] - J \B \nabla \left[ \nabla^2\psi +   \kappa_e^{2} \psi -  \kappa_o^2 \phi \right] =0. 
\end{equation}
\end{widetext}

Note that the last equation can be written in a more compact form
\begin{equation}
\B	\nabla L - J \B \nabla M =0, 
\end{equation}
with,
\begin{align}
	L &= \alpha \nabla^2\phi + \kappa_e^2 \phi + \kappa_o^2\psi, \nonumber \\
	M &= \nabla^2\psi +   \kappa_e^{2} \psi -  \kappa_o^2 \phi,
	\label{LM}
\end{align}
where $\alpha= (\tilde{\lambda} + 2)$.

\subsection{Gauge Transformation}
Next we want to show that with a suitable choice of gauge transformations of the functions $\phi$ and $\psi$, the two terms $L$ and $M$ of Eq.~(\ref{LM}) can be separately reduced to zero. 

Before proceeding, we show in Appendix \ref{harmonic} that if the equation $\B \nabla A - J \B \nabla B =0$ holds for any two scalar functions $A$, and $B$, then $A$ and $B$ are harmonic; that is, $\nabla^2 A=0$, and $\nabla^2 B=0$. This identity will help us to show that a suitable gauge transformation makes $L=0$ and $M=0$ separately.

So let us define a gauge transformation that keeps the displacement field vector invariant. Let \begin{equation}
	\phi \rightarrow \phi + \chi\ , \quad \psi \rightarrow \psi+ \eta \ .
	\label{gauge}
	\end{equation} 
Then, the displacement field transforms according to
\begin{equation}
\tilde {\B  d} = \left( \B \nabla \phi - J \B \nabla \psi \right) + \left( \B \nabla \chi -J \B 
\nabla \eta  \right)  \ .
\end{equation}
For the displacement field to be invariant we must require that
\begin{equation}
\B \nabla \chi -J \B \nabla \eta = 0
\end{equation}
This immediately gives, according to the proof of Appendix~\ref{harmonic}
\begin{equation}
\nabla^2\chi =0, ~~~\text{and}~~~ \nabla^2\eta= 0. 
\end{equation}
Presently we use of this fact to show that under a suitable gauge transformations, $L$ and $M$ can be reduced to zero {\bf separately}, keeping the displacement field invariant. Under the gauge transformation Eq.~(\ref{gauge}),
\begin{align}
\tilde L &= \alpha \nabla^2\phi + \kappa_e^2 \phi + \kappa_o^2 \psi + \alpha \nabla^2\chi + \kappa_e^2 \chi + \kappa_o^2 \eta \nonumber \\
 &= L + \kappa_e^2 \chi + \kappa_o^2 \eta 
\end{align}
Similarly,
\begin{align}
\tilde	M &=   \nabla^2\psi + \kappa_e^2 \psi - \kappa_o^2 \phi +  \nabla^2\chi + \kappa_e^2 \eta - \kappa_o^2 \chi \nonumber \\
	&= M + \kappa_e^2 \eta - \kappa_o^2 \chi 
\end{align}
We can choose $\eta$ and $\chi$ such that $\tilde L=0$ and $\tilde M =0$: 
\begin{align}
\kappa_e^2 \chi + \kappa_o^2 \eta &= -L ,\nonumber \\
\kappa_e^2 \eta - \kappa_o^2 \chi &= -M.
\end{align}

Solving for $\chi$ and $\eta$ we find 
\begin{equation}
\chi = -\frac{L\kappa_e^2 - M\kappa_o^2}{\kappa_e^4+\kappa_o^4}; ~~~~~ \eta=  -\frac{M\kappa_e^2 + L \kappa_o^2}{\kappa_e^4+\kappa_o^4}.
\end{equation}
Thus there always exists a suitable gauge transformation that leads to $\tilde L=0$ and $\tilde M=0$. But we could choose from the start the gauge transformation Eq.(\ref{gauge}). Therefore, under the proper gauge transformations 
\begin{align}
\alpha \nabla^2\phi + \kappa_e^2 \phi + \kappa_o^2 \psi &=0, \nonumber \\
 \nabla^2\psi + \kappa_e^2 \psi - \kappa_o^2 \phi &= 0.
\end{align}
These two partial differential equations represent the coupled Helmholtz equations for two scalar potentials $\phi$ and $\psi$. Note that $\phi$ and $\psi$ are real functions. 

\subsection{Fourier Decomposition}

Since the coefficients of the coupled PDE do not depend on the angular positions $\theta$, each angular Fourier mode $n$ is independent from the other. Thus for a general mode $m$, we can represent the scalar potentials $\phi$ and $\psi$ as they emanate from more general complex potentials, i.e.
\begin{align}
\phi(r,\theta) &= \mathfrak{Re} \left[ \Phi(r)~e^{i m \theta} \right], \nonumber \\
\psi(r,\theta) &= \mathfrak{Re} \left[ \Psi(r)~e^{i m \theta} \right] .
\end{align}
Note that, $\Phi(r) = \Phi_{R} + i \Phi_{I}$, where $\Phi_R$ and $\Phi_I$ are real and imaginary components respectively. Similarly $\Psi = \Psi_{R} + i \Psi_{I}$. Since we are interested in the Eshelby problem, we anticipate an elliptic deformation at the inner boundary. The boundary condition that corresponds to this will involve the functions $cos2\theta$ and $\sin 2\theta$, and see below for more details (appendix \ref{rotated}). Thus for the elliptic boundary we need to take $m=2$.
\begin{eqnarray}
\phi(r,\theta) &= \mathfrak{Re} \left[ \Phi(r)~e^{i 2 \theta} \right] \nonumber \\
\psi(r,\theta) &= \mathfrak{Re} \left[ \Psi(r)~e^{i 2 \theta} \right] \ . \label{Fourdec}
\end{eqnarray}
Note that the differential operator $\nabla^2$ in polar coordinates, $\nabla^2 = \frac{\partial^2}{\partial r^2} + \frac{1}{r}\frac{\partial }{\partial r} + \frac{1}{r^2} \frac{\partial^2}{\partial \theta^2}$, therefore
\begin{equation}
\nabla^2 \left[ \Phi(r)~e^{i 2 \theta} \right] = \left[ \Phi^{\prime\prime} + \frac{1}{r}\Phi^{\prime} - \frac{4}{r^2} \right]e^{i 2 \theta}
\end{equation}

For simplicity, we define 
\begin{equation}
\L_2 =   \Phi^{\prime\prime} + \frac{1}{r}\Phi^{\prime} - \frac{4}{r^2}.
\end{equation}
 Thus, the two coupled Helmholtz equations in terms of complex potentials $\Phi$ and $\Psi$ take the following form:
\begin{eqnarray}
	\left( \alpha \L_{2} + \kappa_e^2 \right) \Phi + \kappa_o^2 \Psi &= 0 \ , \nonumber \\
	\left( \L_2 + \kappa_e^2\right) \Psi -\kappa_o^2\Phi &= 0 \ . \label{simple}
\end{eqnarray}

Now we can show that the above two coupled Helmholtz equations can be solved by a common Radial function. 
To this aim we multiply by $(\L_2 +\kappa_e^2)$ the first equation of (\ref{simple}), and by $\kappa_o^2$ the second equation of (\ref{simple}). Then subtract the second from the first:
\begin{equation}\label{phi}
 	\left[ \left( \alpha \L_{2} + \kappa_e^2 \right) \left( \L_2 + \kappa_e^2\right)  + \kappa_o^4 \right]\Phi =0.
\end{equation}
Similarly,  we multiply by $(\alpha L_2 +\kappa_e^2)$ the second equation of (\ref{simple}), and by $\kappa_o^2$ the first equation of (\ref{simple}). Then sum them together:
\begin{equation}\label{psi}
 	\left[ \left( \alpha \L_{2} + \kappa_e^2 \right) \left( \L_2 + \kappa_e^2\right)  + \kappa_o^4 \right]\Psi =0.
\end{equation}
From Eqs. (\ref{phi}) and (\ref{psi}) it is obvious that $\Phi$ and $\Psi$ share the same basis function, because they are solution of the same differential operator. Thus the common function that solves the differential operator in bracket in the above equations uniquely determines $\Phi$ and $\Psi$. 

The operator in the bracket can be expanded, and both $\Phi$ and $\Psi$ solve the same equation,
\begin{align}\label{common}
\left[ 	\alpha \L_2^2 + \kappa_e^2(\alpha+1)\L_2 + \kappa_e^4 + \kappa_o^4\right] \Phi &= 0 \ , \nonumber \\
\left[ 	\alpha \L_2^2 + \kappa_e^2(\alpha+1)\L_2 + \kappa_e^4 + \kappa_o^4\right] \Psi &= 0 \ .
\end{align}

To solve the above equations, we look for the factorization of the operator in the square brackets , such as
\begin{equation}\label{factor}
\alpha \left( \L_2 + \kappa_A^2\right) \left( \L_2 + \kappa_B^2\right),
\end{equation}
which when expanded gives,
\begin{equation}\label{expand}
	\alpha \left[ \L_2^2 +  (\kappa_A^2 + \kappa_B^2)\L_2   + \kappa_A^2\kappa_B^2  \right] 
\end{equation}
Comparing this equation with 
\begin{equation}
\alpha \L_2^2 + \kappa_e^2(\alpha+1)\L_2 + \kappa_e^4 + \kappa_o^4,
\end{equation}
we obtain,
\begin{align}\label{roots}
\alpha 	(\kappa_A^2 + \kappa_B^2) = \kappa_e^2(\alpha +1) \nonumber \\
\alpha \kappa_A^2\kappa_B^2 = \kappa_e^4 + \kappa_o^4 \ .
\end{align}
The conclusion is that $\kappa_A^2$ and $\kappa_B^2$ are the roots of following equation,
\begin{equation}
	\alpha \kappa^4 - \kappa_e^2(\alpha +1)\kappa^2 + \kappa_e^4 + \kappa_o^4 =0 \ ,
\end{equation}
which are given by,
\begin{equation}
	\kappa_A^2, \kappa_B^2 = \frac{\kappa_e^2(\alpha +1) \pm \sqrt{\kappa_e^4(\alpha-1)^2 - 4\alpha \kappa_o^4 }}{2\alpha} .
\end{equation}
Thus, with equations (\ref{common}, and \ref{factor}), the fourth order equation becomes
\begin{equation}\label{Dphi}
\alpha \left( \L_2 + \kappa_A^2\right) \left( \L_2 + \kappa_B^2\right)\Phi(r) =0.
\end{equation}
and 
\begin{equation}\label{Dpsi}
\alpha \left( \L_2 + \kappa_A^2\right) \left( \L_2 + \kappa_B^2\right)\Psi(r) =0.
\end{equation}
From equation (\ref{Dphi}) it is clear that complete solution is a sum of two functions each of which solve the $\kappa_A^2$ and $\kappa_B^2$ parts separately. Thus $\Phi = \Phi_{A} + \Phi_{B}$, and 
\begin{eqnarray}
\left( \L_2 + \kappa_A^2\right) \Phi_{A} &= 0 \nonumber \\
\left( \L_2 + \kappa_B^2\right) \Phi_{B} &= 0 \ . \label{breakPhi}
\end{eqnarray}
The first part with $\Phi_{A}$ of the above equation (\ref{breakPhi}) in expanded form is 
\begin{equation}
\Phi_{A}^{\prime\prime} + \frac{1}{r}\Phi_{A}^{\prime} + \left( \kappa_A^2 - \frac{4}{r^2}\right) )\Phi_{A} =0
\end{equation}
This is a second-order differential equation whose solution is given by the Bessel functions of second kind, thus we have
\begin{equation}
\Phi_{A} = C_A J_2(\kappa_A r) + D_A Y_2(\kappa_A r) .
\end{equation}
Similarly, solving for the $\Phi_B$ part, we obtain
\begin{equation}
\Phi_{B} = C_B J_2(\kappa_B r) + D_B Y_2(\kappa_B r) .
\end{equation}
Thus the complete solution for $\Phi= \Phi_{A} + \Phi_{B}$ is,
\begin{widetext}
\begin{equation}\label{PHIR}
\Phi(r) = C_A J_2(\kappa_A r) + D_A Y_2(\kappa_A r) + C_B J_2(\kappa_B r) + D_B Y_2(\kappa_B r).
\end{equation}
\end{widetext}

Now let us solve for the function $\Psi$. To this purpose we use the following coupled equations from equation (\ref{simple}),
\begin{equation}
	\left( \alpha \L_{2} + \kappa_e^2 \right) \Phi + \kappa_o^2 \Psi = 0,
\end{equation}
Thus, $\Psi$ is given by
\begin{equation}
 \Psi = - \frac{1}{\kappa_o^2}\left( \alpha \L_{2} + \kappa_e^2 \right) \Phi 
\end{equation}
Since $\Phi = \Phi_{A} + \Phi_{B}$, therefore $\Psi$ also breaks into $\Psi = \Psi_A + \Psi_B$, where the components $\Psi_A$ and $\Psi_B$ are 
\begin{align}\label{breakPsi}
	\Psi_A &= - \frac{1}{\kappa_o^2}\left( \alpha \L_{2} + \kappa_e^2 \right) \Phi_A \nonumber \\
	\Psi_B &= - \frac{1}{\kappa_o^2}\left( \alpha \L_{2} + \kappa_e^2 \right) \Phi_B
\end{align}

From equation (\ref{breakPhi}), we have $\L_2 \Phi_{A} = -\kappa_A^2\Phi_{A}$; and $\L_2 \Phi_{B} = -\kappa_B^2\Phi_{B}$, thus the components of $\Psi$ are
\begin{align}\label{breakPsi}
	\Psi_A &= - \frac{1}{\kappa_o^2}\left( -\alpha \kappa_A^2 + \kappa_e^2 \right) \Phi_A \nonumber \\
	\Psi_B &= - \frac{1}{\kappa_o^2}\left( -\alpha \kappa_B^2 + \kappa_e^2 \right) \Phi_B
\end{align}
Let
\begin{align}
	l_A &= - \frac{1}{\kappa_o^2}\left( -\alpha \kappa_A^2 + \kappa_e^2 \right) = \frac{\alpha\kappa_{A}^2 -\kappa_{e}^2}{\kappa_{o}^2} \nonumber \\
	l_B &=  -\frac{1}{\kappa_o^2}\left( -\alpha \kappa_B^2 + \kappa_e^2 \right) = \frac{\alpha\kappa_{B}^2 -\kappa_{e}^2}{\kappa_{o}^2}
\end{align}
Thus finally, 
\begin{align}
	\Psi_A &= l_A \Phi_A = l_A \left[ C_A J_2(\kappa_A r) +   D_A Y_2(\kappa_A r)\right) \nonumber \\
	\Psi_B &= l_B \Phi_B = l_B \left[ C_B J_2(\kappa_B r) +   D_B Y_2(\kappa_B r)\right] 
\end{align}
Thus $\Psi= \Psi_A + \Psi_B$, 
\begin{widetext}
\begin{equation}\label{PSIR}
\Psi(r) = l_A \left[ C_A J_2(\kappa_A r) +   D_A Y_2(\kappa_A r)\right]  +  l_B \left[ C_B J_2(\kappa_B r) +   D_B Y_2(\kappa_B r)\right]  \ .
\end{equation}
\end{widetext}
The solution presented above applies to the generic case of nonzero odd screening and for distinct characteristic roots. For completeness, Appendix~\ref{limiting_cases} discusses the limiting case of vanishing odd screening, and the construction of generalized Bessel solutions when the characteristic roots coincide. These extra cases do not require a modification of the governing equations but may require a different representation of their solutions.

\section{Radial and tangential displacement fields}\label{radtan}
At this point we are in a position to derive one of the central results of this paper, i.e. the radial and tangential components of the displacement field. Using Eqs.~(\ref{E0B}) and (\ref{Fourdec}) the components of the displacement field vector are,
\begin{equation}\label{radD}
d_r(r,\theta)=\Re\left[\left(\Phi'(r)+\frac{2i}{r}\Psi(r)\right)e^{2i\theta}\right],
\end{equation}
and
\begin{equation}\label{tangD}	d_\theta(r,\theta)=\Re\left[\left(\frac{2i}{r}\Phi(r)-\Psi'(r)\right)e^{2i\theta}\right].
\end{equation}
Here $\Phi(r)^{\prime}$ and $\Psi(r)^{\prime}$ represent the differentiation of $(\Phi)(r)$ and $(\Psi)(r)$ with respect to $r$.
\subsection{Radial displacement field}
\begin{align}
	d_r(r,\theta)
	=\Re\Bigg\{
	\Bigg[
	&C_A
	\left(
	\kappa_A J_2'(\kappa_A r)
	+
	 \frac{2il_A}{r}J_2(\kappa_A r)
	\right) \nonumber	\\
	+{}&
	D_A
	\left(
	\kappa_A Y_2'(\kappa_A r)
	+
	\frac{2il_A}{r}Y_2(\kappa_A r)
	\right) \nonumber	\\
	+{}&
	C_B
	\left(
	\kappa_B J_2'(\kappa_B r)
	+
	\frac{2il_B}{r}J_2(\kappa_B r)
	\right) \nonumber	\\
	+{}&
	D_B
	\left(
	\kappa_B Y_2'(\kappa_B r)
	+
	\frac{2il_B}{r}Y_2(\kappa_B r)
	\right)
	\Bigg]e^{2i\theta}
	\Bigg\}.
\end{align}

\subsection{Tangential displacement field}

\begin{align}
	d_\theta(r,\theta)
	=\Re\Bigg\{
	\Bigg[
	&C_A
	\left(
	\frac{2i}{r}J_2(\kappa_A r)
	-
	l_A  \kappa_A J_2'(\kappa_A r)
	\right) \nonumber	\\
	+{}&
	D_A
	\left(
	\frac{2i}{r}Y_2(\kappa_A r)
	-
	l_A \kappa_A Y_2'(\kappa_A r)
	\right) \nonumber	\\
	+{}&
	C_B
	\left(
	\frac{2i}{r}J_2(\kappa_B r)
	-
	l_B \kappa_B J_2'(\kappa_B r)
	\right)\nonumber	\\
	+{}&
	D_B
	\left(
	\frac{2i}{r}Y_2(\kappa_B r)
	-
	l_B \kappa_B Y_2'(\kappa_B r)
	\right)
	\Bigg]e^{2i\theta}
	\Bigg\}.
\end{align}

In these equations,

\begin{equation}
	J_2'(z)=\frac{dJ_2(z)}{dz}, \qquad Y_2'(z)=\frac{dY_2(z)}{dz},
\end{equation}
Thus, 
\begin{equation}
	\frac{d}{dr}J_2(\kappa_A r) = \kappa_A J_2'(\kappa_A r), ~~\text{and}~~ \frac{d}{dr}Y_2(\kappa_B r) = \kappa_B Y_2'(\kappa_B r).
\end{equation}

\section{Boundary conditions}
\label{boundary}
In this section we write the boundary conditions at the inner and outer circular boundaries for the Eshelby problem. The inner and outer radii are represented by $r_{in}$ and $r_{out}$ respectively.
Suppose the inner boundary is deformed elliptically with orientation $(\theta_0)$:

\begin{equation}
d_r(r_{in},\theta)=\epsilon r_{in} \cos\left[2(\theta-\theta_0)\right],
\end{equation}
\begin{equation}
	d_\theta(r_{in},\theta)=0.
\end{equation}
At the fixed outer boundary,
\begin{equation}
d_r(r_{out},\theta)=0, \qquad d_\theta(r_{out},\theta)=0.
\end{equation}
Since
\begin{equation}
\epsilon r_{in} \cos\left[2(\theta-\theta_0)\right]= \Re\left[ \epsilon r_{in} e^{2i(\theta-\theta_0)} \right],
\end{equation}
Let us define only the boundary amplitude by

\begin{equation}
G_0=\epsilon r_{in} e^{-2i\theta_0}.
\end{equation}
For an ellipse aligned with the $(x)$-axis $\theta_0=0$, and $G_0 =\epsilon r_{in} $.

In Appendix \ref{Boundary} we detail the solution of the boundary conditions. The coefficients that appear in the displacement field are obtained, 
after solving Eq. (\ref{BM}). Then we substitute the four coefficients directly into Eqs. (\ref{radD}) and (\ref{tangD}). Those two equations are the complete physical solution for $(d_r(r,\theta))$ and $(d_\theta(r,\theta))$. The compact form is therefore:

	\begin{align}
		d_r
		&=
		\Re\left[
		\left(
		\Phi'+\frac{2i}{r}\Psi
		\right)e^{2i\theta}
		\right],\\[1mm]
		d_\theta
		&=
		\Re\left[
		\left(
		\frac{2i}{r}\Phi-\Psi'
		\right)e^{2i\theta}
		\right],
	\end{align}
with $(\Phi,\Psi)$ given by Eqs. (\ref{PHIR} and \ref{PSIR}), and the coefficients obtained from the explicit matrix above.

Note that, $\Phi^{\prime}(r)$ and $\Psi^{\prime}(r)$ represent the differentiation of $\Phi(r)$ and $\Psi(r)$ with respect to $r$. Now we define two complex radial amplitudes
\begin{equation}
	X(r)=\Phi'(r)+\frac{2i}{r}\Psi(r),
\end{equation}
\begin{equation}
	Y(r)=\frac{2i}{r}\Phi(r)-\Psi'(r).
\end{equation}
Then the real physical displacement components are, 
\begin{equation}
	d_r(r,\theta) =	\Re\!\left[X(r)e^{2i\theta}\right],
\end{equation}
\begin{equation}
	d_\theta(r,\theta) = \Re\!\left[Y(r)e^{2i\theta}\right].
\end{equation}
The $X$ and $Y$ can be decomposed in Fourier representation: 
\begin{equation}
	X=X_R+iX_I,
	\qquad
	Y=Y_R+iY_I.
\end{equation}
Since
\begin{equation}
	\Re[(X_R+iX_I)e^{2i\theta}]
	=
	X_R\cos2\theta-X_I\sin2\theta,
\end{equation}
and 
\begin{equation}
	\Re[(Y_R+iY_I)e^{2i\theta}] =
	-Y_I\sin2\theta + Y_R\cos2\theta,
\end{equation}
the displacement field components are decomposed as
\begin{equation}
	d_r=a(r)\cos2\theta+A_r(r)\sin2\theta,
\end{equation}
\begin{equation}
	d_\theta=b(r)\sin2\theta+B_\theta(r)\cos2\theta,
\end{equation}
where
\begin{equation}
	a=\Re X,\qquad A_r=-\Im X,
\end{equation}
\begin{equation}
	b=-\Im Y, \qquad B_\theta=\Re Y.
\end{equation}
We note that the odd screening generally makes both $A$ and $B$ nonzero.

To finish this section, we note that the ellipse that we had in mind so far is aligned such that its major axis is parallel to the $x$ axis, and the solution of the displacement field reflects 
this choice. In general this is not always the case, and for completeness with present in Appendix~\ref{rotated} the general case of an ellipse with an arbitrary angle between the major axis and the $x$ axis.

\section{Strain components and their representation in terms of complex strain amplitudes}
\label{strain}

Having obtained the displacement field of the outer Eshelby problem, we are now ready to compute the associated strain and stress fields. As said above, this is important to assess how stress is distributed in space when an Eshelby-like plastic event is taking place. 

We use the results of the last section for infinitesimal strain in polar coordinates: by,
\begin{equation}
	\epsilon_{rr} =\frac{\partial d_r}{\partial r} =\Re\!\left[ X^{\prime}(r)e^{2i\theta}\right] ,
\end{equation}
\begin{equation}
	\epsilon_{\theta\theta} = \frac{d_r}{r} + \frac1r\frac{\partial d_\theta}{\partial\theta} = \Re \left[  \left( \frac{X+2i Y}{r} \right) e^{2i\theta}\right],
\end{equation}
\begin{align}
	\epsilon_{r\theta} &=\frac12 \left[\frac1r\frac{\partial d_r}{\partial\theta} + \frac{\partial d_\theta}{\partial r}	- \frac{d_\theta}{r} \right] \nonumber  \\
	&=  \Re\left[ \frac12 \left( Y' - \frac{Y}{r}+ \frac{2i X}{r}\right) e^{2i\theta} \right]
\end{align}

These results can be written more elegantly in compact notations with the help of complex strain amplitudes. Let, 
\begin{equation}
	\epsilon_{rr}=\Re\!\left[E_{rr}(r)e^{2i\theta}\right],
\end{equation}
\begin{equation}
	\epsilon_{\theta\theta}=\Re\!\left[E_{\theta\theta}(r)e^{2i\theta}\right],
\end{equation}
\begin{equation}
	\epsilon_{r\theta} =\Re\!\left[E_{r\theta}(r)e^{2i\theta}\right].
\end{equation}
Then
\begin{equation}
	E_{rr}=X',
\end{equation}
\begin{equation}
	E_{\theta\theta}=\frac{X+2iY}{r},
\end{equation}
\begin{equation}
	E_{r\theta}=\frac12\left(Y'-\frac Yr+\frac{2iX}{r}\right).
\end{equation}

Shortly,
\begin{equation}
	\epsilon_{ij}(r,\theta) = \Re\!\left[E_{ij}(r)e^{2i\theta}\right],
\end{equation}

\subsection{Strains in terms of $\Phi$ and $\Psi$}
Substituting
\begin{equation}
	X=\Phi'+\frac{2i}{r}\Psi, \qquad Y=\frac{2i}{r}\Phi-\Psi',
\end{equation}
gives explicit expressions. Radial strain
\begin{equation}
	E_{rr}=\Phi''+2i\left(\frac{\Psi'}{r}-\frac{\Psi}{r^2}\right).
\end{equation}
Therefore,
\begin{equation}
	\epsilon_{rr}=\Re\!\left[
	\left(\Phi''+2i\left(\frac{\Psi'}r-\frac{\Psi}{r^2}\right)\right)e^{2i\theta}\right].
\end{equation}
Tangential strain
\begin{equation}
	E_{\theta\theta}
	=	\frac{\Phi'}r-\frac{4\Phi}{r^2}+2i\left(\frac{\Psi}{r^2}-\frac{\Psi'}r\right).
\end{equation}
Thus,
\begin{equation}
	\epsilon_{\theta\theta}=\Re\!\left[\left(\frac{\Phi'}r-\frac{4\Phi}{r^2}+2i\left(\frac{\Psi}{r^2}-\frac{\Psi'}r\right)\right)e^{2i\theta}\right].
\end{equation}
Polar shear strain
\begin{equation}
	E_{r\theta}=2i\left(\frac{\Phi'}r-\frac{\Phi}{r^2}\right)-\frac12\Psi''
	+\frac{\Psi'}{2r}-\frac{2\Psi}{r^2}.
\end{equation}
Hence,
\begin{equation}
	\epsilon_{r\theta}=\Re\!\left[\left(2i\left(\frac{\Phi'}r-\frac{\Phi}{r^2}\right)-\frac12\Psi''
	+\frac{\Psi'}{2r}-\frac{2\Psi}{r^2}\right)e^{2i\theta}\right].
\end{equation}

\subsection{Sum of radial and tangential strains (trace of the strain tensor matrix)}
The trace of the strain matrix is
\begin{equation}
	\epsilon_{kk}=\epsilon_{rr}+\epsilon_{\theta\theta}.
\end{equation}
At the amplitude level,
\begin{equation}
	E_{rr}+E_{\theta\theta}=\Phi''+\frac{\Phi'}r-\frac{4\Phi}{r^2}.
\end{equation}
Thus,
\begin{equation}
	E_{rr}+E_{\theta\theta}=\L_2\Phi.
\end{equation}
Where,
\begin{equation}
	\L_2 = \nabla^2 - \frac{1}{r}\nabla - \frac{4}{r^2}
\end{equation}
Therefore,
\begin{equation}
	\epsilon_{rr}+\epsilon_{\theta\theta}=\Re\!\left[\L_2\Phi\,e^{2i\theta}\right].
\end{equation}
The $\Psi$-terms cancel exactly. This also provides a check for
\begin{equation}
	\nabla\cdot\mathbf d=\nabla^2\phi.
\end{equation}
For the two branches,
\begin{equation}
	\L_2\Phi_A=-\kappa_A^2\Phi_A,
\end{equation}
\begin{equation}
	\L_2\Phi_B=-\kappa_B^2\Phi_B,
\end{equation}
so
\begin{equation}
	\epsilon_{kk}=-\Re\!\left[\left(\kappa_A^2\Phi_A+\kappa_B^2\Phi_B\right)e^{2i\theta}\right].
\end{equation}

\subsection{Stress constitutive relations and its components}
We note that the screening term changes the force balance, but it does not by itself changes the local isotropic elastic constitutive law. Thus we start with 
\begin{equation}
	\sigma_{ij}=\lambda\epsilon_{kk}\delta_{ij}+2\mu\epsilon_{ij}.
\end{equation}
Therefore,
\begin{equation}
	\sigma_{rr}=\lambda(\epsilon_{rr}+\epsilon_{\theta\theta})+2\mu\epsilon_{rr},
\end{equation}
\begin{equation}
	\sigma_{\theta\theta}=\lambda(\epsilon_{rr}+\epsilon_{\theta\theta})+2\mu\epsilon_{\theta\theta},
\end{equation}
\begin{equation}
	\sigma_{r\theta}=2\mu\epsilon_{r\theta}.
\end{equation}
We note that the screening force enters in equilibrium equation as follows
\begin{equation}
	\nabla\cdot\boldsymbol\sigma + \widetilde\kappa_e^2\mathbf d
	+ \widetilde\kappa_o^2J\mathbf d =0.
\end{equation}

\subsection{Complex stress amplitudes}
Let us write
\begin{equation}
	\sigma_{ij} = \Re\!\left[ S_{ij}(r)e^{2i\theta}	\right].
\end{equation}

Then
\begin{equation}
	S_{rr}	=	\lambda L_2\Phi	+	2\mu E_{rr},
\end{equation}

\begin{equation}
	S_{\theta\theta} =	\lambda L_2\Phi	+	2\mu E_{\theta\theta},
\end{equation}
\begin{equation}
	S_{r\theta}	=	2\mu E_{r\theta}.
\end{equation}
Explicitly we can write this,
\begin{equation}
	S_{rr}	=	\lambda	\left(	\Phi''	+\frac{\Phi'}r	-\frac{4\Phi}{r^2}	\right)
	+	2\mu	\left[	\Phi''	+	2i	\left(	\frac{\Psi'}r-\frac{\Psi}{r^2}	\right)	\right].
\end{equation}

Also,
\begin{align}
	S_{\theta\theta}
	={}&
	\lambda
	\left(
	\Phi''
	+\frac{\Phi'}r-\frac{4\Phi}{r^2}
	\right)
	\\
	&+
	2\mu
	\left[
	\frac{\Phi'}r-\frac{4\Phi}{r^2}
	+
	2i
	\left(
	\frac{\Psi}{r^2}-\frac{\Psi'}r
	\right)
	\right].
\end{align}
and finally,
\begin{equation}
	S_{r\theta} = 2\mu
	\left[
	2i
	\left(
	\frac{\Phi'}r-\frac{\Phi}{r^2}
	\right)
	-\frac12\Psi''
	+\frac{\Psi'}{2r}
	-\frac{2\Psi}{r^2}
	\right].
\end{equation}
The actual stresses are their real parts multiplied by $e^{2i\theta}$.

For explicit representation of the stresses we can write the strain coefficients as follows
\begin{equation}
	\epsilon_{rr}=\xi_{rr}^{c}\cos2\theta+\xi_{rr}^{s}\sin2\theta,
\end{equation}
\begin{equation}
	\epsilon_{\theta\theta}=\xi_{\theta\theta}^{c}\cos2\theta+\xi_{\theta\theta}^{s}\sin2\theta,
\end{equation}
where the superscripts $c$ and $s$ in $\xi_{rr}^{c}$ and $\xi_{rr}^{s}$ correspond to $\cos$ and $\sin$ respectively. This will also hold true in the symbols used below. Using the real strain coefficients, we obtain
\begin{equation}
	\sigma_{rr} = \Sigma_{rr}^{c}\cos2\theta + \Sigma_{rr}^{s}\sin2\theta,
\end{equation}
where
\begin{equation}
	\Sigma_{rr}^{c} = (\lambda+2\mu)\xi_{rr}^{c}	+	\lambda\xi_{\theta\theta}^{c},
\end{equation}
\begin{equation}
	\Sigma_{rr}^{s} = (\lambda+2\mu)\xi_{rr}^{s} + \lambda\xi_{\theta\theta}^{s}.
\end{equation}
Similarly,
\begin{equation}
	\sigma_{\theta\theta} = \Sigma_{\theta\theta}^{c}\cos2\theta +	\Sigma_{\theta\theta}^{s}\sin2\theta,
\end{equation}
with
\begin{equation}
	\Sigma_{\theta\theta}^{c} = \lambda\xi_{rr}^{c} +
	(\lambda+2\mu)\xi_{\theta\theta}^{c},
\end{equation}
\begin{equation}
	\Sigma_{\theta\theta}^{s}=\lambda\xi_{rr}^{s}+(\lambda+2\mu)\xi_{\theta\theta}^{s}.
\end{equation}
And
\begin{equation}
	\sigma_{r\theta} = 2\mu\xi_{r\theta}^{c}\cos2\theta + 2\mu\xi_{r\theta}^{s}\sin2\theta.
\end{equation}

\subsection{Cartesian shear strain $(\epsilon_{xy})$}
The polar-to-Cartesian transformation is given by
\begin{equation}
	\epsilon_{xy} =
	\frac12(\epsilon_{rr}-\epsilon_{\theta\theta})\sin2\theta	+ \epsilon_{r\theta}\cos2\theta.
\end{equation}
Let us define
\begin{equation}
	D_c=\xi_{rr}^{c}-\xi_{\theta\theta}^{c},
\end{equation}
\begin{equation}
	D_s	=\xi_{rr}^{s}-\xi_{\theta\theta}^{s}.
\end{equation}
Then
\begin{equation}
	\epsilon_{rr}-\epsilon_{\theta\theta}=D_c\cos2\theta+D_s\sin2\theta.
\end{equation}
Also,
\begin{equation}
	\epsilon_{r\theta}=\xi_{r\theta}^{c}\cos2\theta+\xi_{r\theta}^{s}\sin2\theta.
\end{equation}
With these equations we obtain
\begin{equation}
	\epsilon_{xy}=u_0(r)+u_c(r)\cos4\theta+u_s(r)\sin4\theta,
\end{equation}
where
\begin{equation}
	u_0=\frac14D_s+\frac12\xi_{r\theta}^{c},
\end{equation}
\begin{equation}
	u_c = -\frac14D_s + \frac12\xi_{r\theta}^{c},
\end{equation}
\begin{equation}
	u_s=\frac14D_c+\frac12\xi_{r\theta}^{s}.
\end{equation}

For the special case of the even screening, the odd terms vanish,
\begin{equation}
	A_o(r)=B_o(r)=0.
\end{equation}
Therefore,
\begin{equation}
	\xi_{rr}^{s}=0,\qquad\xi_{\theta\theta}^{s}=0,\qquad\xi_{r\theta}^{c}=0.
\end{equation}
Hence,
\begin{equation}
	D_s=0,\qquad u_0=0,\qquad u_c=0.
\end{equation}
Only
\begin{equation}
	u_s=\frac14\left(\xi_{rr}^{c}-\xi_{\theta\theta}^{c}\right)+\frac12\xi_{r\theta}^{s}
\end{equation}
survives. Thus,
\begin{equation}
	\epsilon_{xy} =
	\left[	\frac12\xi_{r\theta}	+	\frac14	(\xi_{rr}-\xi_{\theta\theta})	\right]\sin4\theta.
\end{equation}
Equivalently,
\begin{equation}
	\epsilon_{xy}	=	\frac12	\left[	\xi_{r\theta}	+	\frac12	(\xi_{rr}-\xi_{\theta\theta})
	\right]\sin4\theta.
\end{equation}

\subsection{Cartesian shear stress}
The stress components are
\begin{equation}
	\sigma_{ij} = \lambda \epsilon_{kk}\delta_{ij} + 2\mu\epsilon_{ij}.
\end{equation}
For $i=x, j=y$, 
\begin{equation}
	\sigma_{xy} = \lambda \epsilon_{kk}\delta_{xy} + 2\mu\epsilon_{xy}.
\end{equation}
Since $\delta_{xy}=0,$ the Cartesian shear stress is
\begin{equation}
	\sigma_{xy}=2\mu\epsilon_{xy}.
\end{equation}
Therefore,
\begin{equation}
	\sigma_{xy}=2\mu\left[u_0(r)+u_c(r)\cos4\theta+u_s(r)\sin4\theta\right].
\end{equation}

\subsection{Phase shift in Odd screening}

If we had only even screening such that all the odd screening terms being zero, then the strain components have following angular structure,
\begin{equation}
	\epsilon_{rr},\epsilon_{\theta\theta}\propto\cos2\theta,
\end{equation}
\begin{equation}
	\epsilon_{r\theta}\propto\sin2\theta.
\end{equation}
Consequently,
\begin{equation}
	\epsilon_{xy}\propto\sin4\theta.
\end{equation}
But the odd screening introduces the crossed terms such as 
\begin{equation}
	\sin2\theta \quad\text{in}\quad \epsilon_{rr},\epsilon_{\theta\theta},
\end{equation}
and
\begin{equation}
	\cos2\theta
	\quad\text{in}\quad
	\epsilon_{r\theta}.
\end{equation}
Therefore the Cartesian shear becomes
\begin{equation}
	\epsilon_{xy}=u_0+u_c\cos4\theta+u_s\sin4\theta.
\end{equation}
Thus the fourfold part can be written as
\begin{equation}
	u_c\cos4\theta+u_s\sin4\theta = U_4(r)\cos[4\theta-\delta(r)],
\end{equation}
where
\begin{equation}
	U_4(r)=\sqrt{u_c^2+u_s^2},
\end{equation}
and
\begin{equation}
	\delta(r)=\operatorname{atan2}(u_s,u_c).
\end{equation}
Thus the quadrupolar orientation is
\begin{equation}
	\theta_{\rm quad}(r)=\frac{\delta(r)}4.
\end{equation}
Odd screening therefore does not only modify the radial decay. It can rotate the quadrupolar strain pattern, and that rotation may depend on $r$.

\subsection{Detailed Computation}
We will first derive the strain produced by considering any of the generic Bessel terms
\begin{equation}
	C_AJ_2(\kappa_A r),\quad	D_AY_2(\kappa_A r),\quad C_BJ_2(\kappa_B r),\quad D_BY_2(\kappa_B r),
\end{equation}
and then substitute separately. For simplicity, we will use the following notations:
\begin{equation}
	Z_n(\xi)=J_n(\xi) ~\text{or}~ Z_n(\xi)=Y_n(\xi)	
\end{equation}
with $\xi=\kappa r; ~ Z_n\equiv Z_n(\kappa r)$.
The first derivative is

\begin{equation}
	Z_2'(\xi)=\frac12\left[Z_1(\xi)-Z_3(\xi)\right].
\end{equation}
Therefore differentiating once we get the odd terms,
\begin{equation}
	\frac{d}{dr}Z_2(\kappa r)=\frac{\kappa}{2}
	\left[Z_1(\kappa r)-Z_3(\kappa r)\right].
\end{equation}

Differentiating twice we obtain the even terms:
\begin{equation}
	Z_2^{\prime\prime}(z)= \frac14 \left[Z_0(z)-2Z_2(z)+Z_4(z)\right].
\end{equation}

Therefore,
\begin{equation}
	\frac{d^2}{dr^2}Z_2(\kappa r)
	=\frac{\kappa^2}{4}\left[Z_0(\kappa r)-2Z_2(\kappa r)+Z_4(\kappa r)\right].
\end{equation}
Thus $Z_0, Z_2, Z_4$ correspond to $(J_0,J_2,J_4)$ when $Z=J$; and $(Y_0,Y_2,Y_4)$ when $Z=Y$.
\\ \\
To avoid any confusion we will deal with each term separately. Let us consider one term first, 
\begin{equation}
	\Phi^{(Z)}=C\,Z_2(\kappa r),
\end{equation}

and
\begin{equation}
	\Psi^{(Z)}=\ell~\Phi^{(Z)} = \ell\,C\,Z_2(\kappa r).
\end{equation}

where
\begin{equation}
	Z=J\quad\text{or}\quad Y,
\end{equation}

and $C, \kappa, \ell$ will later be replaced by
\begin{equation}
	(C_A,\kappa_A,\ell_A),~
	(D_A,\kappa_A,\ell_A),~
	(C_B,\kappa_B,\ell_B),~
	(D_B,\kappa_B,\ell_B).
\end{equation}

Now we will compute the displacement produced by one Bessel term. Note that we have following complex radial displacement fields in terms of two scalar potentials,
\begin{align}
	X(r)&=\Phi'+\frac{2i}{r}\Psi, \nonumber \\
	Y(r)&=\frac{2i}{r}\Phi-\Psi'.
\end{align}

Furthermore, for any general single term we write the derivative of scalar potentials $\Phi$ and $\Psi$, given as
\begin{equation}
	\Phi'=C~\frac{\kappa}{2}(Z_1-Z_3),
\end{equation}
and
\begin{equation}
	\Psi'=\ell~C~\frac{\kappa}{2}(Z_1-Z_3).
\end{equation}
Therefore $X$ and $Y$, can be written in terms of Bessel functions $Z$
\begin{equation}
	X^{(Z)}=C~\left[\frac{\kappa}{2}(Z_1-Z_3)+\frac{2i\ell}{r}Z_2\right].
\end{equation}
Similarly,
\begin{equation}
	Y^{(Z)}=C~\left[\frac{2i}{r}Z_2-\frac{\ell\kappa}{2}(Z_1-Z_3)\right].
\end{equation}

The physical displacement contribution is
\begin{equation}
	d_r^{(Z)}=\Re\!\left[X^{(Z)}e^{2i\theta}\right],
\end{equation}
\begin{equation}
	d_\theta^{(Z)}=\Re\!\left[Y^{(Z)}e^{2i\theta}\right].
\end{equation}

Note that complex radial strain is derivative of complex displacement field $X$, and hence the radial strain amplitude is
\begin{equation}
	E_{rr}=X'.
\end{equation}
We write complex displacement field $X$ in terms of $Z$, 
\begin{equation}
	X^{(Z)} = C ~\left[\kappa Z_2'+\frac{2i\ell}{r}Z_2\right],
\end{equation}
and differentiate it to get,
\begin{equation}
	E_{rr}^{(Z)}=C \left[\kappa^2 Z_2''+2i\ell\left(\frac{\kappa Z_2'}{r}-\frac{Z_2}{r^2}\right)\right].
\end{equation}

Using the full expansion of $Z_1'$, and $Z_2''$, we have
\begin{widetext}
\begin{align}
	E_{rr}^{(Z)} =C \Bigg\{ \frac{\kappa^2}{4}\left(Z_0-2Z_2+Z_4\right) +
	i\ell\left[\frac{\kappa}{r}\left(Z_1-Z_3\right)-\frac{2Z_2}{r^2}\right]\Bigg\}.
\end{align}
\end{widetext}
Similarly, we compute \textit{tangential strain} from one Bessel term. The complex tangential amplitude is
\begin{equation}
	E_{\theta\theta} = \frac{X+2iY}{r}.
\end{equation}
First we compute,
\begin{equation}
	2iY = C \left[ -\frac{4}{r}Z_2 -i\ell\kappa(Z_1-Z_3) \right].
\end{equation}
Therefore,
\begin{widetext}
\begin{equation}
	X+2iY = C \left[ \frac{\kappa}{2}(Z_1-Z_3) -\frac{4Z_2}{r} +
	\frac{2i\ell Z_2}{r} -i\ell\kappa(Z_1-Z_3) \right].
\end{equation}
Dividing by $r$, we obtain
\begin{align}
	E_{\theta\theta}^{(Z)}=C \Bigg\{\frac{\kappa}{2r}\left(Z_1-Z_3\right)-\frac{4Z_2}{r^2} 
	+i\ell\left[\frac{2Z_2}{r^2}-\frac{\kappa}{r}\left(Z_1-Z_3\right)\right]\Bigg\}.
\end{align}
\end{widetext}

Finally we compute \textit{polar shear strain} from one Bessel term. The complex shear amplitude is
\begin{equation}
	E_{r\theta} = \frac12 \left(Y'-\frac Yr+\frac{2iX}{r}\right).
\end{equation}
We compute each part separately. First term,
\begin{equation}
	Y =C \left[ \frac{2iZ_2}{r} -\ell\kappa Z_2' \right],
\end{equation}
which after differentiation gives,
\begin{equation}
	Y'=C \left[ 2i \left( \frac{\kappa Z_2'}r-\frac{Z_2}{r^2} \right) -\ell\kappa^2Z_2''\right].
\end{equation}
The second term is,
\begin{equation}
	-\frac Yr=C \left[-\frac{2iZ_2}{r^2}+\frac{\ell\kappa Z_2'}r\right].
\end{equation}
The third terms is,
\begin{equation}
	\frac{2iX}{r} =C\left[\frac{2i\kappa Z_2'}r -\frac{4\ell Z_2}{r^2} \right].
\end{equation}
Adding the three terms we obtain,
\begin{widetext}
\begin{align}
	Y'-\frac Yr+\frac{2iX}{r}=C \Bigg[ 4i\left(\frac{\kappa Z_2'}r-\frac{Z_2}{r^2}\right)
	-\ell\kappa^2Z_2''+\frac{\ell\kappa Z_2'}r-\frac{4\ell Z_2}{r^2}\Bigg].
\end{align}
Therefore,
\begin{align}
	E_{r\theta}^{(Z)}=C\Bigg[2i\left(\frac{\kappa Z_2'}r-\frac{Z_2}{r^2}\right)
	-\frac{\ell\kappa^2}{2}Z_2''+\frac{\ell\kappa}{2r}Z_2'-\frac{2\ell Z_2}{r^2}\Bigg].
\end{align}
Using the Bessel identities,
\begin{align}
	E_{r\theta}^{(Z)}=C\Bigg\{ i\left[\frac{\kappa}{r}(Z_1-Z_3)-\frac{2Z_2}{r^2}\right]
	+\ell\left[-\frac{\kappa^2}{8}\left(Z_0-2Z_2+Z_4\right)   +\frac{\kappa}{4r}\left(Z_1-Z_3\right)-\frac{2Z_2}{r^2}\right]
	\Bigg\}.
\end{align}

Now let us define the three generic Bessel kernels (for either $Z=J$ or $Y$). For simplicity we name the radial kernels by defining

\begin{equation}
	\mathbb{R}_Z(\kappa,\ell;r)=\frac{\kappa^2}{4}(Z_0-2Z_2+Z_4)+i\ell\left[\frac{\kappa}{r}(Z_1-Z_3)
	-\frac{2Z_2}{r^2}\right],
\end{equation}
\begin{align}
	\mathbb{T}_Z(\kappa,\ell;r)= \frac{\kappa}{2r}(Z_1-Z_3)-\frac{4Z_2}{r^2}
	+
	i\ell\left[\frac{2Z_2}{r^2}-\frac{\kappa}{r}(Z_1-Z_3)\right],
\end{align}
and
\begin{align}
	\mathbb{S}_Z(\kappa,\ell;r)
	= i\left[\frac{\kappa}{r}(Z_1-Z_3)-\frac{2Z_2}{r^2}\right]
	+\ell\left[-\frac{\kappa^2}{8}(Z_0-2Z_2+Z_4)+\frac{\kappa}{4r}(Z_1-Z_3)
	-\frac{2Z_2}{r^2}\right].
\end{align}
Note that all $Z_n$ in each kernel have argument $\kappa r$.

\subsection{Complete radial strain amplitude $E_{rr}$}
The four Bessel contributions are
\begin{align}
	E_{rr}(r)=
	C_A\mathbb{R}_J(\kappa_A,\ell_A;r)+D_A\mathbb{R}_Y(\kappa_A,\ell_A;r)
	+C_B\mathbb{R}_J(\kappa_B,\ell_B;r)+D_B\mathbb{R}_Y(\kappa_B,\ell_B;r).
\end{align}
\end{widetext}
Expanding the above equation fully,
\begin{align}
	E_{rr}
	={}&
	C_A\Bigg\{
	\frac{\kappa_A^2}{4}
	\left[
	J_0(\kappa_A r)-2J_2(\kappa_A r)+J_4(\kappa_A r)
	\right]
	\nonumber	\\
	&\qquad
	+i\ell_A
	\left[
	\frac{\kappa_A}{r}
	\left(
	J_1(\kappa_A r)-J_3(\kappa_A r)
	\right)
	-\frac{2J_2(\kappa_A r)}{r^2}
	\right]
	\Bigg\}
	\nonumber \\
	&+
	D_A\Bigg\{
	\frac{\kappa_A^2}{4}
	\left[
	Y_0(\kappa_A r)-2Y_2(\kappa_A r)+Y_4(\kappa_A r)
	\right]
	\nonumber	\\
	&\qquad
	+i\ell_A
	\left[
	\frac{\kappa_A}{r}
	\left(
	Y_1(\kappa_A r)-Y_3(\kappa_A r)
	\right)
	-\frac{2Y_2(\kappa_A r)}{r^2}
	\right]
	\Bigg\}
	\nonumber	\\
	&+
	C_B\Bigg\{
	\frac{\kappa_B^2}{4}
	\left[
	J_0(\kappa_B r)-2J_2(\kappa_B r)+J_4(\kappa_B r)
	\right]
	\nonumber	\\
	&\qquad
	+i\ell_B
	\left[
	\frac{\kappa_B}{r}
	\left(
	J_1(\kappa_B r)-J_3(\kappa_B r)
	\right)
	-\frac{2J_2(\kappa_B r)}{r^2}
	\right]
	\Bigg\}
	\nonumber	\\
	&+
	D_B\Bigg\{
	\frac{\kappa_B^2}{4}
	\left[
	Y_0(\kappa_B r)-2Y_2(\kappa_B r)+Y_4(\kappa_B r)
	\right]
	\nonumber	\\
	&\qquad
	+i\ell_B
	\left[
	\frac{\kappa_B}{r}
	\left(
	Y_1(\kappa_B r)-Y_3(\kappa_B r)
	\right)
	-\frac{2Y_2(\kappa_B r)}{r^2}
	\right]
	\Bigg\}.
\end{align}
Therefore the physical radial strain is
\begin{equation}
	\epsilon_{rr}(r,\theta)= \Re\!\left[E_{rr}(r)e^{2i\theta}\right].
\end{equation}
\subsection{The complete tangential strain amplitude $E_{\theta\theta}$ is given by}
\begin{align}
	E_{\theta\theta}(r)=
	C_A\mathbb{T}_J(\kappa_A,\ell_A;r)
	+
	D_A\mathbb{T}_Y(\kappa_A,\ell_A;r)
	+
	C_B\mathbb{T}_J(\kappa_B,\ell_B;r)
	+
	D_B\mathbb{T}_Y(\kappa_B,\ell_B;r).
\end{align}
We can write it explicitly as follows,
\begin{align}
	E_{\theta\theta}
	={}&
	C_A\Bigg\{
	\frac{\kappa_A}{2r}
	\left[
	J_1(\kappa_A r)-J_3(\kappa_A r)
	\right]
	-\frac{4J_2(\kappa_A r)}{r^2}
	\nonumber	\\
	&\qquad
	+i\ell_A
	\left[
	\frac{2J_2(\kappa_A r)}{r^2}
	-\frac{\kappa_A}{r}
	\left(
	J_1(\kappa_A r)-J_3(\kappa_A r)
	\right)
	\right]
	\Bigg\}
	\nonumber	\\
	&+
	D_A\Bigg\{
	\frac{\kappa_A}{2r}
	\left[
	Y_1(\kappa_A r)-Y_3(\kappa_A r)
	\right]
	-\frac{4Y_2(\kappa_A r)}{r^2}
	\nonumber	\\
	&\qquad
	+i\ell_A
	\left[
	\frac{2Y_2(\kappa_A r)}{r^2}
	-\frac{\kappa_A}{r}
	\left(
	Y_1(\kappa_A r)-Y_3(\kappa_A r)
	\right)
	\right]
	\Bigg\}
	\nonumber	\\
	&+
	C_B\Bigg\{
	\frac{\kappa_B}{2r}
	\left[
	J_1(\kappa_B r)-J_3(\kappa_B r)
	\right]
	-\frac{4J_2(\kappa_B r)}{r^2}
	\nonumber	\\
	&\qquad
	+i\ell_B
	\left[
	\frac{2J_2(\kappa_B r)}{r^2}
	-\frac{\kappa_B}{r}
	\left(
	J_1(\kappa_B r)-J_3(\kappa_B r)
	\right)
	\right]
	\Bigg\}
	\nonumber	\\
	&+
	D_B\Bigg\{
	\frac{\kappa_B}{2r}
	\left[
	Y_1(\kappa_B r)-Y_3(\kappa_B r)
	\right]
	-\frac{4Y_2(\kappa_B r)}{r^2}
	\nonumber	\\
	&\qquad
	+i\ell_B
	\left[
	\frac{2Y_2(\kappa_B r)}{r^2}
	-\frac{\kappa_B}{r}
	\left(
	Y_1(\kappa_B r)-Y_3(\kappa_B r)
	\right)
	\right]
	\Bigg\}.
\end{align}
Therefore the physical tangential strain is
\begin{equation}
	\epsilon_{\theta\theta}(r,\theta)=\Re\!\left[E_{\theta\theta}(r)e^{2i\theta}\right].
\end{equation}

\subsection{The complete polar shear-strain amplitude}
Further, we can write  $E_{r\theta}$ as follows
\begin{widetext}
\begin{align}
	E_{r\theta}(r)=C_A\mathbb{S}_J(\kappa_A,\ell_A;r)+D_A\mathbb{S}_Y(\kappa_A,\ell_A;r)
	+
	C_B\mathbb{S}_J(\kappa_B,\ell_B;r)+D_B\mathbb{S}_Y(\kappa_B,\ell_B;r).
\end{align}
Now we compute each term one by one. The $C_A\mathbb{S}_J$ contribution is
\begin{align}
	C_A\mathbb{S}_J
	=C_A\Bigg\{&
	i\left[
	\frac{\kappa_A}{r}
	\left(
	J_1(\kappa_A r)-J_3(\kappa_A r)
	\right)
	-\frac{2J_2(\kappa_A r)}{r^2}
	\right]
	\nonumber	\\
	&+\ell_A\left[
	-\frac{\kappa_A^2}{8}
	\left(
	J_0(\kappa_A r)-2J_2(\kappa_A r)+J_4(\kappa_A r)
	\right)
	\right.
	\nonumber	\\
	&\hspace{21mm}\left.
	+\frac{\kappa_A}{4r}
	\left(
	J_1(\kappa_A r)-J_3(\kappa_A r)
	\right)
	-\frac{2J_2(\kappa_A r)}{r^2}
	\right]
	\Bigg\}.
\end{align}
The $D_A\mathbb{S}_Y$ contribution is
\begin{align}
	D_A\mathbb{S}_Y
	=D_A\Bigg\{&
	i\left[
	\frac{\kappa_A}{r}
	\left(
	Y_1(\kappa_A r)-Y_3(\kappa_A r)
	\right)
	-\frac{2Y_2(\kappa_A r)}{r^2}
	\right]
	\nonumber \\
	&+\ell_A\left[
	-\frac{\kappa_A^2}{8}
	\left(
	Y_0(\kappa_A r)-2Y_2(\kappa_A r)+Y_4(\kappa_A r)
	\right)
	\right.
	\nonumber \\
	&\hspace{21mm}\left.
	+\frac{\kappa_A}{4r}
	\left(
	Y_1(\kappa_A r)-Y_3(\kappa_A r)
	\right)
	-\frac{2Y_2(\kappa_A r)}{r^2}
	\right]
	\Bigg\}.
\end{align}
The $C_B\mathbb{S}_J$, and $D_B\mathbb{S}_Y$ contribution is obtained by replacing
\begin{equation}
	C_A,\kappa_A,\ell_A \longrightarrow C_B,\kappa_B,\ell_B,
\end{equation}
thus,
\begin{align}
	C_B\mathbb{S}_J
	=C_B\Bigg\{&
	i\left[
	\frac{\kappa_B}{r}
	\left(
	J_1(\kappa_B r)-J_3(\kappa_B r)
	\right)
	-\frac{2J_2(\kappa_B r)}{r^2}
	\right]
	\nonumber \\
	&+\ell_B\left[
	-\frac{\kappa_B^2}{8}
	\left(
	J_0(\kappa_B r)-2J_2(\kappa_B r)+J_4(\kappa_B r)
	\right)
	\right.
	\nonumber \\
	&\hspace{21mm}\left.
	+\frac{\kappa_B}{4r}
	\left(
	J_1(\kappa_B r)-J_3(\kappa_B r)
	\right)
	-\frac{2J_2(\kappa_B r)}{r^2}
	\right]
	\Bigg\}.
\end{align}
The $D_BY_2$ contribution is
\begin{align}
	D_B\mathbb{S}_Y
	=D_B\Bigg\{&
	i\left[
	\frac{\kappa_B}{r}
	\left(
	Y_1(\kappa_B r)-Y_3(\kappa_B r)
	\right)
	-\frac{2Y_2(\kappa_B r)}{r^2}
	\right]
	\nonumber \\
	&+\ell_B\left[
	-\frac{\kappa_B^2}{8}
	\left(
	Y_0(\kappa_B r)-2Y_2(\kappa_B r)+Y_4(\kappa_B r)
	\right)
	\right.
	\nonumber \\
	&\hspace{21mm}\left.
	+\frac{\kappa_B}{4r}
	\left(
	Y_1(\kappa_B r)-Y_3(\kappa_B r)
	\right)
	-\frac{2Y_2(\kappa_B r)}{r^2}
	\right]
	\Bigg\}.
\end{align}
\end{widetext}
Then
\begin{equation}
	E_{r\theta}= C_A\mathbb{S}_J + D_A\mathbb{S}_Y+ C_B\mathbb{S}_J + D_B\mathbb{S}_Y.
\end{equation}
The physical shear strain is
\begin{equation}
	\epsilon_{r\theta}(r,\theta)=\Re\!\left[E_{r\theta}(r)e^{2i\theta}\right].
\end{equation}

\subsection{Separation into $\cos2\theta$ and $\sin2\theta$}
We note that the coefficients and screening numbers may be complex leading to complex strain amplitudes. Therefore we write
\begin{equation}
	E_{ij}(r)=E_{ij}^{R}(r)+iE_{ij}^{I}(r).
\end{equation}
Then
\begin{equation}
	\Re\!\left[ E_{ij}e^{2i\theta} \right] =E_{ij}^{R}\cos2\theta - E_{ij}^{I}\sin2\theta.
\end{equation}
Therefore we define the six real radial strain functions:
\begin{equation}
	\xi_{rr}^{c}=\Re E_{rr},\qquad \xi_{rr}^{s}= -\Im E_{rr},
\end{equation}
\begin{equation}
	\xi_{\theta\theta}^{c}=\Re E_{\theta\theta},\qquad\xi_{\theta\theta}^{s}=-\Im E_{\theta\theta},
\end{equation}
\begin{equation}
	\xi_{r\theta}^{c}=\Re E_{r\theta},\qquad\xi_{r\theta}^{s}=-\Im E_{r\theta}.
\end{equation}
With these notations, the physical strain components can be written as
\begin{equation}
	\epsilon_{rr}=\xi_{rr}^{c}\cos2\theta+\xi_{rr}^{s}\sin2\theta,
\end{equation}
\begin{equation}
	\epsilon_{\theta\theta}=\xi_{\theta\theta}^{c}\cos2\theta+\xi_{\theta\theta}^{s}\sin2\theta,
\end{equation}
\begin{equation}
	\epsilon_{r\theta}=
	\xi_{r\theta}^{c}\cos2\theta+\xi_{r\theta}^{s}\sin2\theta.
\end{equation}

Let us compute the trace of strain which corresponds to the sum of radial strain $\epsilon_{rr}$ and tangential strain $\epsilon_{\theta\theta}$. For one branch,
\begin{equation}
	E_{rr,s}^{(Z)}+E_{\theta\theta,s}^{(Z)}=C \left[\kappa^2Z_2''+\frac{\kappa Z_2'}r-\frac{4Z_2}{r^2}
	\right].
\end{equation}
Using the Bessel equation,
\begin{equation}
	\kappa^2Z_2'' + \frac{\kappa Z_2'}r + \left(\kappa^2-\frac4{r^2} \right)Z_2=0,	 
\end{equation}
we obtain
\begin{equation}
	E_{rr,s}^{(Z)}  + E_{\theta\theta,s}^{(Z)} = -C \kappa^2Z_2(\kappa r).
\end{equation}
Notice that all $\ell$-dependent terms cancel. Therefore,
\begin{widetext}
\begin{align}
	E_{kk}=-\kappa_A^2\left[C_AJ_2(\kappa_A r)+D_AY_2(\kappa_A r) \right]
	- \kappa_B^2\left[C_BJ_2(\kappa_B r)+D_BY_2(\kappa_B r)\right].
\end{align}
\end{widetext}
Thus,
\begin{equation}
	\epsilon_{rr}+\epsilon_{\theta\theta}
	=
	\Re\!\left[E_{kk}e^{2i\theta}\right].
\end{equation}
This is an important exact consistency check. The transverse part does not participate in the trace.

\subsection{Bessel-expanded stresses}
For isotropic elasticity,
\begin{equation}
	\sigma_{rr}=\lambda\epsilon_{kk}+2\mu\epsilon_{rr},
\end{equation}
\begin{equation}
	\sigma_{\theta\theta} = \lambda\epsilon_{kk} + 2\mu\epsilon_{\theta\theta},
\end{equation}
\begin{equation}
	\sigma_{r\theta}=2\mu\epsilon_{r\theta}.
\end{equation}
At the complex-amplitude level,
\begin{equation}
	\mathscr{S}_{rr}
	=
	\lambda E_{kk}+2\mu E_{rr},	 
\end{equation}
\begin{equation}
	\mathscr{S}_{\theta\theta}
	=
	\lambda E_{kk}+2\mu E_{\theta\theta},	
\end{equation}
\begin{equation}
	\mathscr{S}_{r\theta}=2\mu E_{r\theta}.
\end{equation}

For one generic $CZ_2(\kappa r)$ term,
\begin{widetext}
\begin{align}
	\mathscr{S}_{rr}^{(Z)}=C \Bigg\{-\lambda\kappa^2Z_2+\frac{\mu\kappa^2}{2}(Z_0-2Z_2+Z_4)
	+2i\mu\ell\left[\frac{\kappa}{r}(Z_1-Z_3)-\frac{2Z_2}{r^2}\right]\Bigg\}.
\end{align}
For the tangential stress,
\begin{align}
	\mathscr{S}_{\theta\theta}^{(Z)} =C \Bigg\{-\lambda\kappa^2Z_2 +\mu\frac{\kappa}{r}(Z_1-Z_3)
	-\frac{8\mu Z_2}{r^2}
	+2i\mu\ell\left[\frac{2Z_2}{r^2}-\frac{\kappa}{r}(Z_1-Z_3)\right]\Bigg\}.
\end{align}
For the shear stress,
\begin{align}
	\mathscr{S}_{r\theta}^{(Z)} =C \Bigg\{ 2i\mu \left[\frac{\kappa}{r}(Z_1-Z_3)-\frac{2Z_2}{r^2}\right] +
	2\mu\ell\left[-\frac{\kappa^2}{8}(Z_0-2Z_2+Z_4)+\frac{\kappa}{4r}(Z_1-Z_3)-\frac{2Z_2}{r^2}\right]\Bigg\}.
\end{align}
The complete stress amplitudes are obtained by the same four-term sum:
\begin{align}
	\mathscr{S}_{ij}=
	C_A \mathscr{S}_{ij,J}(\kappa_A,\ell_A)+D_A \mathscr{S}_{ij,Y}(\kappa_A,\ell_A) 
	+
	C_B \mathscr{S}_{ij,J}(\kappa_B,\ell_B)+D_B \mathscr{S}_{ij,Y}(\kappa_B,\ell_B).
\end{align}
\end{widetext}
Then
\begin{equation}
	\sigma_{ij}(r,\theta)
	=\Re\!\left[\mathscr{S}_{ij}(r)e^{2i\theta}\right].	
\end{equation}

\subsection{Cartesian shear strain}
Now after obtaining the complete Bessel expanded terms we can write all the components of the strain tensor. Again, the most important component that is used in computation is the shear strain which is, 
\begin{equation}
	\epsilon_{xy}=\frac12(\epsilon_{rr}-\epsilon_{\theta\theta})
	\sin2\theta+\epsilon_{r\theta}\cos2\theta.	
\end{equation}
Although we have explored the shear strain before, we gain write here for completeness. As before we define
\begin{equation}
	D_c=\xi_{rr}^{c}-\xi_{\theta\theta}^{c},	
\end{equation}
\begin{equation}
	D_s=\xi_{rr}^{s}-\xi_{\theta\theta}^{s}.	
\end{equation}

Then
\begin{equation}
	\epsilon_{xy}=u_0(r)+u_c(r)\cos4\theta+u_s(r)\sin4\theta,	
\end{equation}

with
\begin{equation}
	u_0=\frac14D_s+\frac12\xi_{r\theta}^{c},	
\end{equation}

\begin{equation}
	u_c=-\frac14D_s+\frac12\xi_{r\theta}^{c},	
\end{equation}

and
\begin{equation}
	u_s=\frac14D_c+\frac12\xi_{r\theta}^{s}.	
\end{equation}
The Cartesian shear stress is
\begin{equation}
	\sigma_{xy}=2\mu\left[u_0+u_c\cos4\theta+u_s\sin4\theta\right].	
\end{equation}
For the special case of even screening when we do not consider the odd screening part, then some of the coefficient vanish and,
\begin{equation}
	\xi_{rr}^{s}=\xi_{\theta\theta}^{s}=\xi_{r\theta}^{c}=0,	
\end{equation}
so
\begin{equation}
	u_0=u_c=0,	
\end{equation}
and
\begin{equation}
	\epsilon_{xy}=\frac12\left[\xi_{r\theta}+\frac12(\xi_{rr}-\xi_{\theta\theta})\right]\sin4\theta,
\end{equation}
The formulas above are the complete term-by-term Bessel expansion for the strain and stress of the odd Eshelby problem.

\section{Numerical Simulations and Discussions}\label{numerics}
 
\begin{figure}
	\centering
	\includegraphics[width=0.45\textwidth]{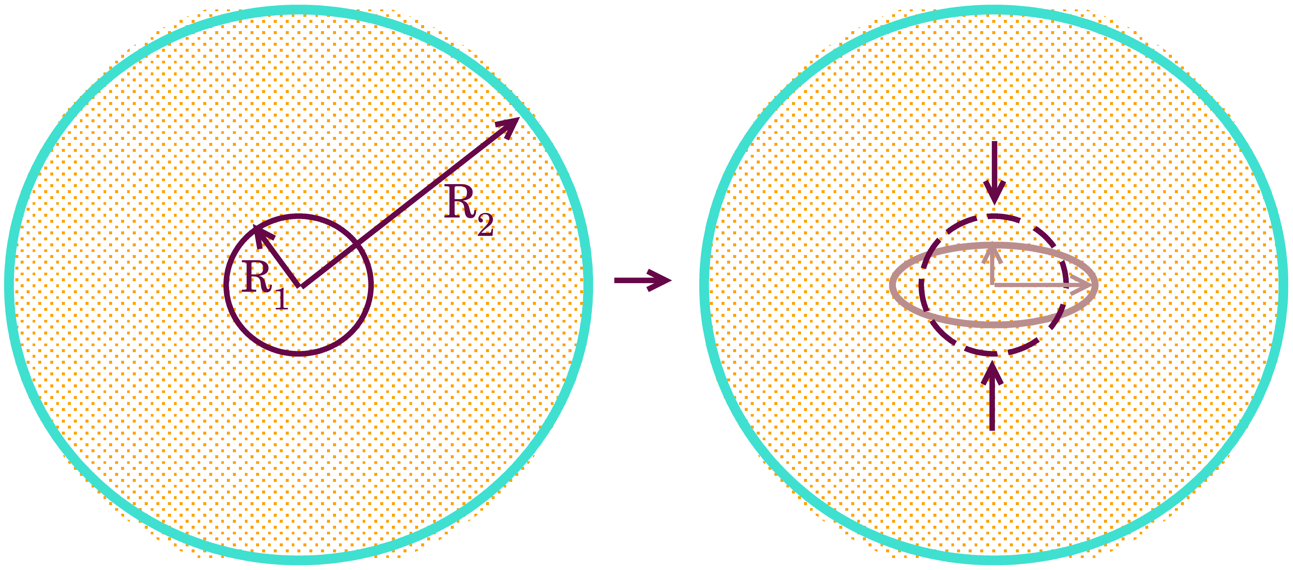}
	\caption{Formation of an elliptical inclusion. Left panel: A two-dimensional amorphous solid is formed within an annular region of inner radius $r_{in}=R_1$ and outer radius $r_{out} =R_2$, with $R_2 \gg R_1$. The amorphous solid is mechanically equilibrated. Right panel: Transformation of inner circle into an ellipse generating an Eshelby inclusion in the amorphous solid with semi-major and minor axes $a$ and $b$ respectively such that $ R_1^2 = ab$. After the transformation the system is equilibrated again before computing the displacement field.}
	\label{fig:Elp}
\end{figure}

 Here, we consider the Eshelby problem in an annular geometry as shown in the left panel of Fig.~\ref{fig:Elp}. The amorphous solid is confined between an inner circular cavity of radius $r_{\mathrm{in}}=R_1$ and an outer circle of radius $r_{\mathrm{out}}=R_2$. The inner circle is then deformed to an ellipse with major semi-axis $a$ and minor semi-axis $b$, such that it has same area, i.e., $ ab =  r_{\mathrm{in}}^2$. The boundary of the ellipse is now
\begin{equation}
\frac{x^2}{a^2} + \frac{y^2}{b^2} = 1.
\end{equation}
Defining $\delta \equiv a/r_{\mathrm{in}}$, the boundary of the ellipse $r(\theta)$ is traced by
\begin{equation}
r(\theta) = \frac{r_{\mathrm{in}}}{\sqrt{\cos^2(\theta)/\delta^2 + \delta^2 \sin^2 \theta}},
\end{equation}
where $\theta = \arctan(y/x)$. We are interested in the displacement field that responds to the change from a circle to an ellipse. Please refer to appendix \ref{rotated} for a detailed discussion. 
 
\textit{Numerical method}. We study the Eshelby problem numerically in a two-dimensional polydispersed model of point particles having equal mass $m=1$, with pair interaction given by shifted and smoothed Lennard-Jones (LJ) potentials $u(r)$ \cite{Ninarello2017},
\begin{equation}\label{PLJ}
	u_{ij}(r) = 
	\begin{cases}
		u_{ij}^{LJ} + A_{ij} + B_{ij}r + C_{ij}r^2, & \text{if } r < R_{ij}^{\mathrm{cut}},\\
		0, & \text{if } r > R_{ij}^{\mathrm{cut}},
	\end{cases}
\end{equation}
where
\begin{equation}\label{LJ}
	u_{ij}^{LJ} = 4\epsilon_{ij}\left[\left(\frac{\sigma_{ij}}{r}\right)^{12} - \left(\frac{\sigma_{ij}}{r}\right)^{6}\right].
\end{equation}
The smoothing of potentials in Eq.~(\ref{PLJ}) is such that it vanish with two zero derivatives at distances $R_{ij}^{\mathrm{cut}} = 2.5\sigma_{ij}$ \cite{Lerner2009}. The interaction lengths $\sigma_{ij}$ are chosen from the probability distribution $P(\sigma)\propto 1/\sigma^3$. The parameters for smoothing the LJ potentials in Eq.~\ref{PLJ} and for $i$ and $j$ particle interactions in Eq.~\ref{LJ} are as follows: $A_{ij} = 0.4526\epsilon_{ij}$, $B_{ij} = -0.3100\epsilon_{ij}/\sigma_{ij}$, $C_{ij} = 0.0542\epsilon_{ij}/\sigma_{ij}^2$. The reduced units for mass, length, energy, and time have been taken as $m$, $\bar{\sigma}$, $\epsilon_{ij}=1$, and $\bar{\sigma}\sqrt{m/\epsilon_{ij}}$, respectively.

We employ 20\,000 particles in an annulus of initial inner radius $r_{\mathrm{in}}=5$ and $r_{\mathrm{out}}=80$. The system is thermalized at some ``mother temperature'' $T_m$ using \textit{swap Monte Carlo} \cite{Grigera2001} and then cooled down to $T=0$, using conjugate gradient methods. The interaction between the point particles and the two walls are of the same form Eq.~(\ref{PLJ}), where $r_{ij}$ and $\sigma_{ij}$ are replaced by the distance to the wall and by $\sigma_i$.

After the system is cooled down to $T=0$, it is mechanically equilibrated with the total force on each particle smaller than $10^{-8}$. Afterward the shear and bulk moduli $\mu$ and $K$ are computed using standard methods \cite{Lutsko1989}, giving the values for the second Lam\'e coefficients $\lambda = K - \mu$. Then we transform the inner radius $r_{\mathrm{in}}$ into an ellipse as shown in Fig.~\ref{fig:Elp}. It should be stressed that this deformation is instantaneous, not quasistatic. After the deformation we mechanically equilibrate the system again by the conjugate gradients, and then measure the displacement field $\mathbf{d}$, comparing the two equilibrated configuration before and after distortion.

Since the density of our particle is always the same, the nature of the mechanical response that we observe depends on two other parameters, which are the mother temperature $T_m$ and the strength of distortion of the initial inner circle that is quantified by $\delta$. The lower is the mother temperature and the smaller is $\delta>1$, the response becomes quasi-elastic, not showing any anomalous nature. Mutatis mutandis, when the mother temperature is higher and $\delta$ gets bigger, the system responds anomalously, displaying the theoretically expected huge deviations from the quasi-classical expectation.  
\begin{figure}
	\centering
	\includegraphics[width=0.32\textwidth]{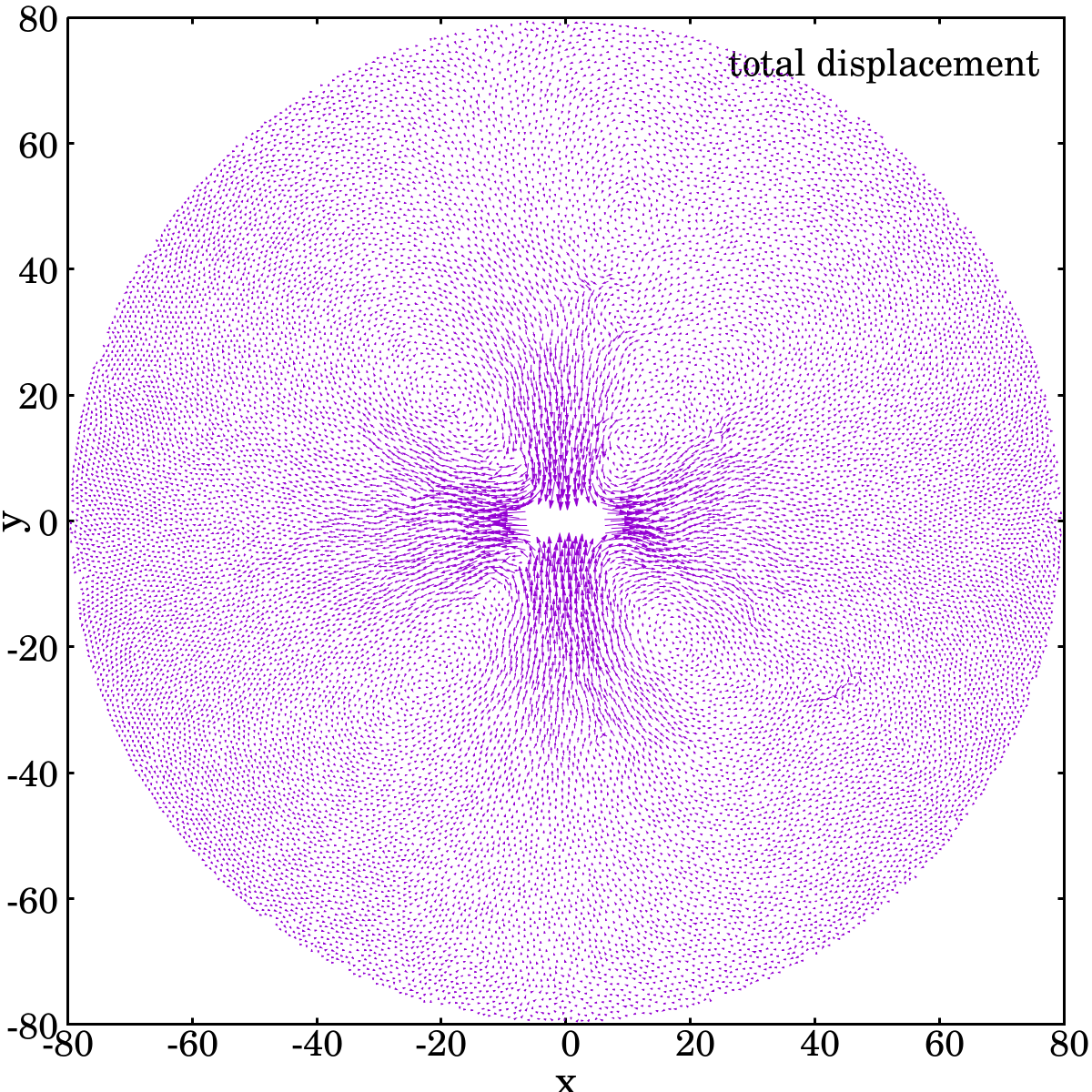}
	\hfill
	\includegraphics[width=0.32\textwidth]{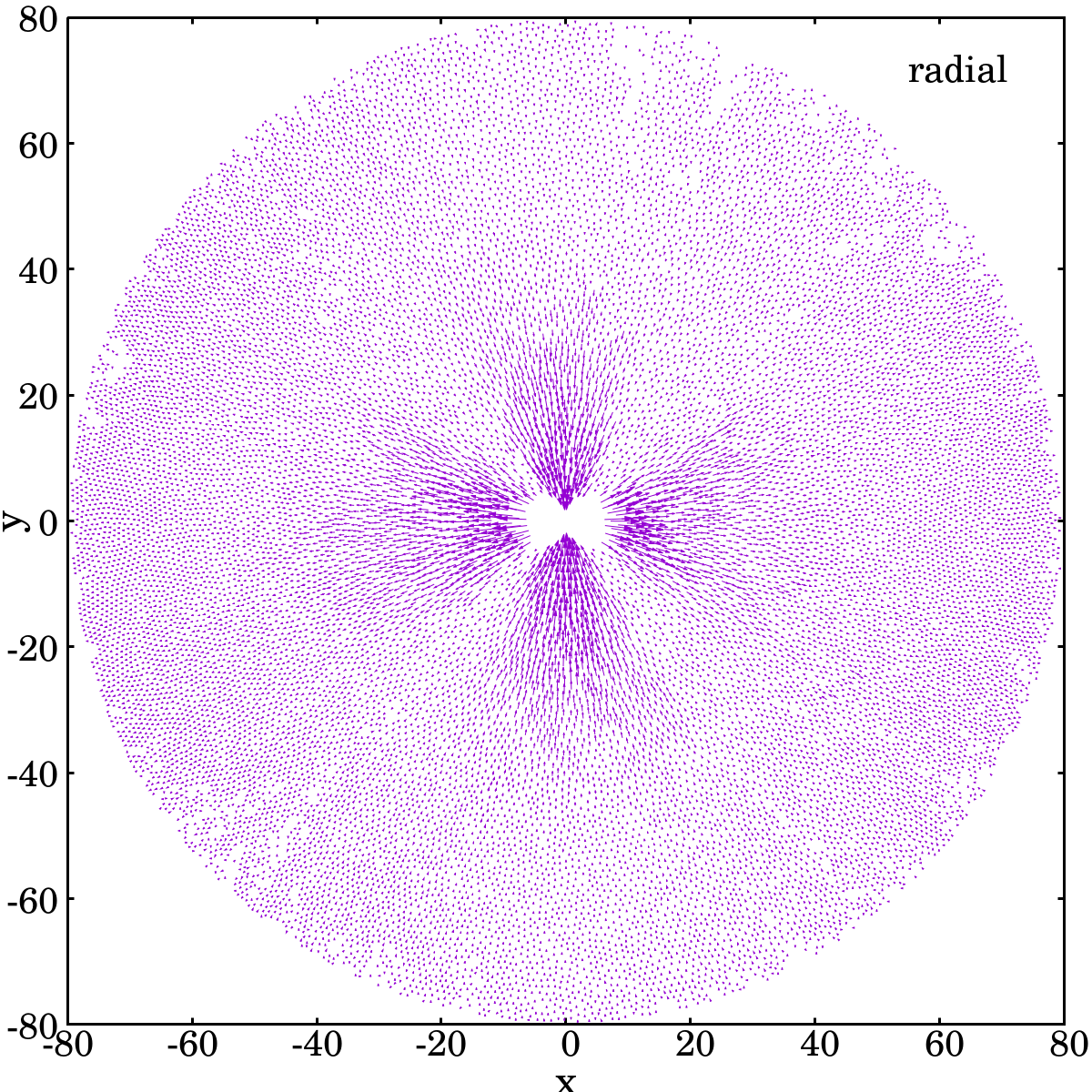}
	\hfill
	\includegraphics[width=0.32\textwidth]{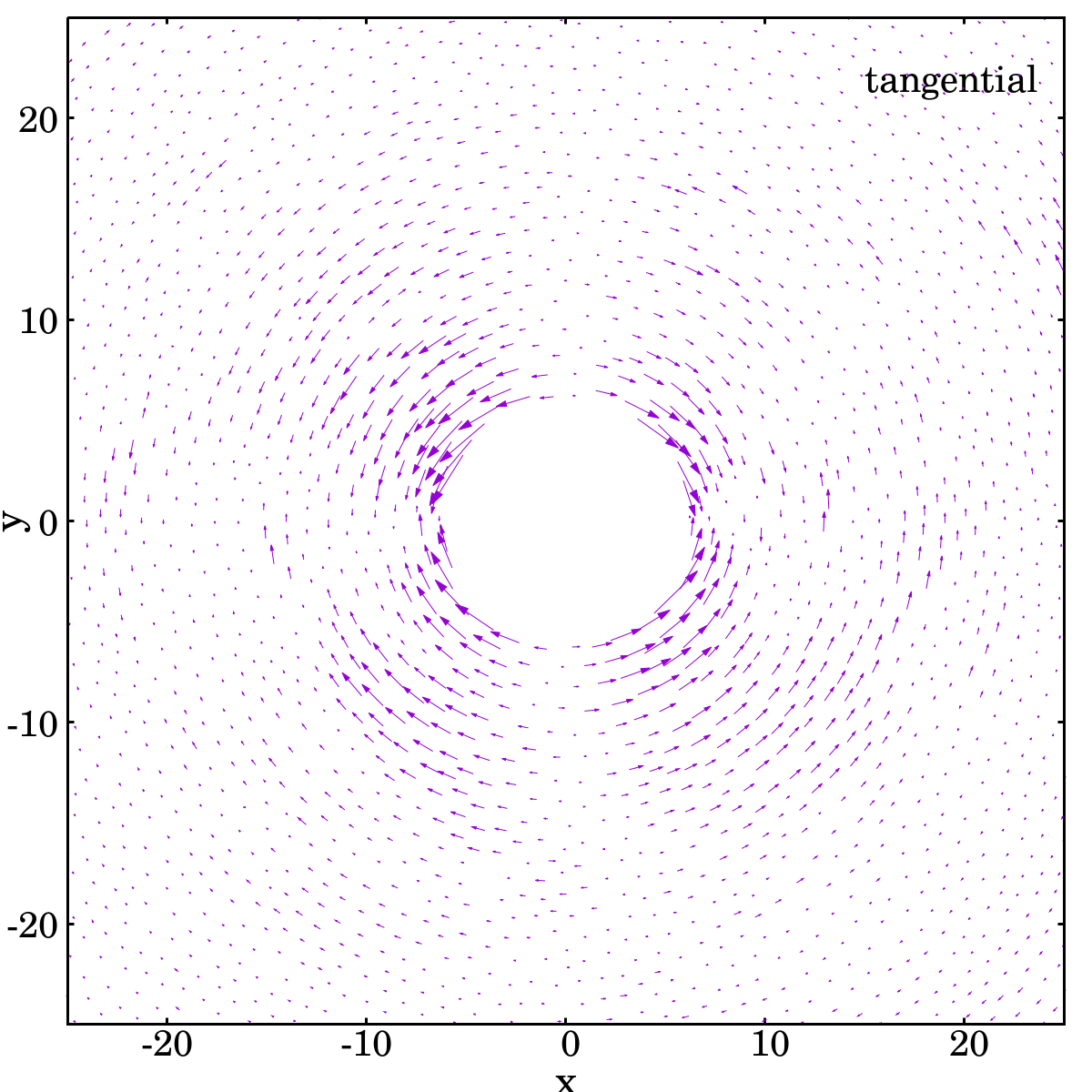}
	\caption{Top panel shows the displacement vector field for a typical glassy system that corresponds to quasi-elastic responses. The system parameters are $r_{in} = 5$, $r_{out} = 80$, $\delta = 1.01, T_m = 1$. Middle and bottom panel correspond to radial and tangential components obtained computed from the total displacement vector field of the top panel.}
	\label{fig:elasticMap}
\end{figure}
Thus for example in Fig.~\ref{fig:elasticMap} we show a typical quasi-elastic response. The top panel in Figure~\ref{fig:elasticMap} corresponds to a well-quenched configuration from $T_m=1$, with a small distortion from circle to ellipse, $\delta=1.01$. We see the classical quadrupolar response that is now well known as the Eshelby solution. Middle and bottom panels show the radial and tangential components respectively, computed from the total displacement vector field from the top panel. 
\begin{figure}
	\centering
	\includegraphics[width=0.3\textwidth]{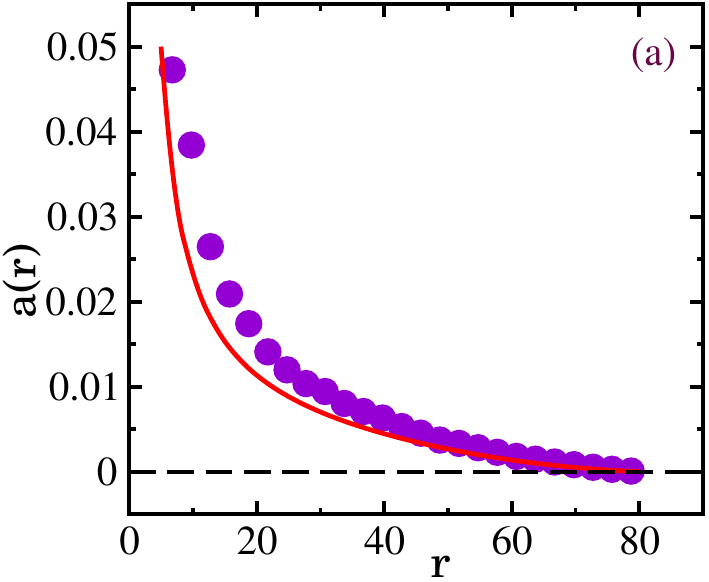}
	\hfill
	\includegraphics[width=0.32\textwidth]{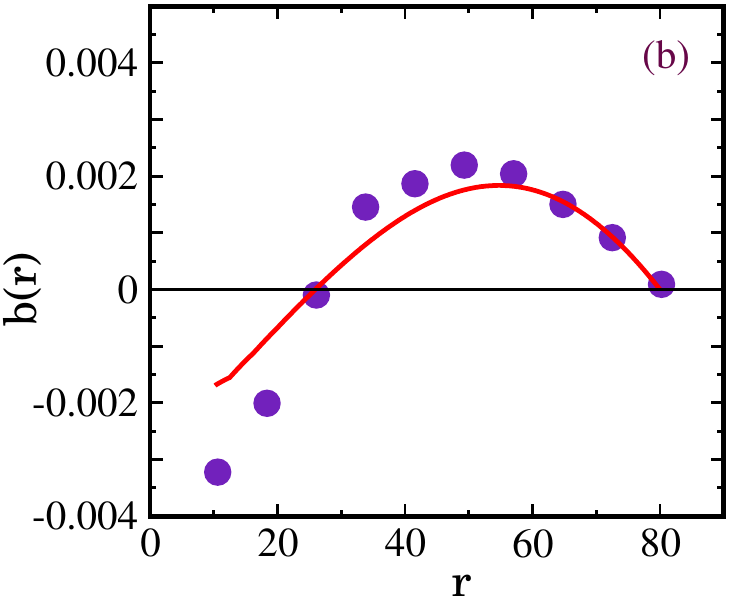}
	\caption{Top panel shows the radial displacement field for a typical glassy system that corresponds to quasi-elastic responses shown in Fig.~\ref{fig:elasticMap}. The system parameters are $r_{in} = 5$, $r_{out} = 80$, $\delta = 1.01, T_m = 1$. Bottom panel corresponds to the tangential displacement field which is obtained from the same set of parameters. Solid lines correspond to the theory and filled circles correspond to the numerical data.}
	\label{fig:elasticField}
\end{figure}
To compare with the theory we compute the angle-averaged radial and tangential components of the displacement field $a(r)$ and $b(r)$. The angle average, which consists of multiplying the displacement field by $\cos 2\theta$ or $\sin 2\theta$ and integrating in $[0,2\pi]$, substitutes the ensemble average for this comparison. Typically we use about 25 concentric annuli, starting with the first one at $r \approx 2r_{\mathrm{in}}$. Angle averaging at smaller values of $r$ is meaningless due to the small number of participating particles. In Fig.~\ref{fig:elasticField} we show the functions $a(r)$ and $b(r)$ when the response is the one shown in Fig.~\ref{fig:elasticMap}. The agreement is quite good, displaying the expected power-law decay of the Eshelby solution, modified by the zero boundary conditions at $r_{\mathrm{out}}$. 
\subsection{Anomalous Response}
Presently we turn to those systems whose response is non-classical anomalous response. 
\begin{figure}
	\centering
	\includegraphics[width=0.325\textwidth]{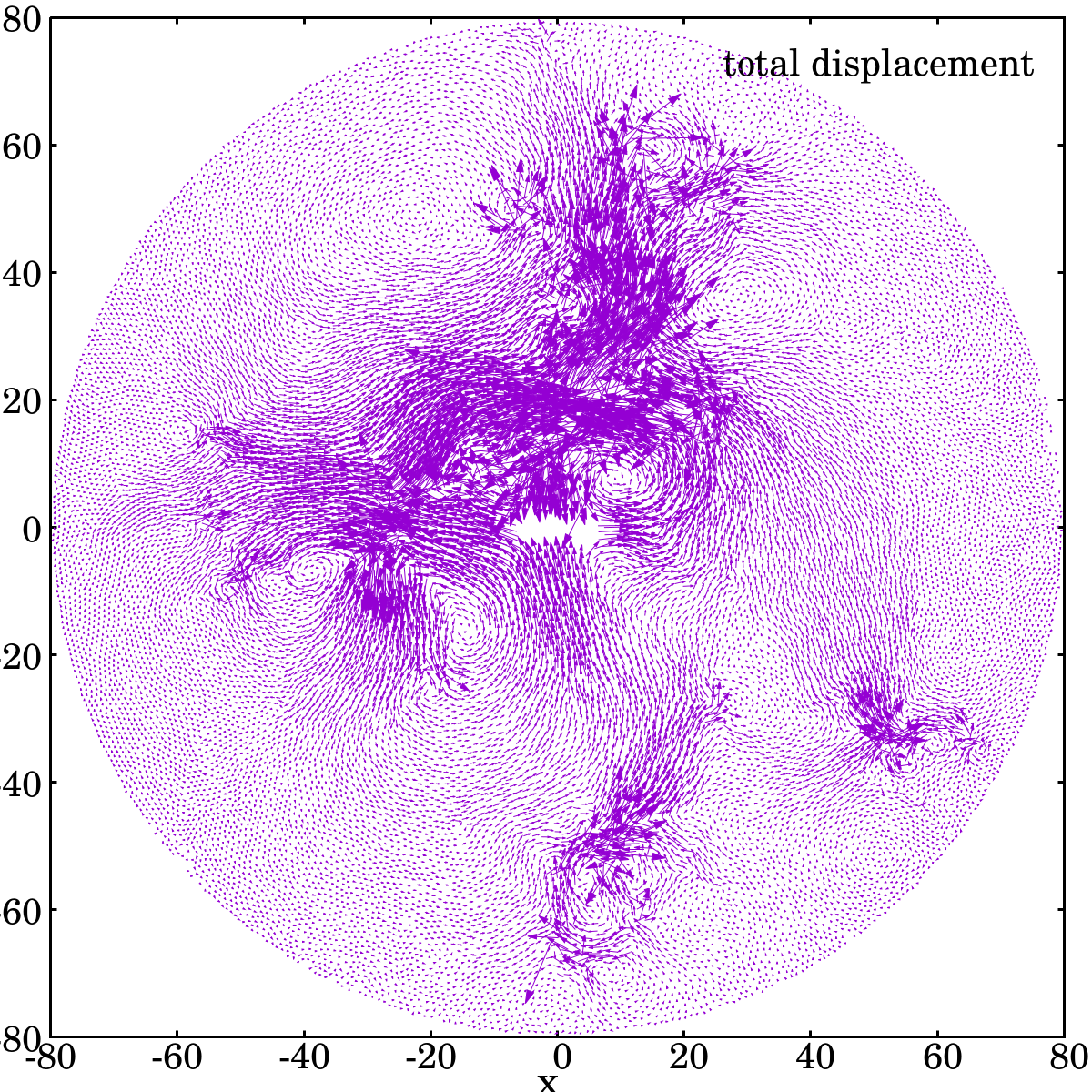}
	\hfill
	\includegraphics[width=0.325\textwidth]{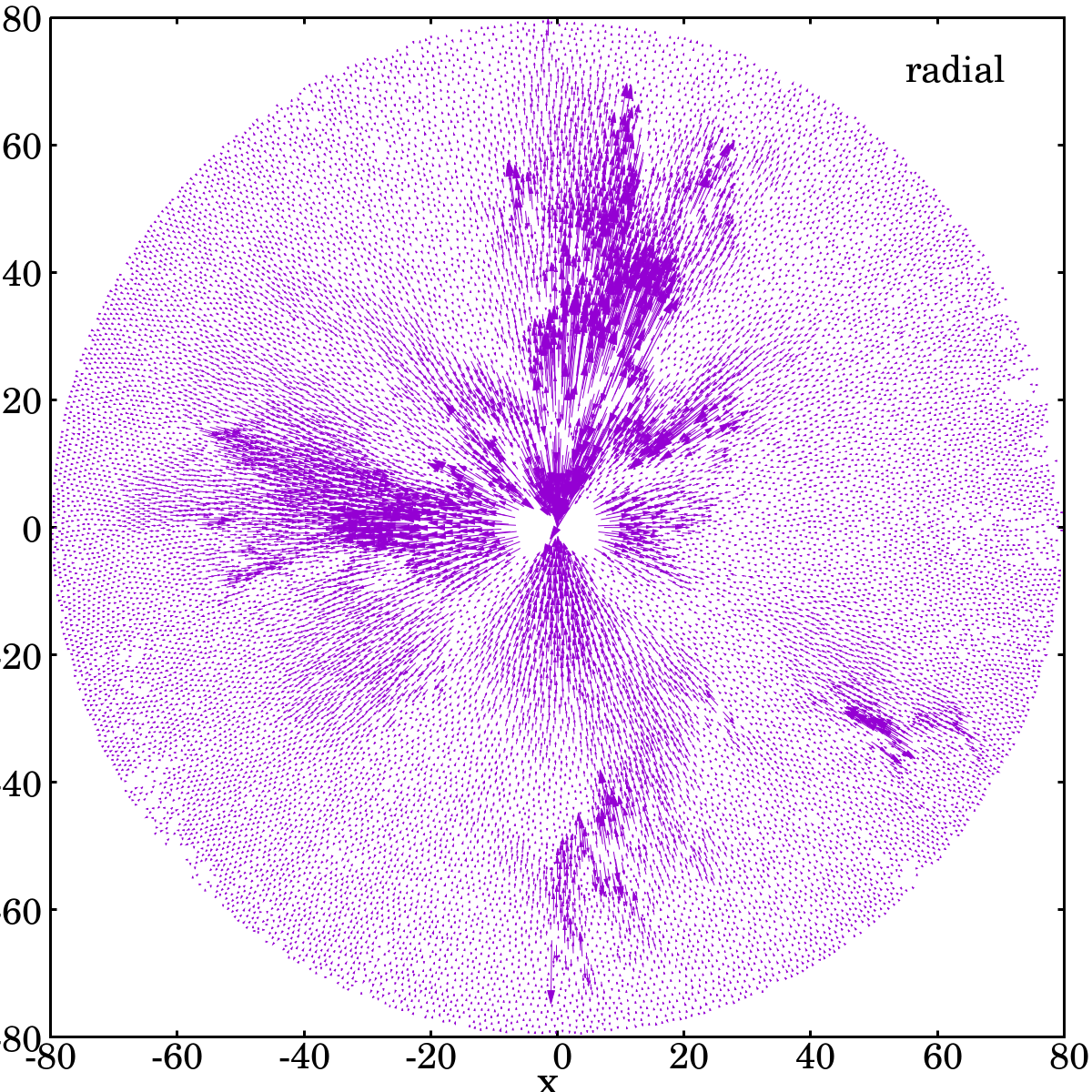}
	\hfill
	\includegraphics[width=0.325\textwidth]{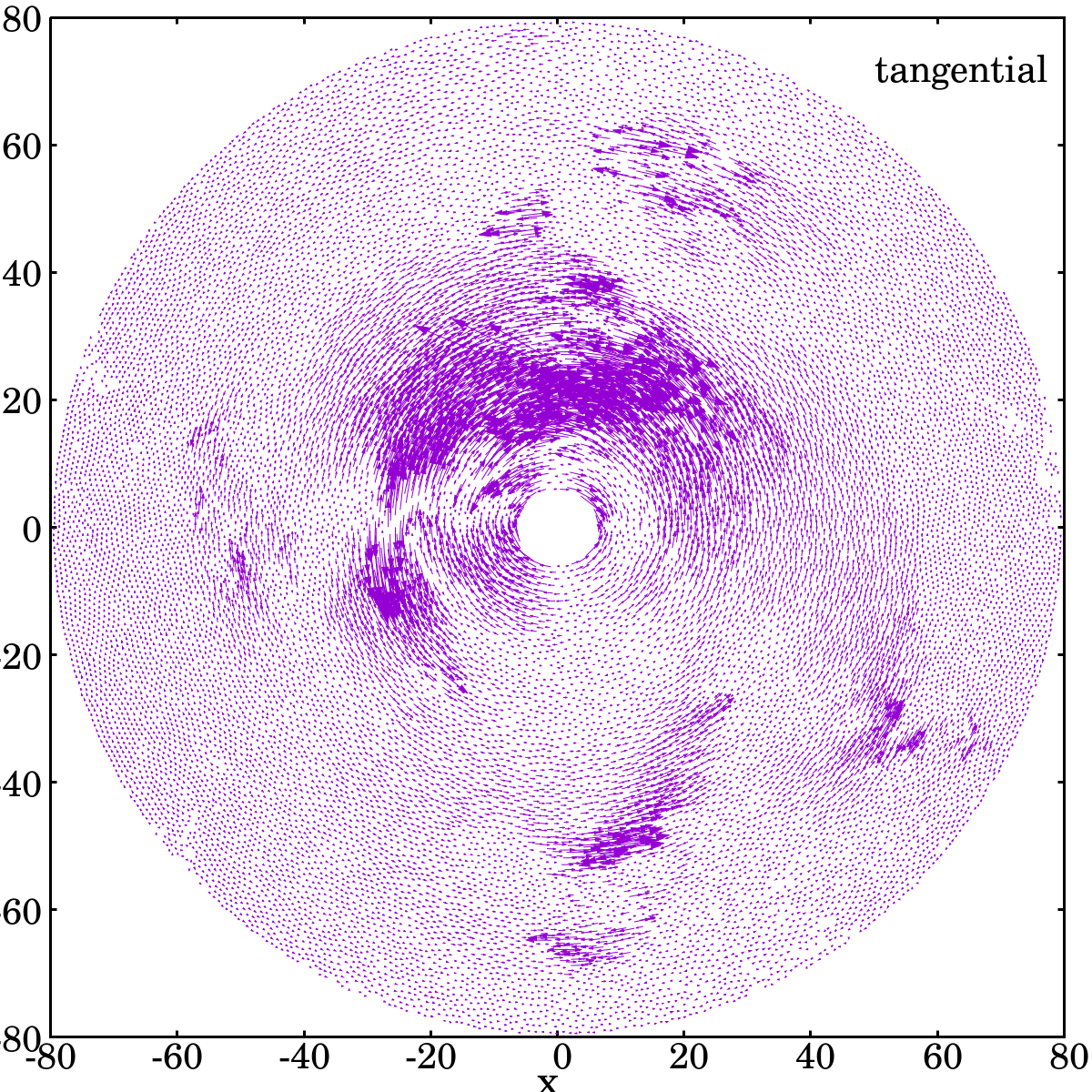}
	\caption{Top panel: a typical displacement vector field that corresponds to anomalous response. The parameters are $r_{in} = 5$, $r_{out} = 80$, $\delta = 1.05, T_m = 3$. Middle panel: the radial component of the displacement field . Bottom panel: the tangential component of the displacement field.}
	\label{fig:anomalousMap}
\end{figure}
The top panel of Fig.~\ref{fig:anomalousMap} corresponds to a system which was quenched from $T_m=3$ and the distortion was higher, $\delta=1.05$. A host of plastic events are taking place, leading to screening. The middle and bottom panel correspond to radial and tangential components of the total vector field computed form the top panel.

\subsection{Fourier Decomposition}
The total displacement vector field in Fig.~\ref{fig:anomalousMap} shows a complex geometric pattern. We note that the imposed transformation from a circle to an ellipse that involves $\cos 2\theta$ corresponds to second harmonics. However, the complexity in the resulting vector field exhibits a richer response. Unlike the quasi-elastic case of Fig.~\ref{fig:elasticMap}, whose response also corresponds to second harmonics;  here we expect many modes. Therefore, to understand and analyze the mechanical response after the inner circular boundary transforms into an ellipse, it is advantageous to  decompose the total displacement vector field into different Fourier modes. In the linear case the displacement field is a superposition of both even and odd harmonics, and therefore the field is naturally written as
\begin{equation}
	\mathbf d(r,\theta) = \sum_{m=0}^{\infty} \mathbf d_m(r,\theta).
\end{equation}

Our simplest analytical theory assumes, essentially,
\begin{equation}
	\mathbf d(r,\theta) = \mathbf d_{m=2}(r,\theta),	
\end{equation}
with radial functions determined by a linear screened elastic operator.

For a strain resulting from transforming a circle to an ellipse, it is expected that the dominant contribution to the resulting displacement  should be the m=2 Fourier mode. This is certainly true in the small $\delta$ limit,  when the system shows a quasi-elastic behavior. However, when the perturbation amplitude is large enough, it may excite other harmonics and their contribution should be accounted for. The numerical simulations indicate the existence of modes corresponding to $m = 1, 3, 4, 5, 6.\cdots$. The displacement fields contain appreciable higher harmonics beyond the leading Eshelby $m=2$ mode. Their appearance can be ascribed structural disorder and, at a finite deformation, to nonlinear mode coupling.

\begin{figure}
	\begin{center}
		\begin{tabular}{cc}
			\includegraphics*[width=0.24\textwidth]{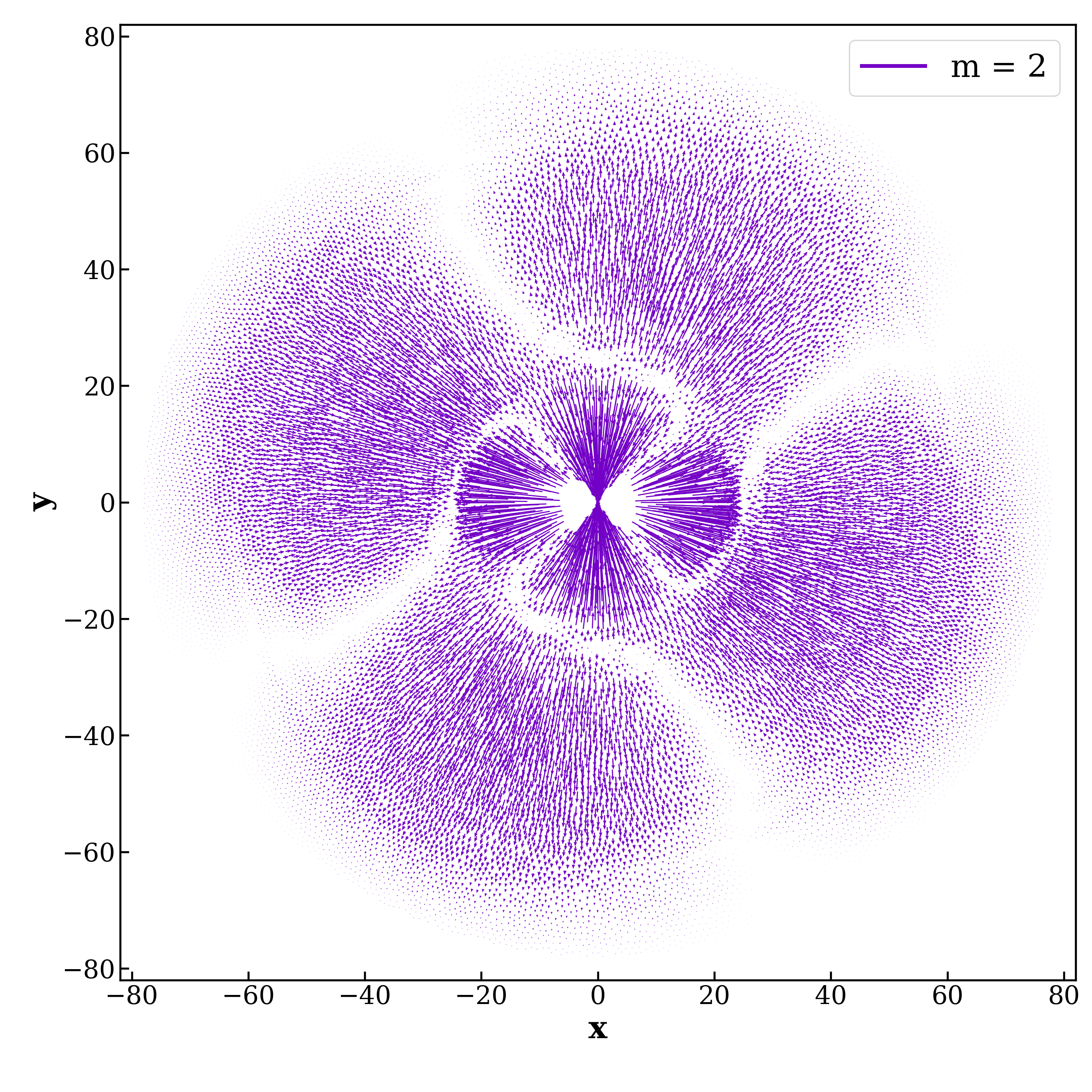} & \includegraphics*[width=0.24\textwidth]{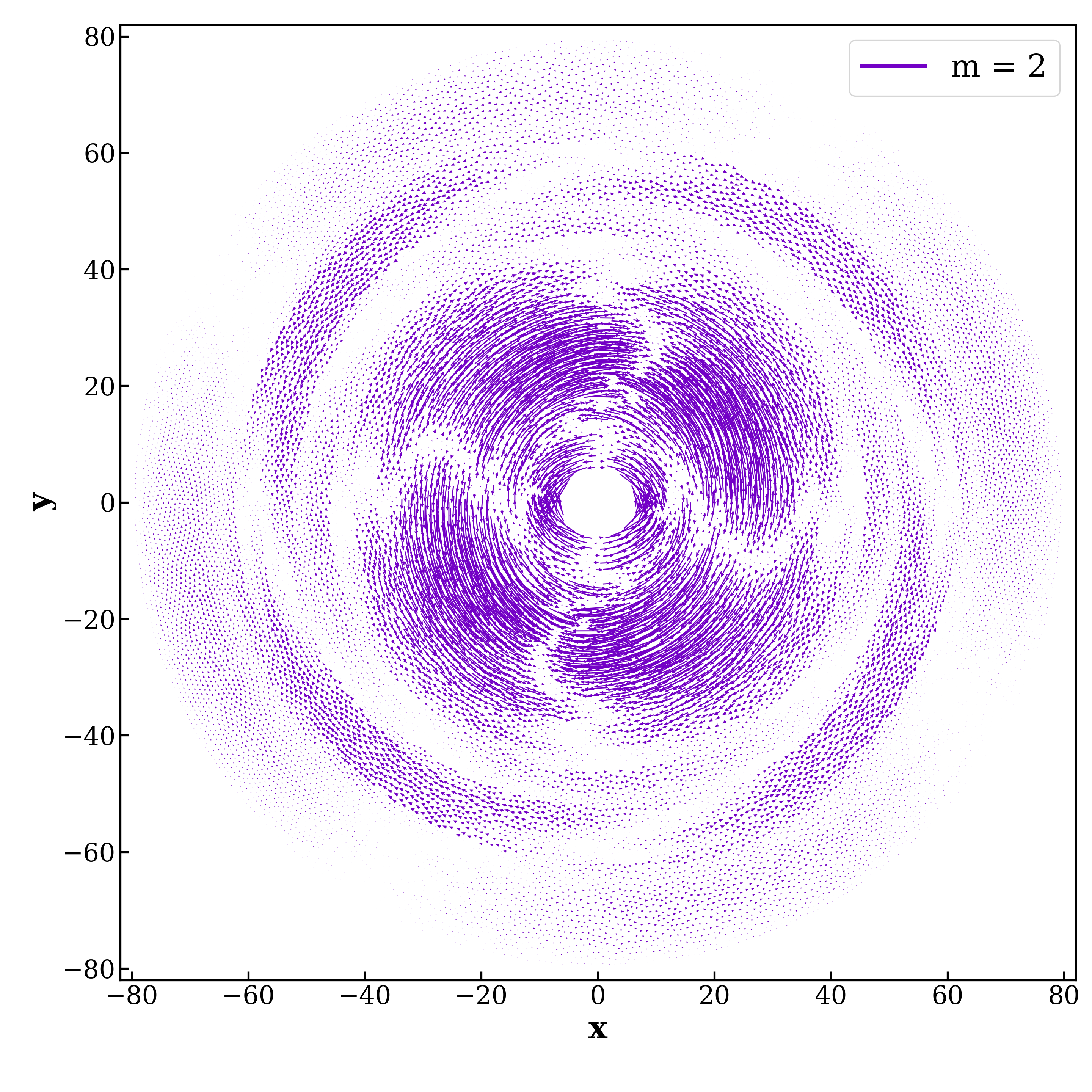} \\ 
			\includegraphics*[width=0.24\textwidth]{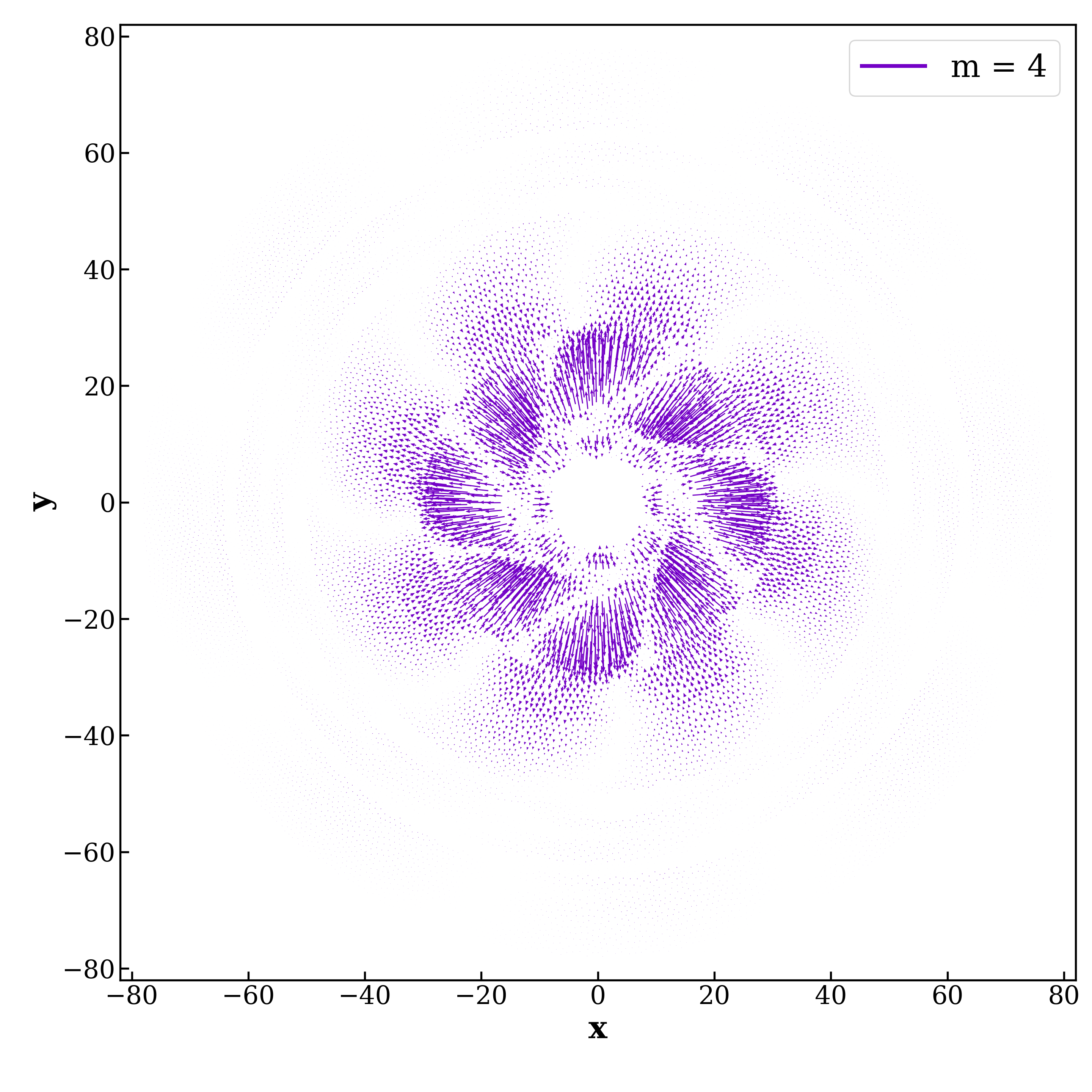} & \includegraphics*[width=0.24\textwidth]{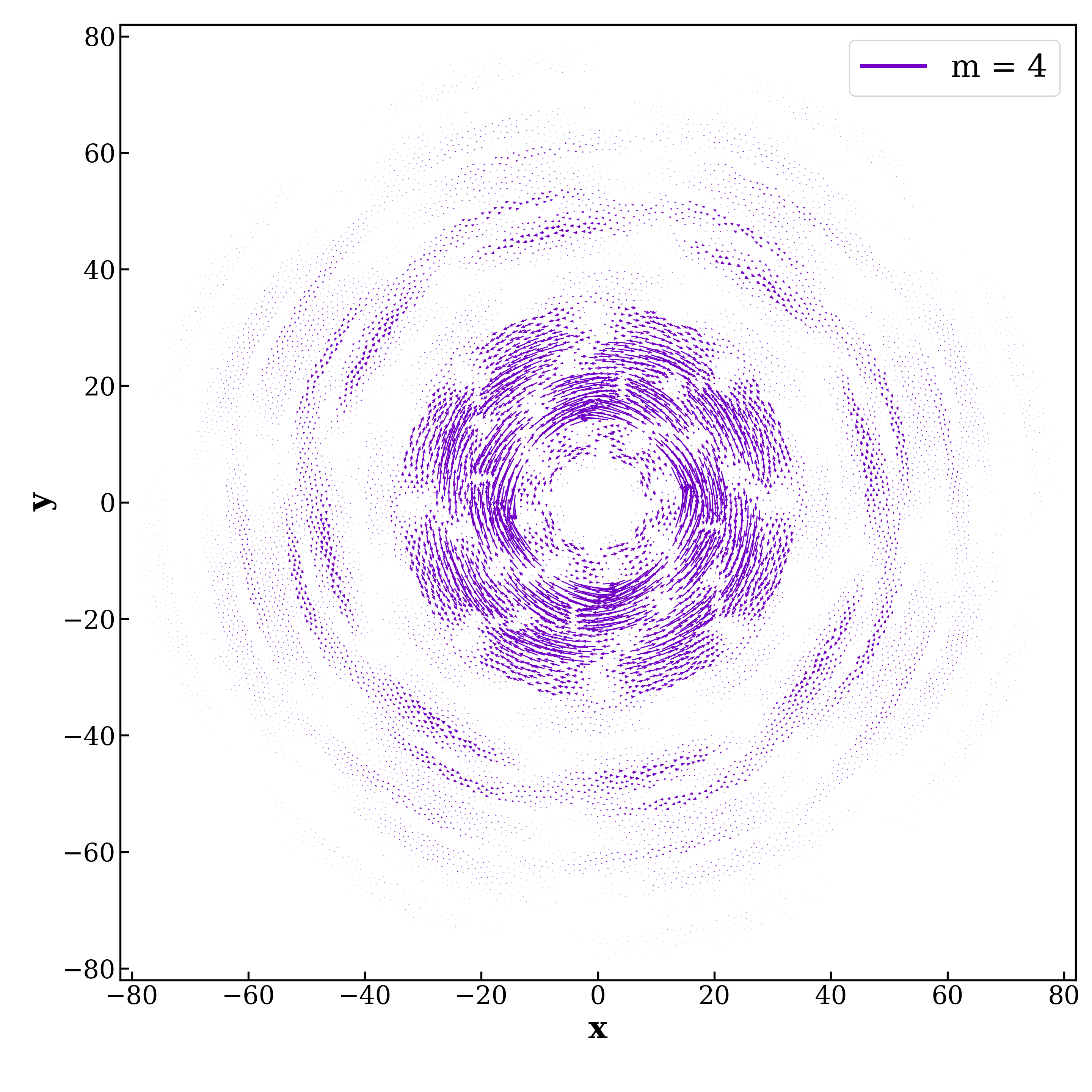} \\ 
			\includegraphics*[width=0.24\textwidth]{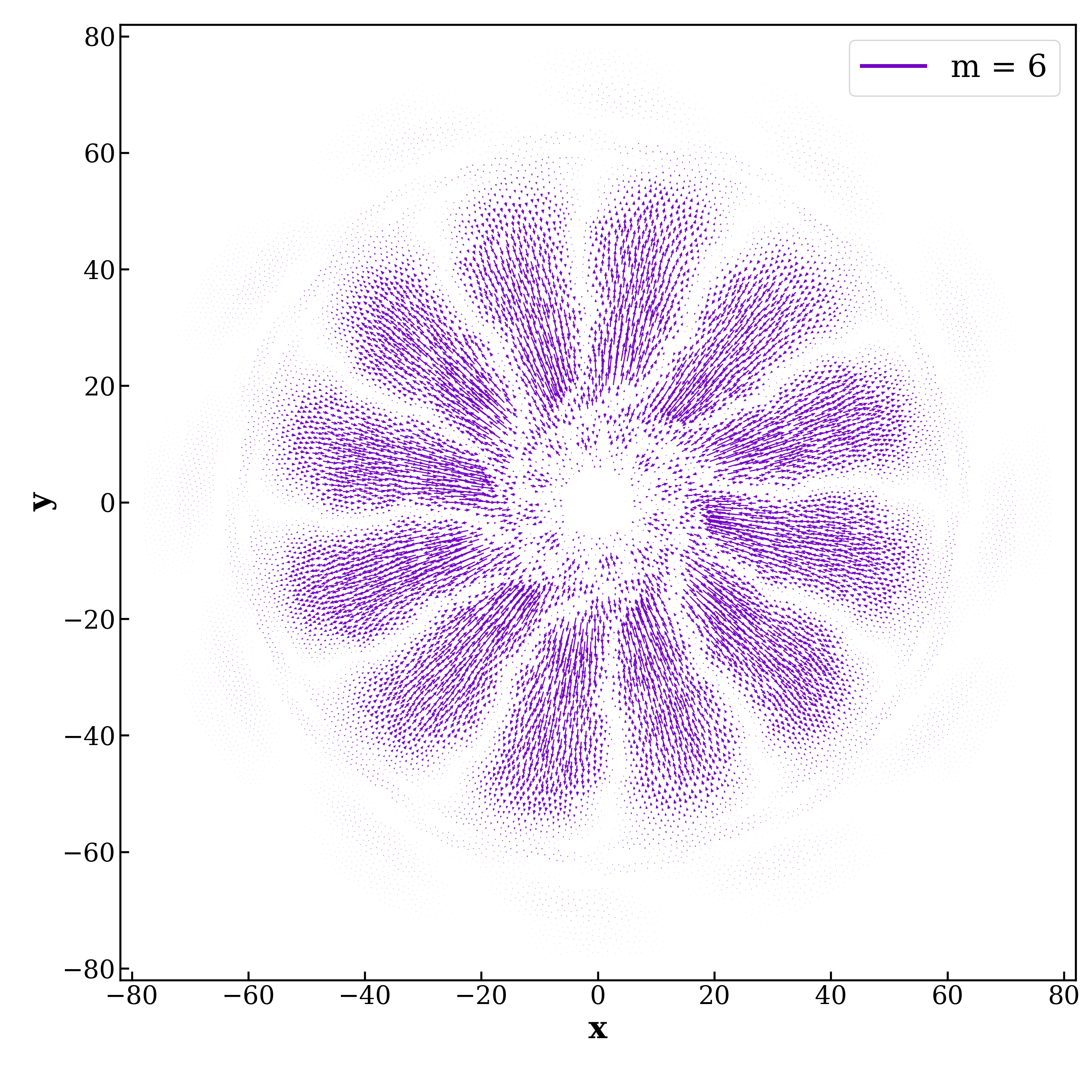} & \includegraphics*[width=0.24\textwidth]{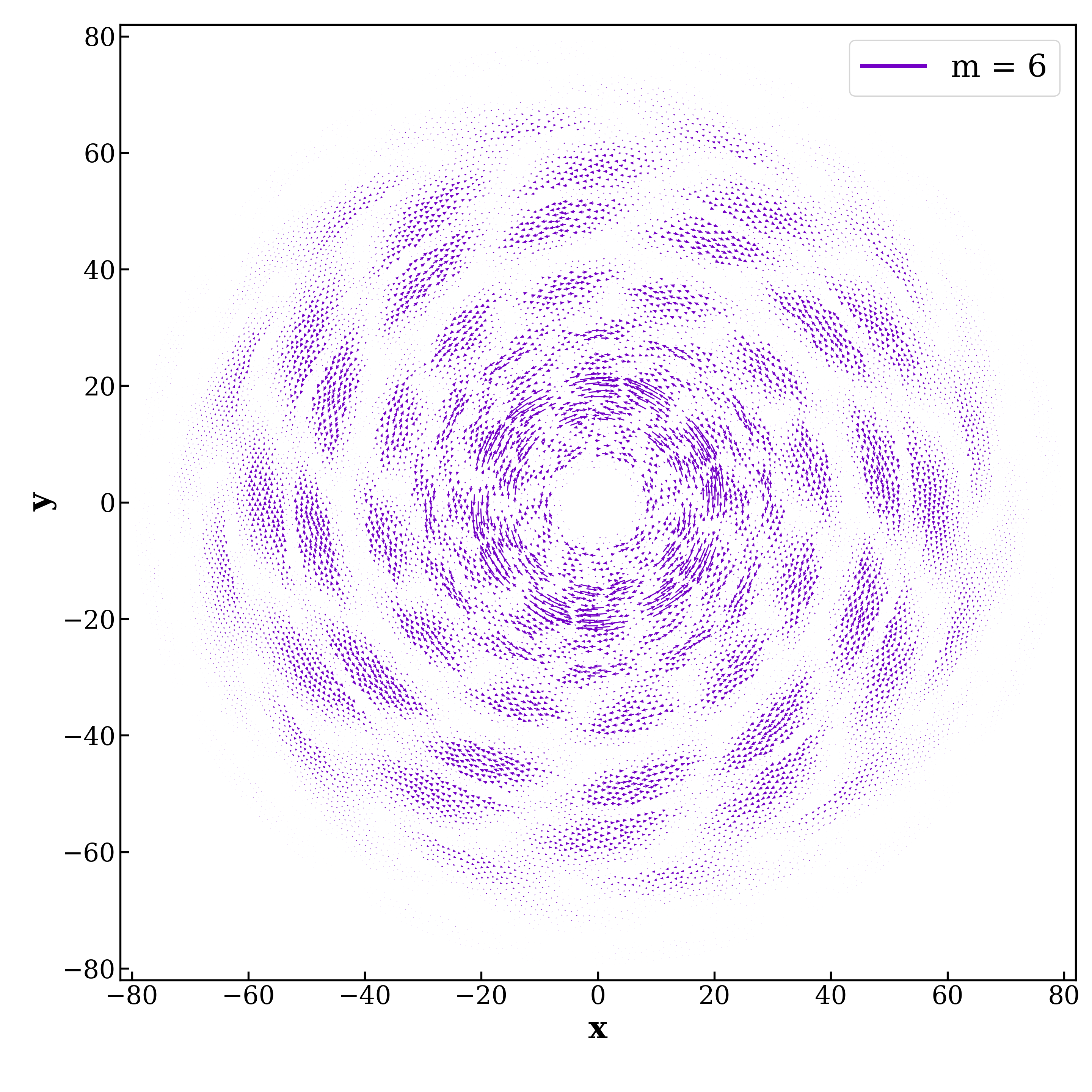} 
		\end{tabular}
	\end{center}
	\caption{Left and right panels correspond to the radial and tangential displacement vector field respectively, for the \textit{even} Fourier harmonics $m=2,4,6$ which are computed from the total displacement field shown in Fig.~\ref{fig:anomalousMap}.}
	\label{fig:evenHarmonic}
\end{figure}

\begin{figure}
	\begin{center}
		\begin{tabular}{cc}
			\includegraphics*[width=0.24\textwidth]{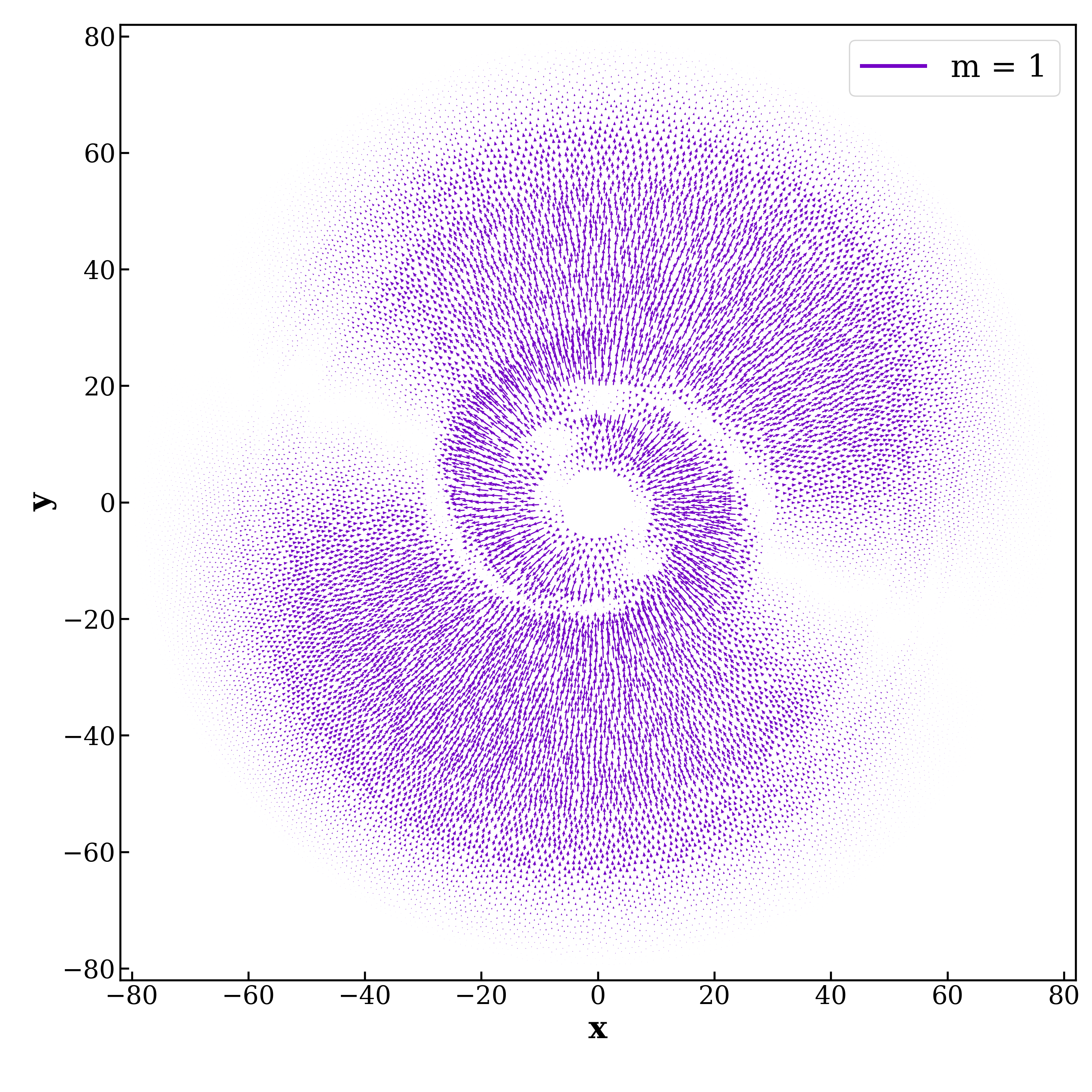} & \includegraphics*[width=0.24\textwidth]{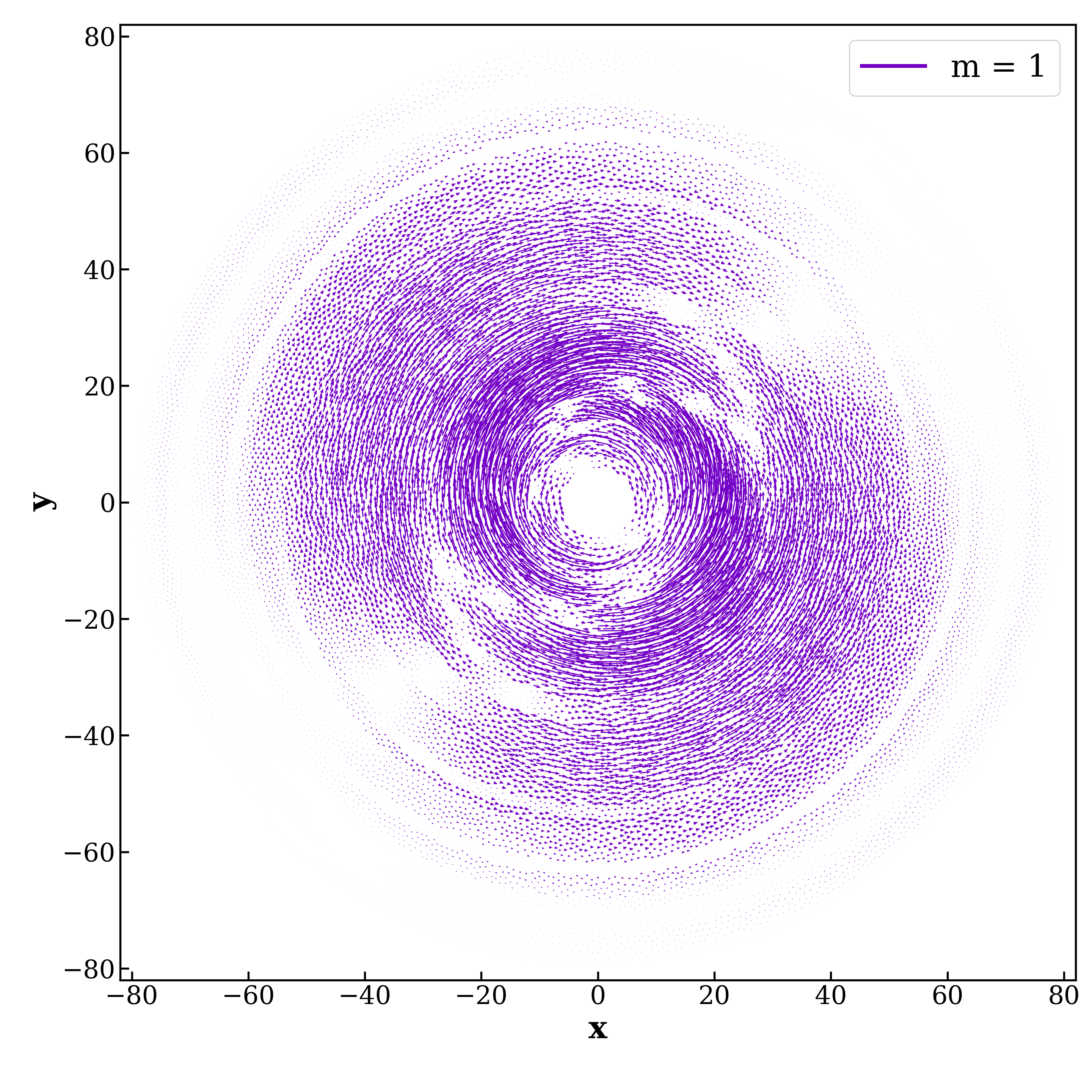} \\ 
			\includegraphics*[width=0.24\textwidth]{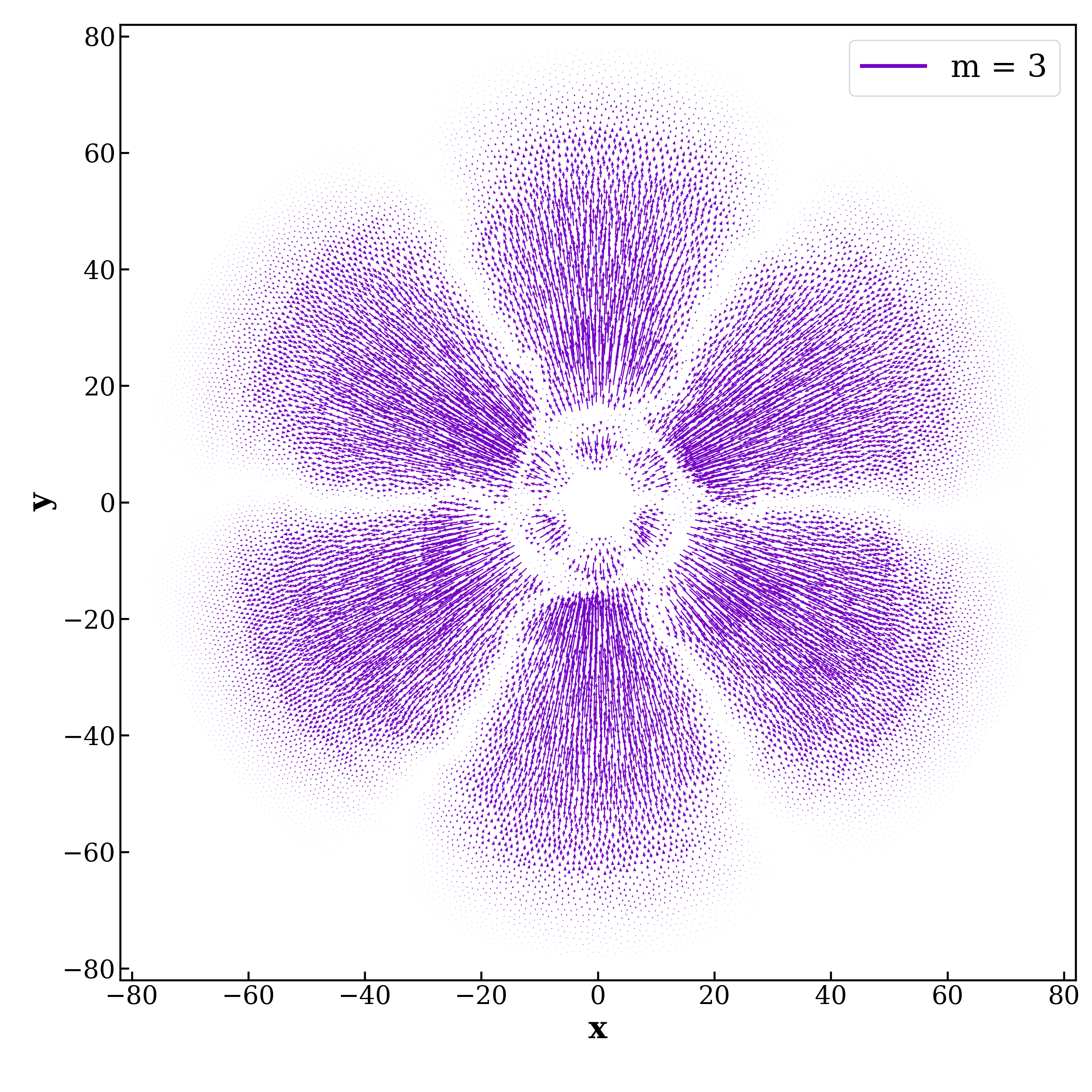} & \includegraphics*[width=0.24\textwidth]{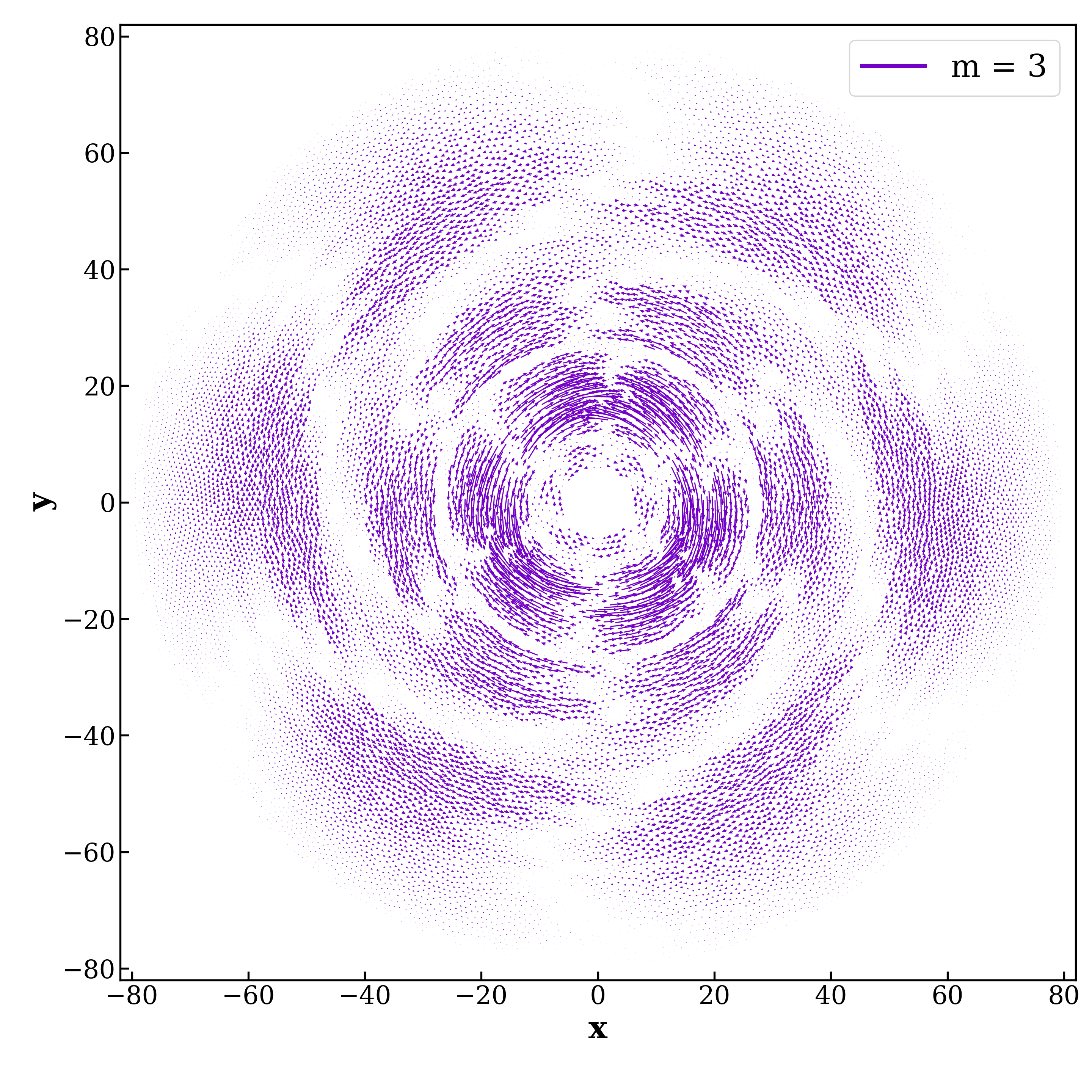} \\ 
			\includegraphics*[width=0.24\textwidth]{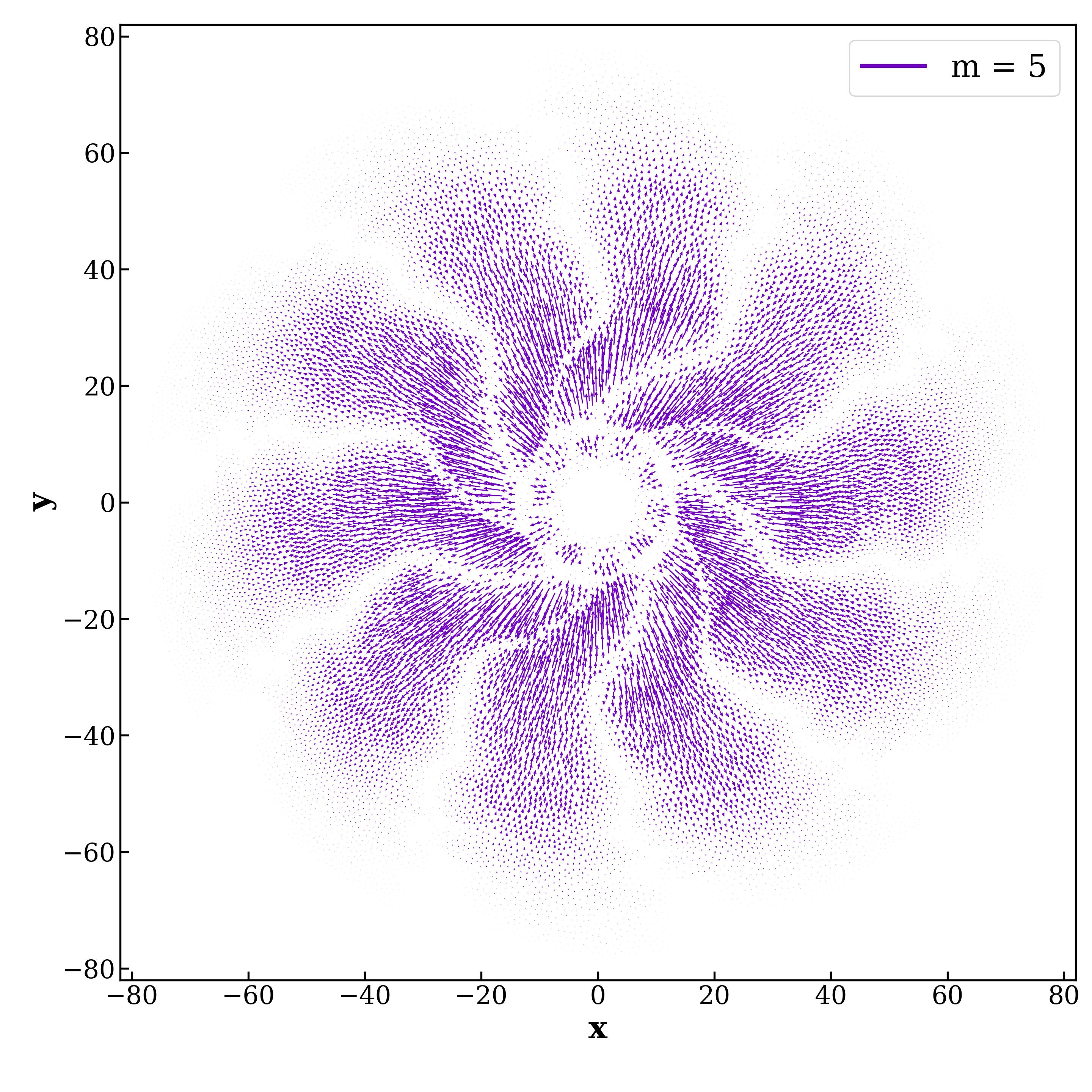} & \includegraphics*[width=0.24\textwidth]{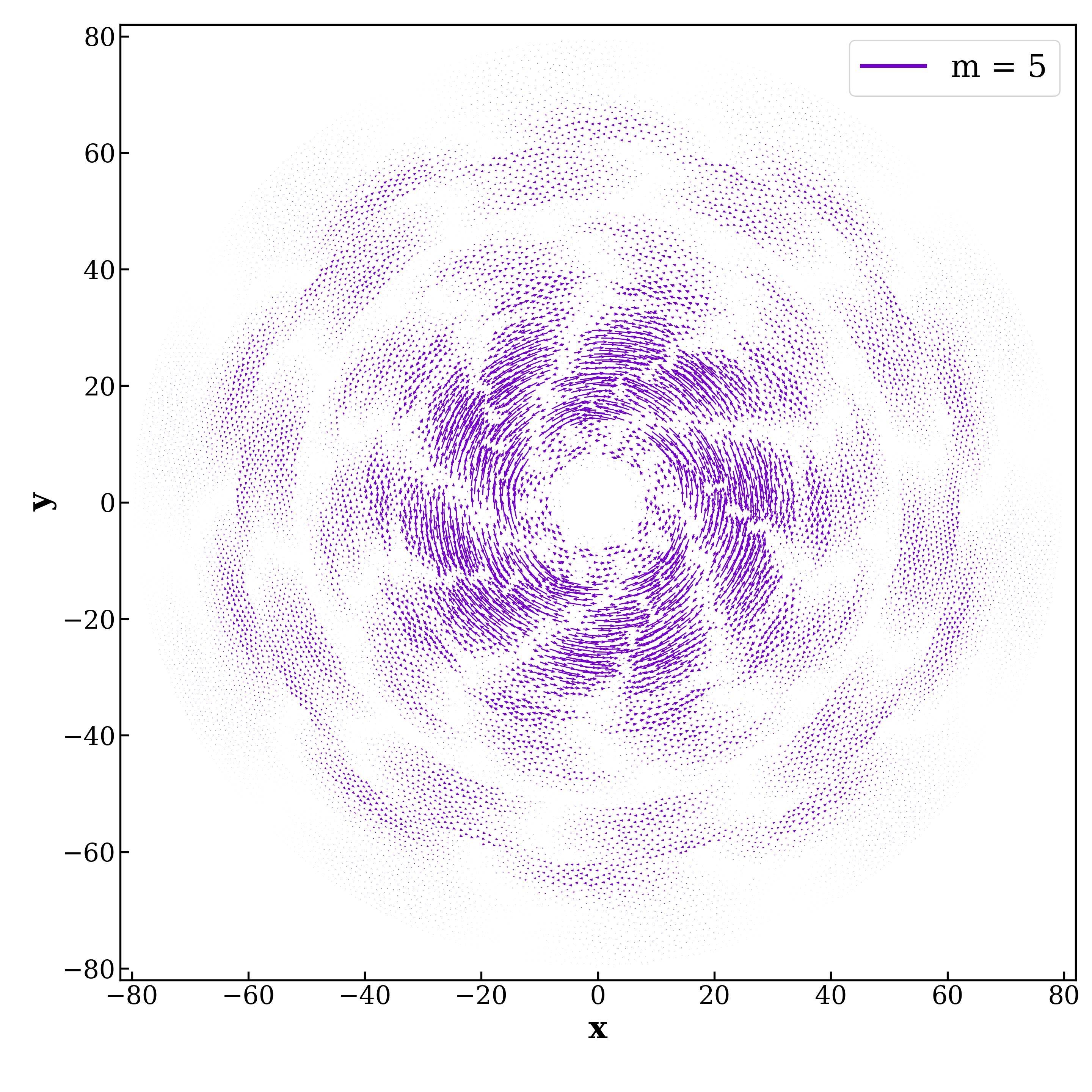}
		\end{tabular}
	\end{center}
	\caption{Left and right panels correspond to the radial and tangential components of the displacement field respectively, for the \textit{odd} Fourier harmonics $m=1,3,5$ which are computed from the total displacement field shown in Fig.~\ref{fig:anomalousMap}.}
	\label{fig:oddHarmonics}
\end{figure}

In Fig.~\ref{fig:evenHarmonic} we show the radial and tangential components corresponding to the even harmonics, i.e., for the Fourier modes $m=2, 4, 6$.  In Fig.~\ref{fig:oddHarmonics} we exhibit the odd harmonics, i.e., the Fourier modes $m=1, 3, 5$. All these modes have been extracted from the total displacement vector field in Fig.~\ref{fig:anomalousMap} representing the anomalous response.

For the present analytical work, the imposed Eshelby geometry selects $m = 2$, so we solve that sector exactly within the screened theory, fit its displacement amplitudes, and use the resulting screening parameters to calculate strain and stress as discussed in subsequent sections. The other measured harmonics demonstrate that a single microscopic amorphous realization contains modes beyond the leading continuum second Fourier mode.

However, it is relevant to ask, how the other modes will influence the dominant Eshelby mode for $m=2$. If the imposed deformation amplitude is $\epsilon$, we can write
\begin{equation}
	\mathbf d = \epsilon\mathbf d^{(1)} +\epsilon^2\mathbf d^{(2)} +\epsilon^3\mathbf d^{(3)} +\cdots .	
\end{equation}
Then $\epsilon\rightarrow0$ gives predominantly
$
	\mathbf d\simeq\epsilon\mathbf d^{(1)}	
$
so linear response dominates and mode coupling becomes negligible. For a leading $m=2$ response,
$
	d^{(1)}\sim\cos2\theta,	
$.
The quadratic nonlinearities give
\begin{equation}
	\epsilon^2\cos^22\theta =	\frac{\epsilon^2}{2}(1+\cos4\theta),
\end{equation}
so they generate $m=0,4$. Cubic nonlinearities give
\begin{equation}
	\epsilon^3\cos^32\theta = \frac{\epsilon^3}{4} (3\cos2\theta+\cos6\theta),	
\end{equation}
so they generate corrections to $m=2$ and create $m=6$. Therefore, schematically,
\begin{equation}
	H_2\sim O(\epsilon), \qquad H_4\sim O(\epsilon^2), \qquad H_6\sim O(\epsilon^3)	
\end{equation}
for this simplest symmetry-preserving expansion. Here $H_2, H_4 \cdots$ correspond to second, fourth \textit{harmonics} etc. As strain increases, the ratios
\begin{equation}
	\frac{H_4}{H_2}\sim O(\epsilon), \qquad \frac{H_6}{H_2}\sim O(\epsilon^2)	
\end{equation}
grow, and nonlinear mode coupling becomes increasingly visible.

Fundamentally, there are really two separate explanations for the complexity of the anomalous response of the displacement field: amorphous heterogeneity + nonlinear mode coupling. The first can exist already in the nominally linear regime of a particular realization; the second grows with deformation. Separating those experimentally/numerically by their amplitude scaling would be our future extension, while allowing the present work to remain a clean analytical $m=2$ theory. \textit{In the next subsection we briefly describe the analytical treatment of higher-order Fourier modes which is analogous to analytic solution obtained for $m=2$ \cite{Kumar2023, Kumar2024}.}

\subsection{Analytical treatment for higher-order Fourier Modes}
In our analytical treatment of the Eshelby problem we focused on $m=2$. For arbitrary integer $m$, we have 
\begin{equation}
	\Phi(r,\theta)=\Phi_m(r)e^{im\theta},
	\qquad
	\Psi(r,\theta)=\Psi_m(r)e^{im\theta},	
\end{equation}
with
\begin{equation}
	\mathcal L_m= \frac{d^2}{dr^2}+ \frac1r\frac{d}{dr} -\frac{m^2}{r^2}.	
\end{equation}
Then \textbf{for either even or odd $m$},
\begin{align}
	\alpha\mathcal L_m\Phi_m +\kappa_e^2\Phi_m	+\kappa_o^2\Psi_m=0, \nonumber \\
	\mathcal L_m\Psi_m	+\kappa_e^2\Psi_m	-\kappa_o^2\Phi_m=0.
\end{align}
These are precisely the general-$m$ equations. Now assuming a radial eigenfunction $Z_m(\kappa r)$ satisfying
\begin{equation}
	\mathcal L_m Z_m(\kappa r) = -\kappa^2Z_m(\kappa r).	
\end{equation}
Substitution gives the same $2\times2$ algebraic problem which was obtained for $m=2$,
\begin{equation}
	\begin{pmatrix}
		\kappa_e^2-\alpha\kappa^2 & \kappa_o^2\\
		-\kappa_o^2 & \kappa_e^2-\kappa^2
	\end{pmatrix}
	\begin{pmatrix}\Phi\\\Psi\end{pmatrix}=0.	
\end{equation}
Therefore the characteristic equation contains \textbf{no $m$}:
\begin{equation}
	(\kappa_e^2-\alpha\kappa^2) (\kappa_e^2-\kappa^2) +\kappa_o^4=0.	
\end{equation}
Hence
\begin{equation}
	\kappa_A,\kappa_B\ \text{are identical for all }m	
\end{equation}
and likewise
\begin{equation}
	\ell_A,\ell_B\ \text{are identical for all }m.
\end{equation}

Therefore, changing $m$ changes the \textbf{Bessel order}, not $\kappa_A,\kappa_B$ or the polarization ratios. So the general solution is
\begin{align}
	\Phi_m(r)=& C_{A,m}J_m(\kappa_A r)+D_{A,m}Y_m(\kappa_A r) \nonumber \\
	& + C_{B,m}J_m(\kappa_B r) +D_{B,m}Y_m(\kappa_B r),
\end{align}	
and
\begin{align}
	\Psi_m(r)=& \ell_A \left[ C_{A,m}J_m(\kappa_A r) +D_{A,m}Y_m(\kappa_A r) \right] \nonumber \\
	&+
	\ell_B \left[ C_{B,m}J_m(\kappa_B r)+D_{B,m}Y_m(\kappa_B r) \right].
\end{align}
Therefore the sequence is simply
\begin{align}
	m=1: J_1,Y_1; ~~m=2: J_2,Y_2, \nonumber \\
	m=3: J_3,Y_3,~~	m=4: J_4,Y_4,	
\end{align}
and so forth. The displacement amplitudes also have exactly the same general structure:
\begin{equation}
	U_m=\Phi_m'+\frac{im}{r}\Psi_m; \qquad	V_m=\frac{im}{r}\Phi_m-\Psi_m'.
\end{equation}
For example, the odd \(m=3\) solution is
\begin{align}
	\Phi_3=& C_{A,3}J_3(\kappa_A r)+D_{A,3}Y_3(\kappa_A r)  \nonumber \\
	&+ C_{B,3}J_3(\kappa_B r)+D_{B,3}Y_3(\kappa_B r),	
\end{align}
with
\begin{equation}
	\Psi_3=
	\ell_A[\cdots]_A+\ell_B[\cdots]_B,	
\end{equation}
and
\begin{equation}
	U_3=\Phi_3'+\frac{3i}{r}\Psi_3, \qquad V_3=\frac{3i}{r}\Phi_3-\Psi_3'.
\end{equation}
So we can now organize the theory as:
\begin{equation}
	\text{general screened solution} = \sum_{m=-\infty}^{\infty} \mathbf d_m(r)e^{im\theta},
\end{equation}
where \textit{every $m$} obeys the same screened radial operator and has the same two screening scales,
\begin{equation}
	\kappa_A^{-1},\qquad \kappa_B^{-1},	
\end{equation}
but different
\begin{equation}
	J_m,\quad Y_m,\quad C_{A,m},D_{A,m},C_{B,m},D_{B,m}.
\end{equation}
The coefficients encode \textit{how strongly each angular sector is actually excited}.

For our ideal ellipse-like inclusion,
\begin{equation}
	\mathbf d(r,\theta+\pi)=\mathbf d(r,\theta)	
\end{equation}
in the local polar basis. Since $e^{im(\theta+\pi)} =(-1)^m e^{im\theta}$, centrosymmetry requires $(-1)^m=1$, so only $m=0,2,4,6,\ldots$ occur for the perfectly symmetric continuum source. Odd modes require that this $\pi$-rotation symmetry be broken.  In a single amorphous realization, the microscopic environment is not exactly centrosymmetric. Consequently, there is no reason for $	m=1,3,5,\ldots$ to vanish identically. They can be excited by structural disorder/asymmetry even though the imposed idealized inclusion is predominantly $m=2$.

\subsection{Boundary Conditions}

\subsubsection{Radial Boundary Condition}
It should be noted that the $\cos2\theta$ deformation is an $m = 2$ or quadrupolar deformation. We employ two boundary conditions to describe the deformation of an initially circular inner boundary of radius $r_{\rm in}$ into an ellipse. The deformation amplitude is defined by $\delta=a/r_{\rm in}=1+\epsilon$, where $a$ is the semi-major axis and $\epsilon$ measures the fractional extension along the major axis. Exact conservation of the enclosed area, $\pi ab=\pi r_{\rm in}^2$, requires $b=r_{\rm in}/(1+\epsilon)=r_{\rm in}(1-\epsilon)+O(\epsilon^2)$. Thus, to linear order, an area-preserving ellipse has semi-axes $a=r_{\rm in}(1+\epsilon)$ and $b=r_{\rm in}(1-\epsilon)$. 

The affine pure-shear boundary condition, $x'=(1+\epsilon)x$ and $y'=(1-\epsilon)y$, generates exactly an ellipse with these semi axes and therefore preserves its area to $O(\epsilon)$, with the first area correction appearing only at $O(\epsilon^2)$. The purely radial condition, $u_r(r_{\rm in},\theta)=\epsilon r_{\rm in}\cos2\theta$, $u_\theta=0$, likewise produces the same elliptical distortion to linear order and has no first-order change in area because the angular average of $\cos2\theta$ vanishes. The distinction is geometric: the affine prescription produces an exact ellipse, whereas the radial curve $r(\theta)=r_{\rm in}[1+\epsilon\cos2\theta]$ represents an ellipse only to linear order. Nevertheless, within the linear theory used here, both boundary conditions describe the same area-preserving $m=2$ deformation.

The radial boundary condition is
\begin{equation}
	r(\theta) = r_{in} + \delta r(\theta), ~~\text{with}~~\delta r(\theta) = \epsilon r_{in}\cos2\theta
\end{equation}
And the proposed boundary displacement is then given by
\begin{equation}
	d_r(r_{in,\theta}) = \epsilon r_{in}\cos2\theta, ~~~~ d_{\theta}(r_{in},\theta) = 0
\end{equation}

\subsubsection{Affine Pure Shear}
Note that the radial boundary condition applied to inner circular boundary assumes that when the inner circle is transformed to an ellipse, the displacement of the boundary and particles associated with it in this transformation is completely radial. Theoretically the inner circular boundary follows this constraint but the particles in its close proximity are not bound to follow this constraint. Moreover, the radial BCs are better choice for the continuum but in discrete particle systems, although this works very well, in actual scenarios there is always a tangential drift and hence tangential displacement to the boundary particles. 
Therefore, we adopt pure affine shear to the inner circle which transforms it to an ellipse, and preserves the area. First, we perform affine Cartesian pure shear 
\begin{align}
	x^{\prime} &= (1+\epsilon)x \nonumber \\
	y^{\prime} &= (1-\epsilon)y.
\end{align}
For a point initially at circular boundary ($x=r_{in}\cos\theta, y=r_{in}\sin\theta$) the above affine transformation gives following radial and tangential displacement $d_x$ and $d_y$,
\begin{align}
	d_x	&=  x^{\prime} - x = \epsilon r_{in} \cos\theta \nonumber \\
	d_y &=  y^{\prime} - y = -\epsilon r_{in} \sin\theta 
\end{align}
In the polar basis, these displacement fields can be written as
\begin{align}
	d_r	&=  d_x \cos\theta + d_y \sin\theta \nonumber \\
	d_{\theta} &= -d_x \sin\theta + d_y \cos\theta
\end{align}
which when substituted the values of ($d_x, d_y$) becomes,
\begin{align}
	d_r	&=  \epsilon r_{in}\cos 2\theta  \nonumber \\
	d_{\theta} &= -\epsilon r_{in} \sin 2\theta 
\end{align}
Thus pure affine shear applied to a circular boundary essentially gives the desired elliptical inclusion with both radial and tangential displacement to the boundary.

Note that to first-order both the imposed boundary conditions preserve the area. A point on the original circle can be written as
$
	x=R\cos\theta, y=R\sin\theta,	
$,
so that after the affine pure shear
\begin{equation}
	X=R(1+\epsilon)\cos\theta,\qquad
	Y=R(1-\epsilon)\sin\theta.	
\end{equation}
This is an ellipse with semi-major and semi-minor axes given by
$
	a=R(1+\epsilon), b=R(1-\epsilon).	
$
Its area is therefore
\begin{equation}
	A_{\rm aff}=\pi ab	=\pi R^2(1-\epsilon^2).	
\end{equation}
Consequently the affine pure-shear deformation preserves area \textit{to first order}.
This is also immediately apparent from
\begin{equation}
	\nabla\cdot\mathbf u = \partial_xu_x+\partial_yu_y = \epsilon-\epsilon=0.
\end{equation}
So the infinitesimal strain is traceless and hence locally area preserving.

\subsection{Displacement-Fields}
\begin{figure}
	\centering
	\includegraphics[width=0.38\textwidth]{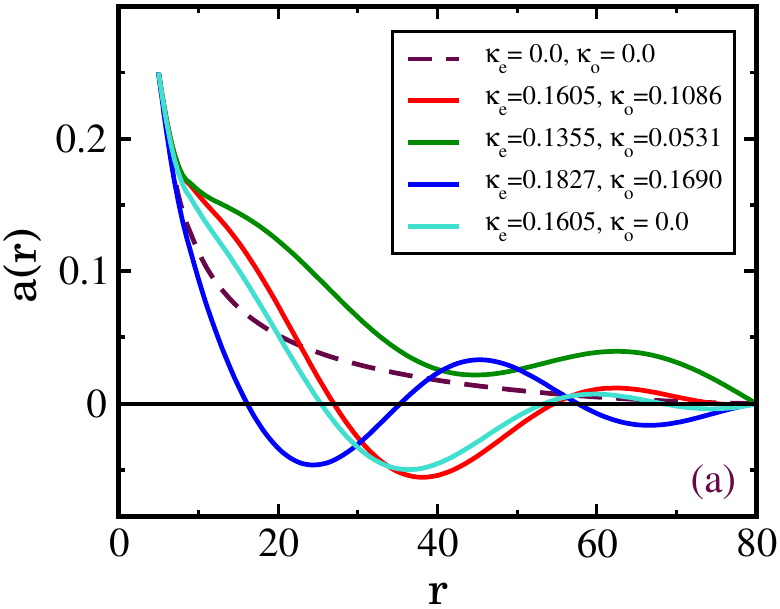}
	\hfill
	\includegraphics[width=0.38\textwidth]{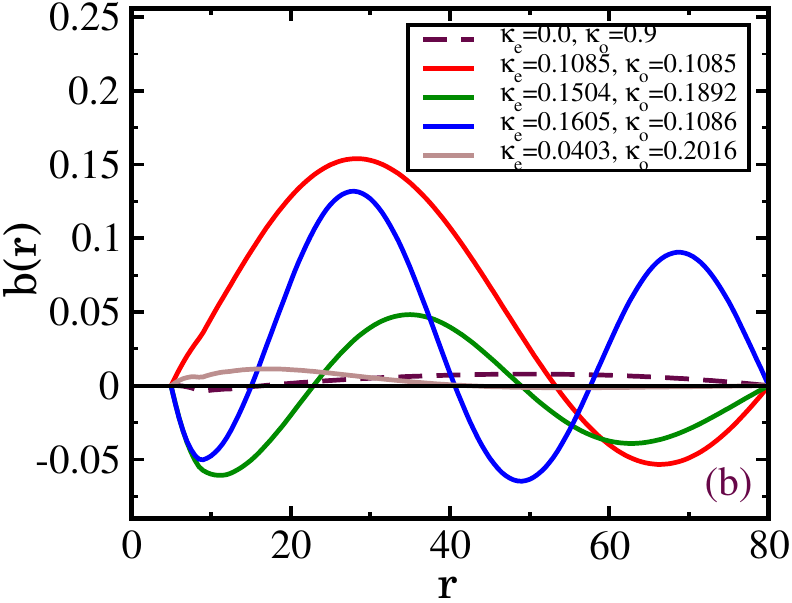}
	\caption{(a) radial displacement field and (b) tangential displacement field, computed from analytical solution. The parameters are $\mu=6.15, \lambda=65$, $r_{in}=5$, $r_{out}=80$, $\epsilon r_{in} =0.25$ using the radial boundary condition.}
	\label{fig:compareSchematic}
\end{figure}

Note that our original imposed transformation corresponds to the $m=2$ Fourier mode which captures the geometry of the Eshelby transformation. Therefore here we present the displacement fields computed essentially for the second harmonic, consistent with our imposed geometry. In Fig.~\ref{fig:compareSchematic}, we show typical angle averaged displacement plots obtained from analytical solution for $m=2$, for different combinations of screening parameters. We clearly observe non monotonic behavior in both radial and tangential displacement vector fields. Note that the odd components $A_o(r)$, and $B_o(r)$ are negligible in the numerically produced samples, so we have not shown here. However, they are significant in frictional granular solids as shown in Sect.~\ref{granular} where we discuss screening in the anomalous response of such solids. 
\begin{figure}
	\centering
	\includegraphics[width=0.32\textwidth]{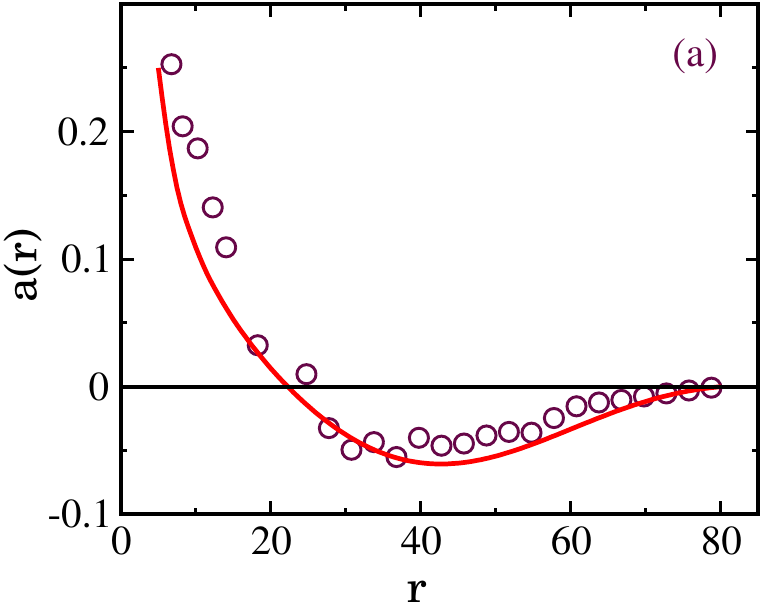}
	\hfill
	\includegraphics[width=0.32\textwidth]{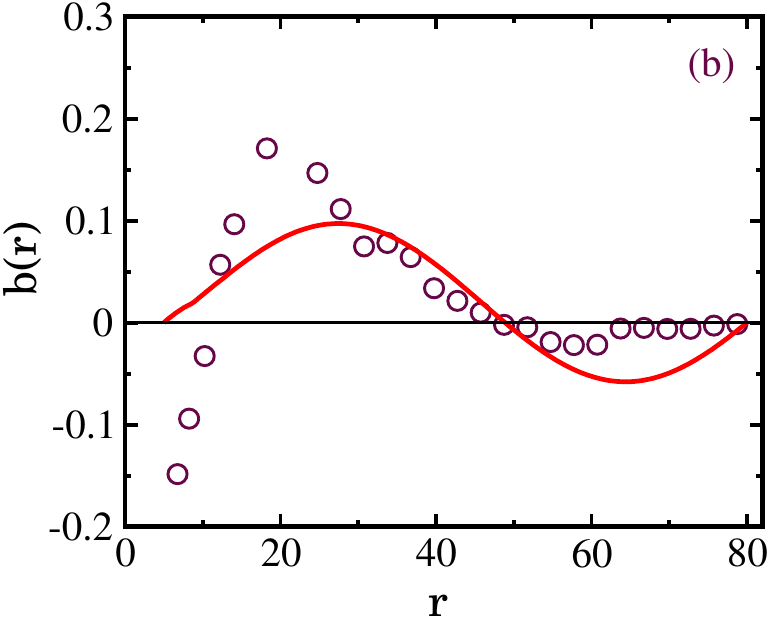}
	\caption{(a) Radial, and (b) tangential displacement corresponding to Fig.~\ref{fig:anomalousMap}. The parameters are $\kappa_e = 0.09677$, $\kappa_o =0.08065$, $\mu=6.15, \lambda=65$, $r_{in}=5$, $r_{out}=80$; and $\epsilon r_{in} =0.25$ using the radial boundary condition.}
	\label{fig:anomalyRadBC}
\end{figure}

\begin{figure}
	\centering
	\includegraphics[width=0.32\textwidth]{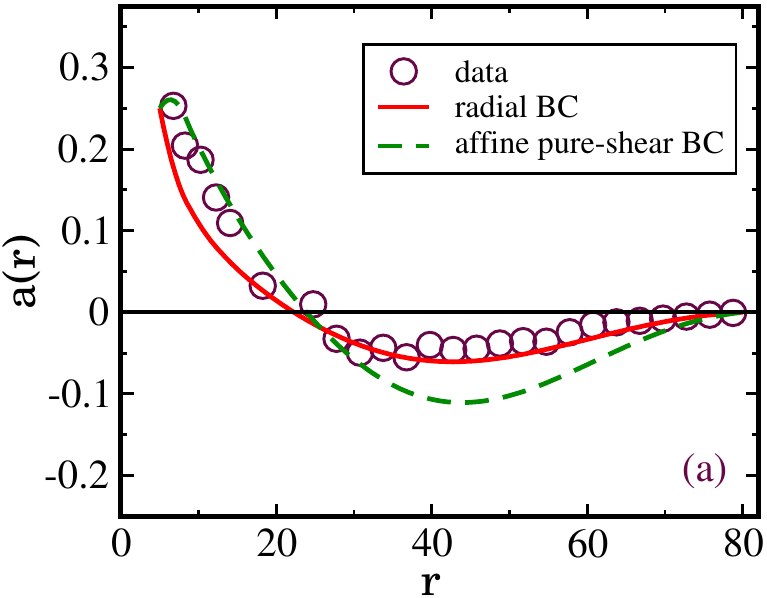}
	\hfill
	\includegraphics[width=0.32\textwidth]{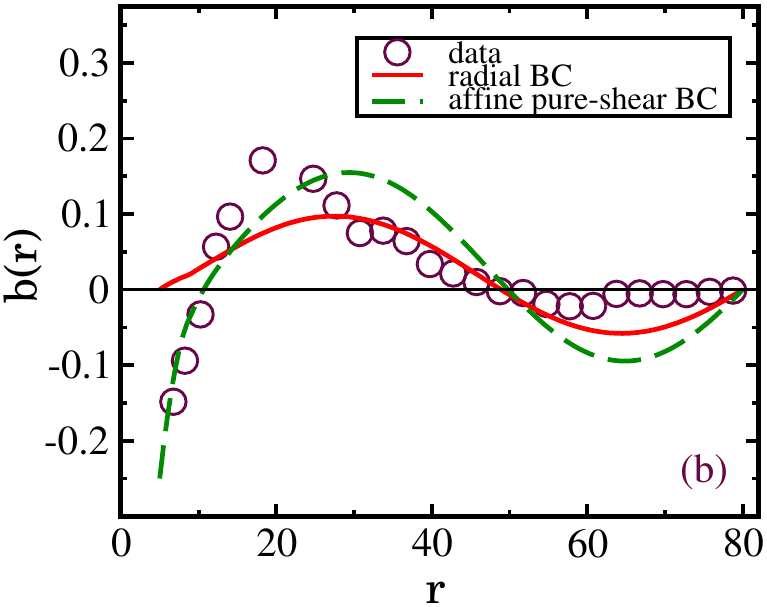}
	\caption{(a) Radial, and (b) tangential displacement corresponding to Fig.~\ref{fig:anomalousMap}, showing the comparison of fitting numerical data with the theory in radial boundary condition with that of affine pure-shear boundary conditions.}
	\label{fig:anomalyCompare}
\end{figure}

Consider now the angle-averaged radial and tangential components of the  displacement vector field shown in Fig.~\ref{fig:anomalousMap}. As can be seen, the data presented in Fig.~\ref{fig:anomalyRadBC} exhibit strong deviations from the predictions of the classical Eshelby solution. In this case the function $a(r)$ crosses zero and comes back from the negative side, showing that the power law expected from the Eshelby solution is strongly screened. We note that the agreement here is still quite good, although we have used only one sample for the comparison. The nature of the deviations indicates that the small $\epsilon$ assumption in the analytic solutions already captures the essential features. 
\begin{figure}
	\centering
	\includegraphics[width=0.32\textwidth]{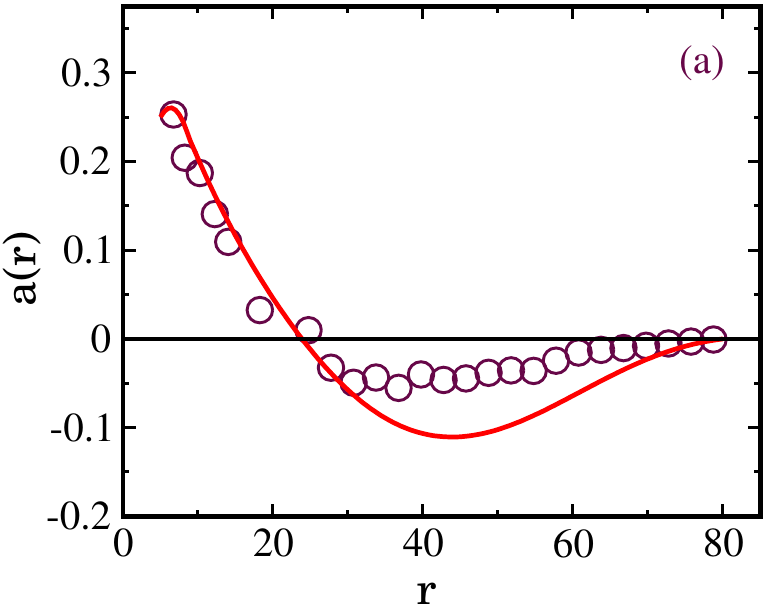}
	\hfill
	\includegraphics[width=0.32\textwidth]{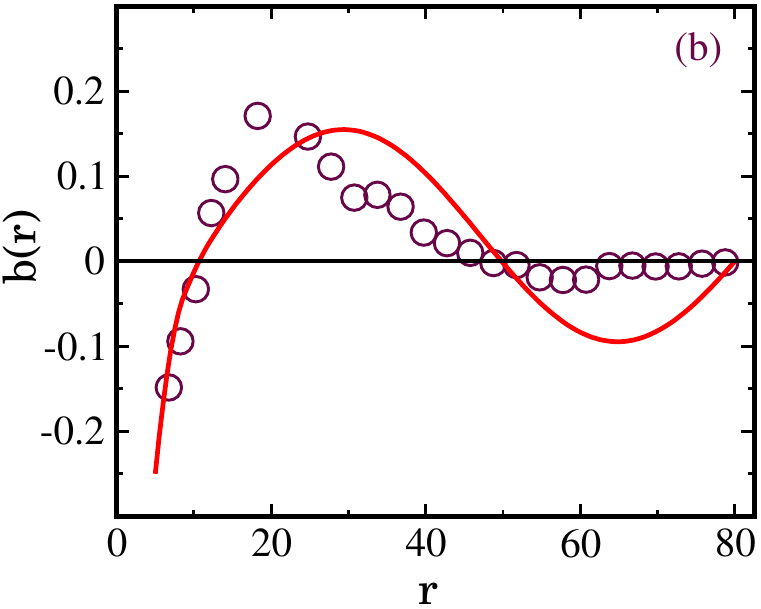}
	\caption{(a) Radial, and (b) tangential displacement corresponding to Fig.~\ref{fig:anomalousMap}. The parameters are $\kappa_e = 0.09778$, $\kappa_o =0.07664$, $\mu=6.15, \lambda=65$, $r_{in}=5$, $r_{out}=80$; and $\epsilon r_{in} =0.25$ using the \textit{affine pure-shear} boundary condition.}
	\label{fig:anomalyAffine}
\end{figure}
In Fig.~\ref{fig:anomalyAffine} we plot the data and use a affine pure-shear boundary condition in the analytical solution. We find that in this case the tangential component is substantially improved near the peak, but the agreement of the radial component suffers from extra assumption about the affine pure-shear boundary conditions.
In Fig.~\ref{fig:anomalyCompare} we present data with both the boundary conditions together to make a  better comparison. It shows that radial boundary condition with analytical solution is naturally more suitable to describe the observed anomalies in the displacement field. 

\subsection{Strain fields}
\begin{figure}[htbp]
	\centering
	\includegraphics[width=0.38\textwidth]{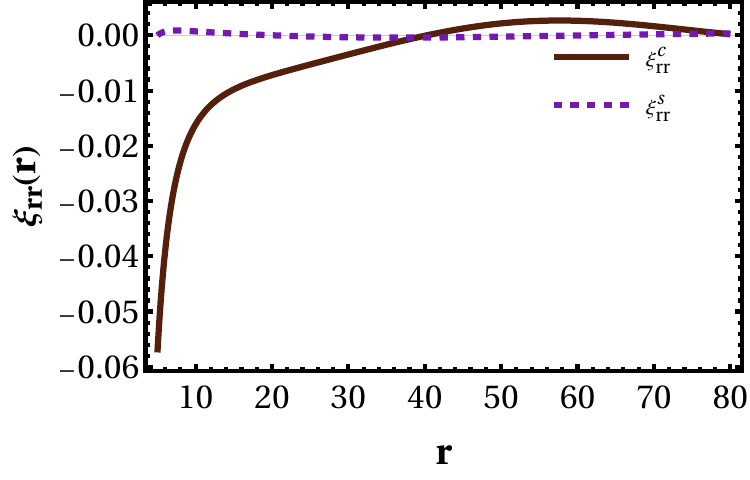}
	\hfill
	\includegraphics[width=0.38\textwidth]{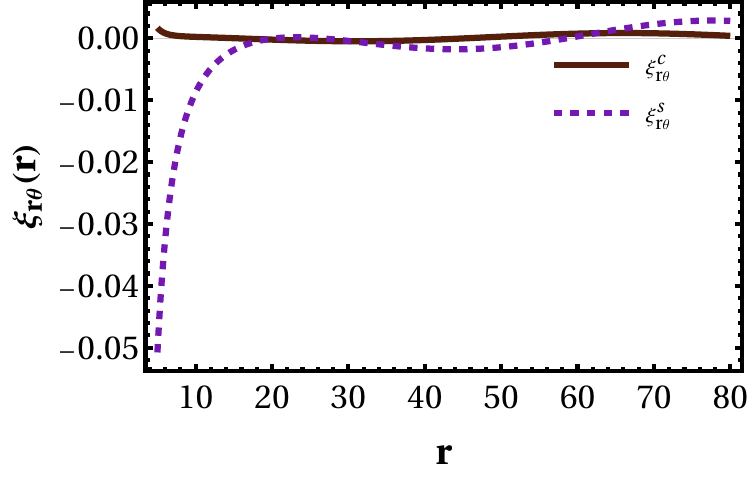}
	\hfill
	\includegraphics[width=0.38\textwidth]{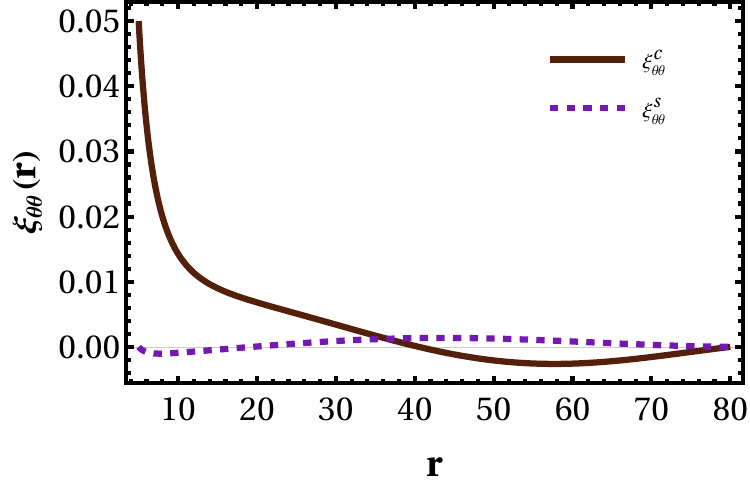}
	\caption{The three strain components $\xi_{rr}$, $\xi_{r\theta}$, and $\xi_{\theta\theta}$ in top, middle, and bottom panels respectively corresponding to the displacement field in Fig.~\ref{fig:anomalousMap}, and Fig.~\ref{fig:anomalyRadBC}. The parameters are $\kappa_e = 0.09677$, $\kappa_o =0.08065$, $\mu=6.15, \lambda=65$, $r_{in}=5$, $r_{out}=80$.}
	\label{fig:strain_fields}
\end{figure}

\begin{figure}[H]
	\centering
	\includegraphics[width=0.32\textwidth]{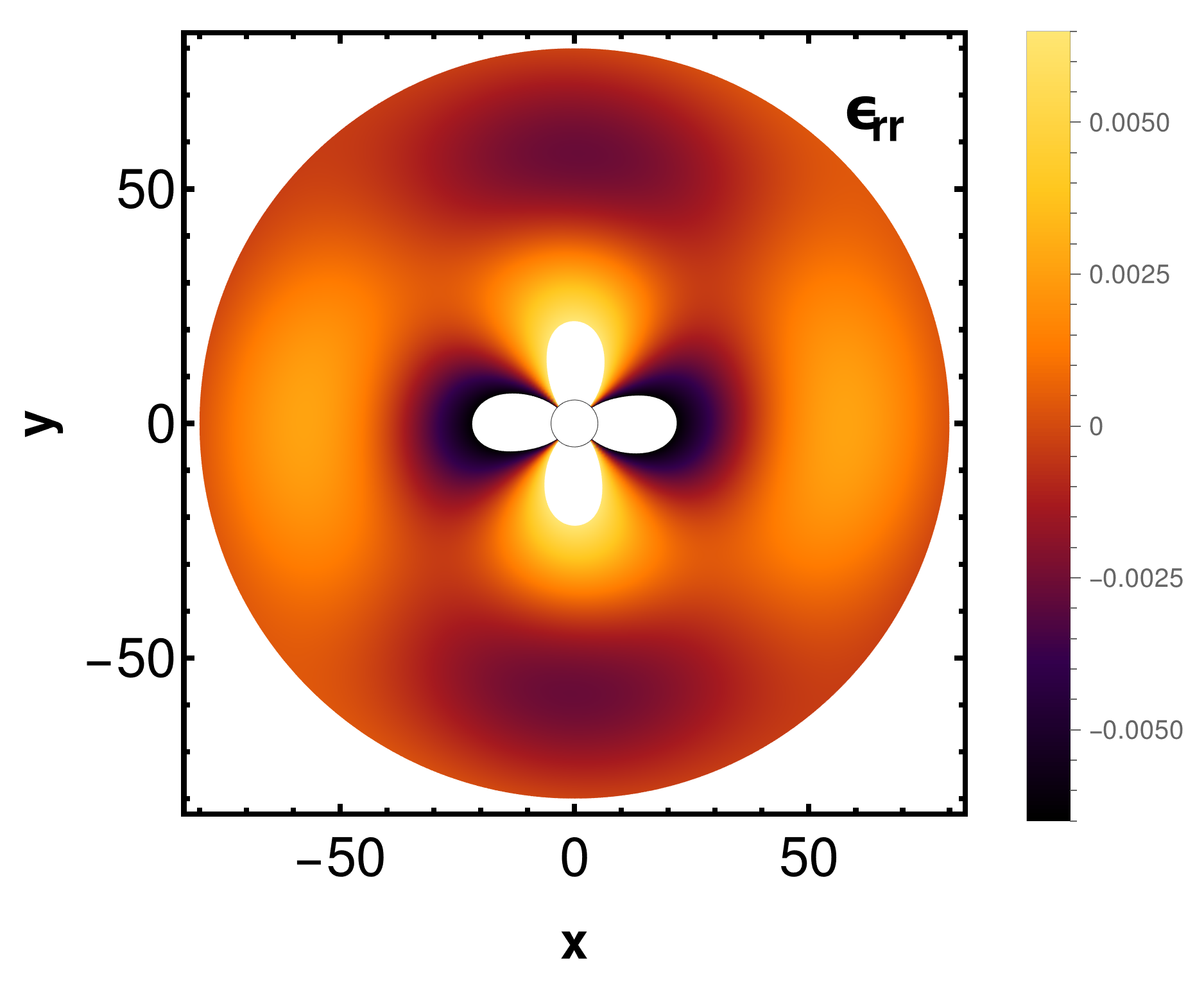}
	\hfill
	\includegraphics[width=0.32\textwidth]{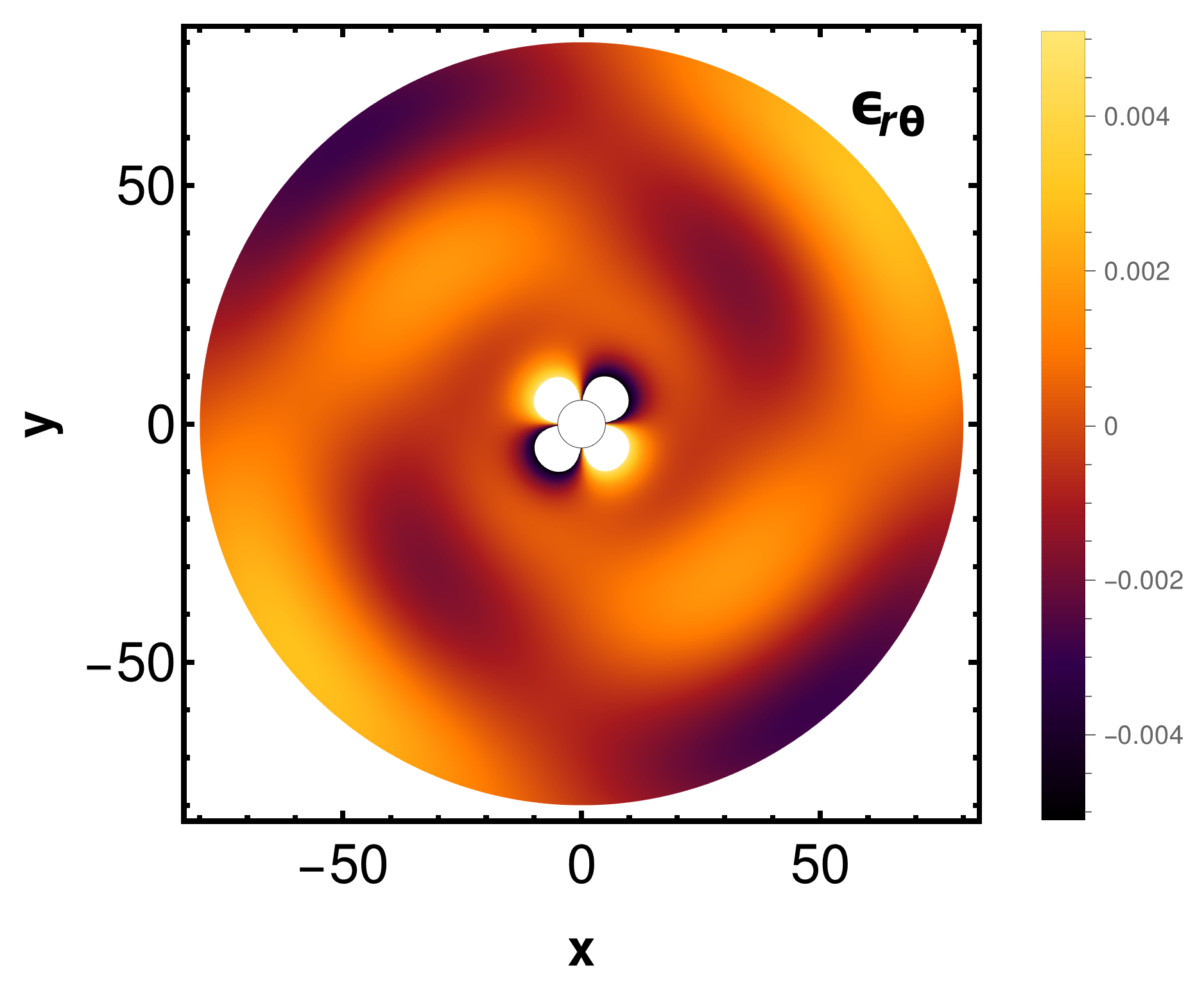}
	\hfill
	\includegraphics[width=0.32\textwidth]{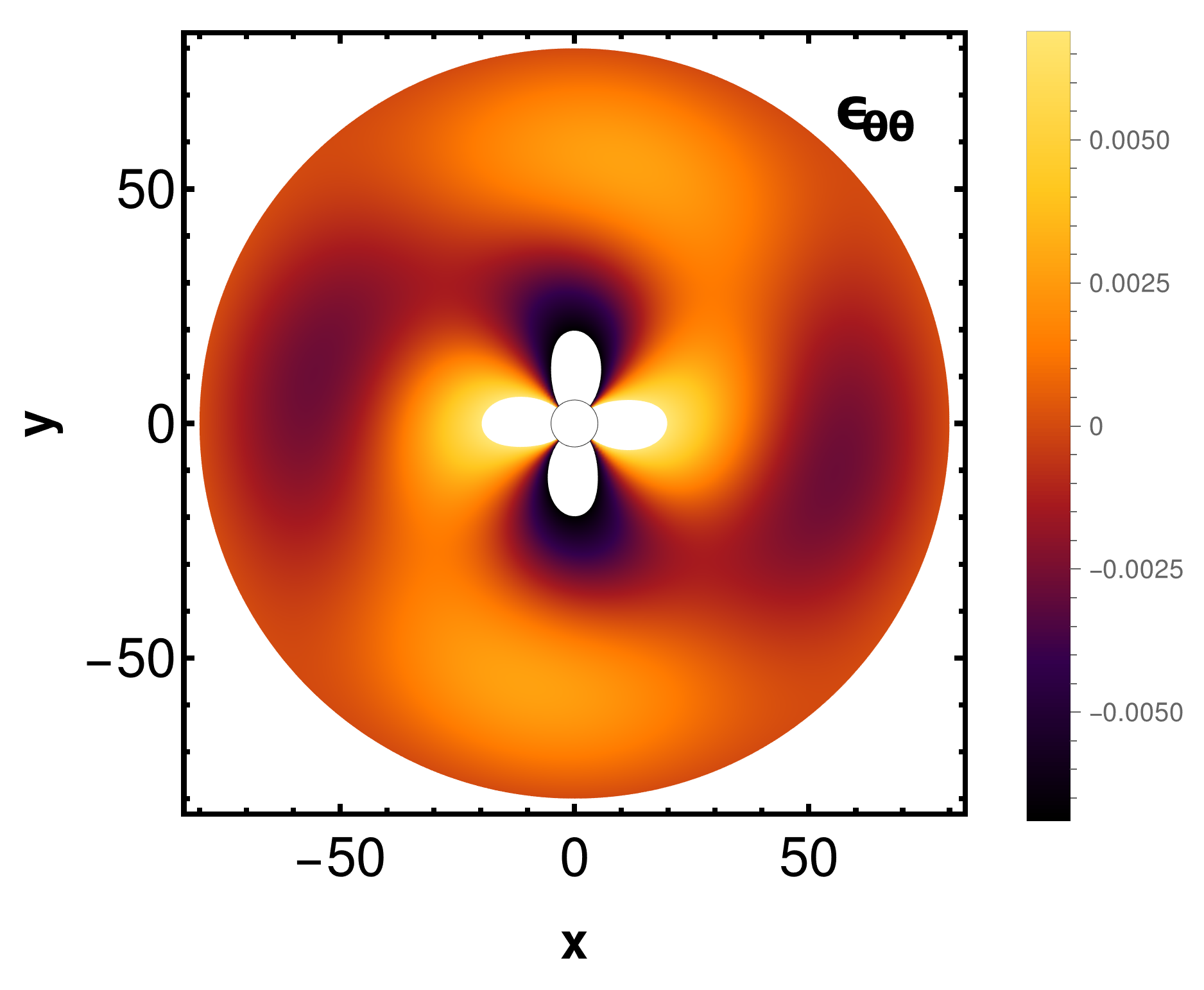}
	\caption{The strain field maps corresponding to three strain components $\xi_{rr}$, $\xi_{r\theta}$, and $\xi_{\theta\theta}$ in top, middle, and bottom panels respectively in Fig.~\ref{fig:strain_fields}.}
	\label{fig:strain_maps}
\end{figure}

\begin{figure}
	\centering
	\includegraphics[width=0.38\textwidth]{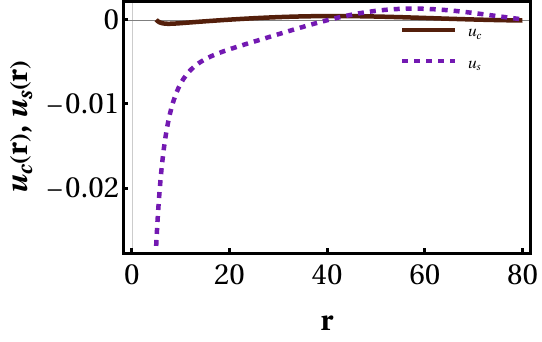}
	\caption{ Radial dependence of the two quadrature components of the fourfold Cartesian shear strain. The Cartesian shear field is written as $\epsilon_{xy}(r,\theta)=u_0(r)+u_c(r)\cos(4\theta)+u_s(r)\sin(4\theta)$. The solid curve shows the cosine quadrature $u_c(r)$, whereas the dotted curve shows the sine quadrature $u_s(r)$. These functions are not separate strain modes; together they specify the amplitude and angular orientation of the same fourfold contribution to $\epsilon_{xy}$. Their changing relative weights encode a radial rotation of the fourfold pattern.}
	\label{fig:Uxy}
\end{figure}

\begin{figure}
	\centering
	\includegraphics[width=0.38\textwidth]{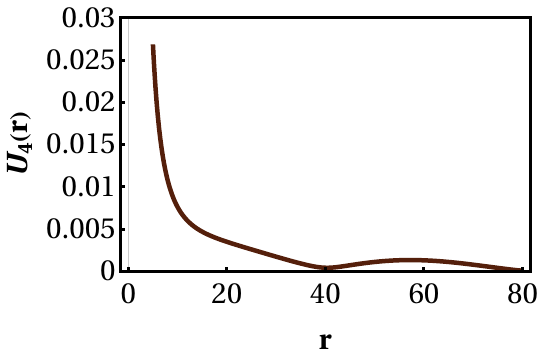}
	\caption{Amplitude of the fourfold Cartesian shear. $U4(r)=[uc2(r)+us2(r)]1/2$ is the rotationally invariant magnitude of the $m=4$ part of $\epsilon_{xy}$. It measures the strength of the fourfold shear pattern independently of its angular orientation.}
	\label{fig:Uabs}
\end{figure}

In this Subsection we examine the three strain components $\xi_{rr}$, $\xi_{r\theta}$, and $\xi_{\theta\theta}$,  shown in Fig.~\ref{fig:strain_fields} in the top, middle, and bottom panels respectively. The corresponding three spatial map are shown in Fig.~\ref{fig:strain_maps} in the same order. Further, in Fig.~\ref{fig:Uxy} we use the Cartesian shear strain 
\begin{align}
	\epsilon_{xy}(r,\theta)&=u_0(r)+u_c(r)\cos(4\theta)+u_s(r)\sin(4\theta) \nonumber\\
	&=u_0(r)+U_4(r)\cos[4\theta-\delta_4(r)]
\end{align}
to show the radial field components. The radial evolution of the Cartesian shear field is conveniently characterized by its fourfold amplitude $U_4(r)=\sqrt{u_c^2(r)+u_s^2(r)}$ and phase $\delta_4(r)=\arg[u_c(r)+iu_s(r)]$.
The amplitude $U_4(r)$ in Fig.~\ref{fig:Uabs} measures the strength of the fourfold shear pattern independently of its orientation. It is largest near the inner boundary, decreases strongly with distance, becomes very small at an intermediate radius, and subsequently develops a weaker outer maximum. The corresponding quantity $\delta_4(r)/4$ specifies the angular orientation of a representative positive shear lobe, modulo $90^\circ$, and therefore reveals how the fourfold pattern reorients with distance from the inclusion. Importantly, the rapid variation of $\delta_4/4$ near the minimum of $U_4$ should not be interpreted as a rapid physical rotation of a well-defined strain pattern: because the fourfold amplitude is nearly zero there, its orientation is intrinsically ill-defined.

\begin{figure}
	\centering
	\includegraphics[width=0.32\textwidth]{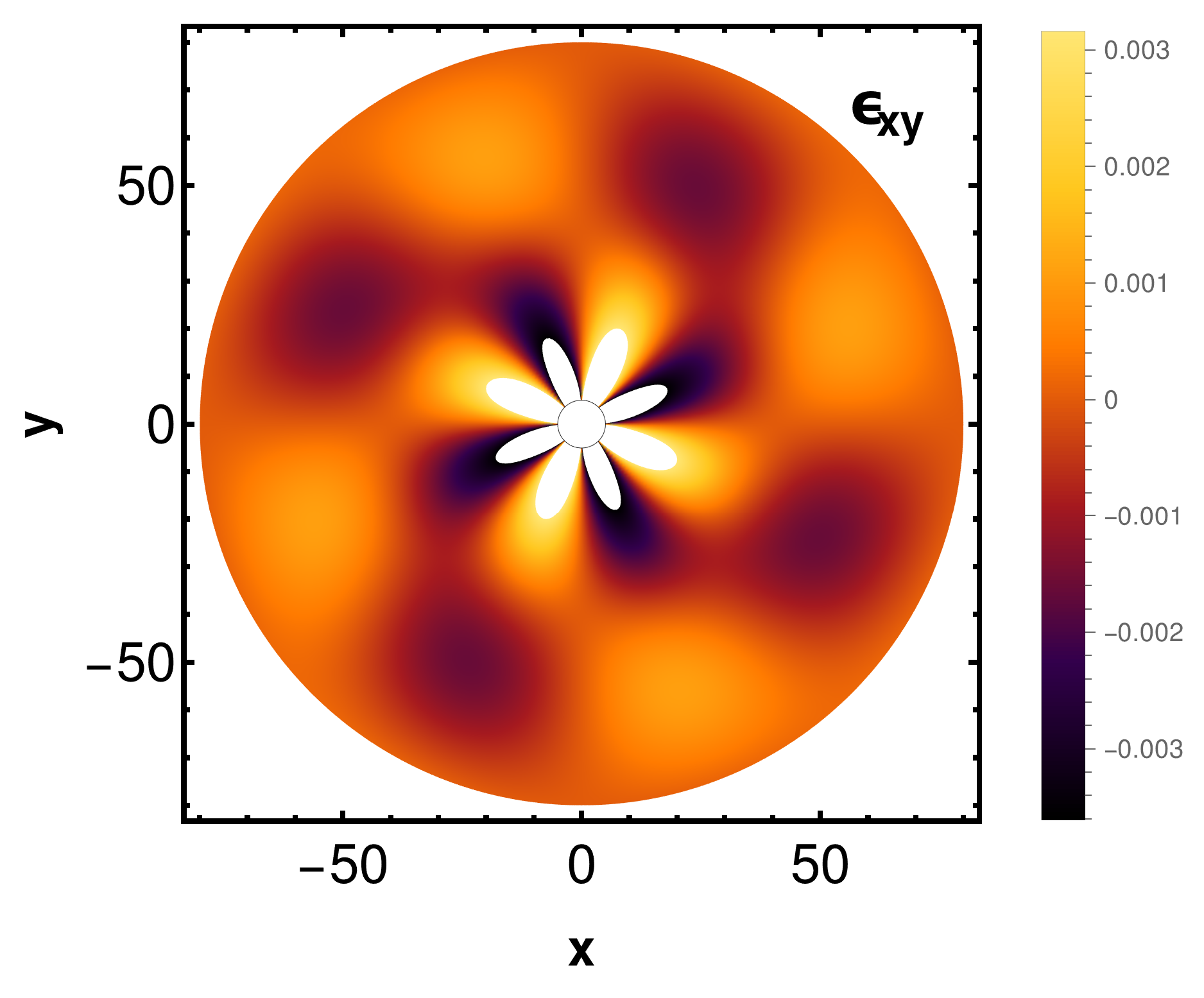}
	\caption{Spatial map of the Cartesian shear strain reconstructed from $\epsilon_{xy}$. The central circular region $r<r_{in}$ is excluded from the bulk. Alternating bright and dark lobes represent positive and negative $\epsilon_{xy}$, respectively revealing the fourfold shear structure surrounding the excluded inner region, while the boundaries between them correspond to values close to zero shear strain. The variation of lobe orientation with radius reflects the phase $\delta_4(r)/4$, while the intermediate weakening and subsequent re-emergence of the pattern correspond to the minimum and outer maximum of $U_4 (r)$.}
	\label{fig:uxyMap}
\end{figure}

\begin{figure}
	\centering
	\includegraphics[width=0.42\textwidth]{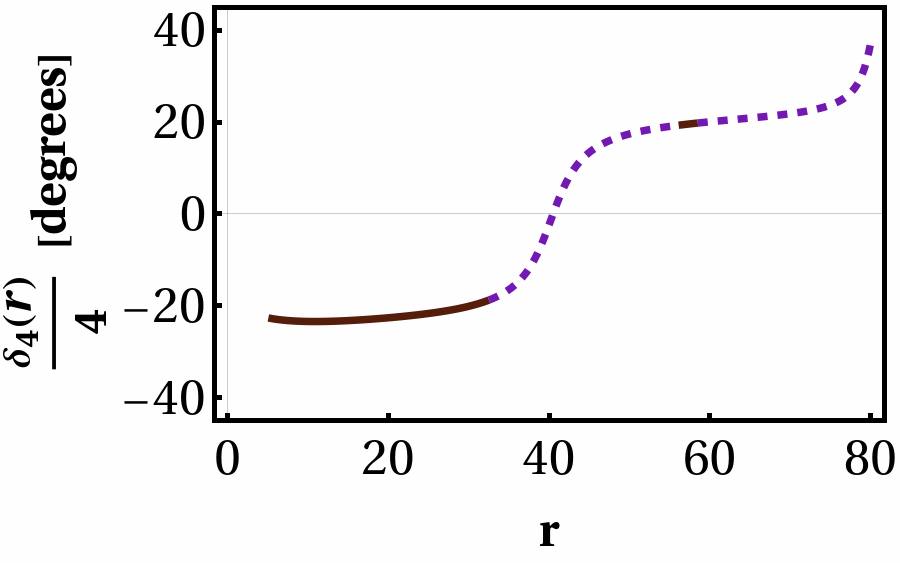}
	\caption{Radial orientation of the fourfold shear pattern. With $\delta_4 (r) = arg[u_c (r) + i u_s(r)]$, the fourfold contribution is $U_4\cos[4\theta -\delta_{4}]$, so that $\delta_{4}/4$ gives a representative positive-lobe direction modulo $90^{\circ}$. Solid segments indicate regions of appreciable $U_4$; the dotted segment corresponds to the low-amplitude region where the phase, although mathematically defined, is not a robust physical orientation.}
	\label{fig:phase}
\end{figure}

The physically meaningful observation is that the finite-amplitude inner and outer fourfold patterns possess different orientations, separated by a radial region in which the fourfold shear is strongly suppressed. This amplitude--phase evolution is directly visible in Fig.~\ref{fig:uxyMap} which represents the two-dimensional $\epsilon_{xy}(x,y)$ map; the alternating positive and negative lobes display the fourfold angular symmetry, their progressive change of direction reflects the variation of $\delta_4(r)/4$ (Fig.~\ref{fig:phase}), and their weakening followed by a weaker re-emergence that mirrors the non-monotonic behavior of $U_4(r)$. Thus, the three representations together show that screening produces not merely a decay of the shear field, but a spatially evolving fourfold strain pattern whose \textbf{magnitude and orientation both vary with radial distance}.

At this point one should add the comment that the displacement and the strain fields can be studied in various additional ways, and in particular it is worthwhile to explore the nature of the points at which the displacement field vanishes, where topological charges can be found. Such an analysis is offered in Appendix \ref{defects}.

\subsection{Granular Simulations}
\label{granular}
 Since granular matter is so prevalent in our world, it is important and interesting to examine whether our long-wavelength field theory applies verbatim to such materials.
For this purpose we simulated a two-dimensional granular model consisting of $N=6750$ equimolar frictionless bi-disperse disks and one central circular inclusion using the LAMMPS package. The disk radii are $R_1=0.5$ and $R_2=0.7$, with the small disk diameter defining the length unit; the inclusion radius is 5 times the small disk radius. Random configurations are generated under periodic boundary conditions in a square box of side length $L$, with the inclusion fixed at the box center. The repulsive Hertzian contact force is
\begin{equation}
	\mathbf{F}_{ij} =
	\begin{cases}
		\dfrac{4}{3}E^*\sqrt{\dfrac{R_iR_j}{(R_i+R_j)}}\,\Delta_{ij}^{3/2}\hat{\mathbf{n}}_{ij}, & \text{if } \Delta_{ij} > 0, \\[1.2em]
		0, & \text{if } \Delta_{ij} \le 0,
	\end{cases}
\end{equation}
where $\Delta_{ij}=R_i+R_j-r_{ij}$ is the overlap, $r_{ij}$ is the inter-particle distance, and $\hat{\mathbf{n}}_{ij}$ is the unit vector from particle $j$ to particle $i$. We set the areal mass density and the effective modulus $E^*$ to unity. Starting from a relaxed configuration at packing fraction $\phi_0=0.700$, the system is compressed by scaling all radii, including the inclusion radius. The packing fraction is increased in steps of $0.005$ up to $0.840$ and $0.001$ thereafter up to $0.900$. After each increment, the system is relaxed to mechanical equilibrium using FIRE with symplectic Euler integration. Coarse relaxation to a maximum particle-force magnitude of $3\times10^{-6}$ is followed by refinement to $5\times10^{-8}$, and compression proceeded only after the force criterion is satisfied.
\begin{figure}
	\centering
	\includegraphics[width=0.48\textwidth]{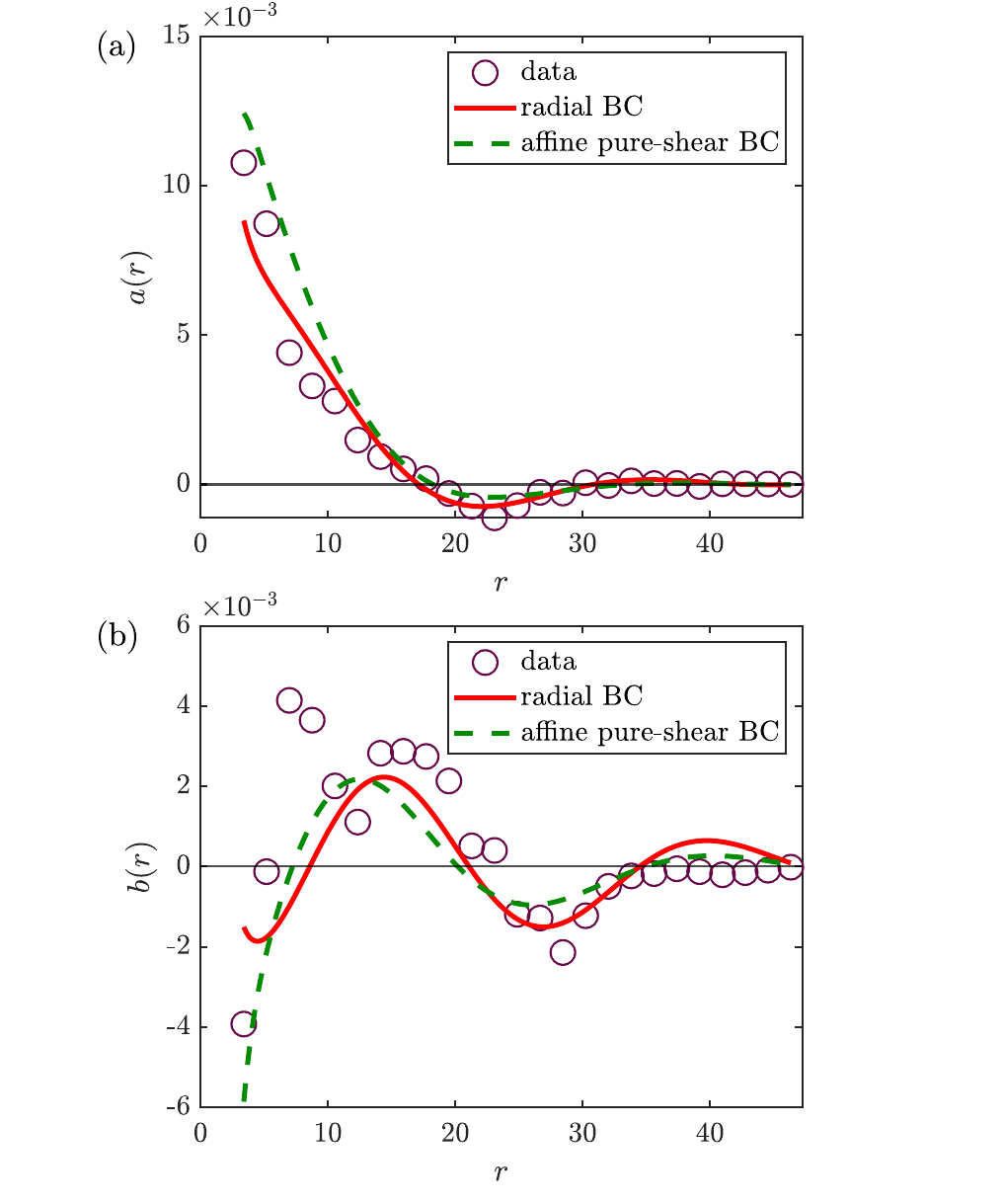}
	\caption{Top and bottom panel correspond to radial and tangential component of displacement field as a function of radial distance $r$. They correspond to the even coupling. }
	\label{fig:GReven}
\end{figure}
\begin{figure}
	\centering
	\includegraphics[width=0.48\textwidth]{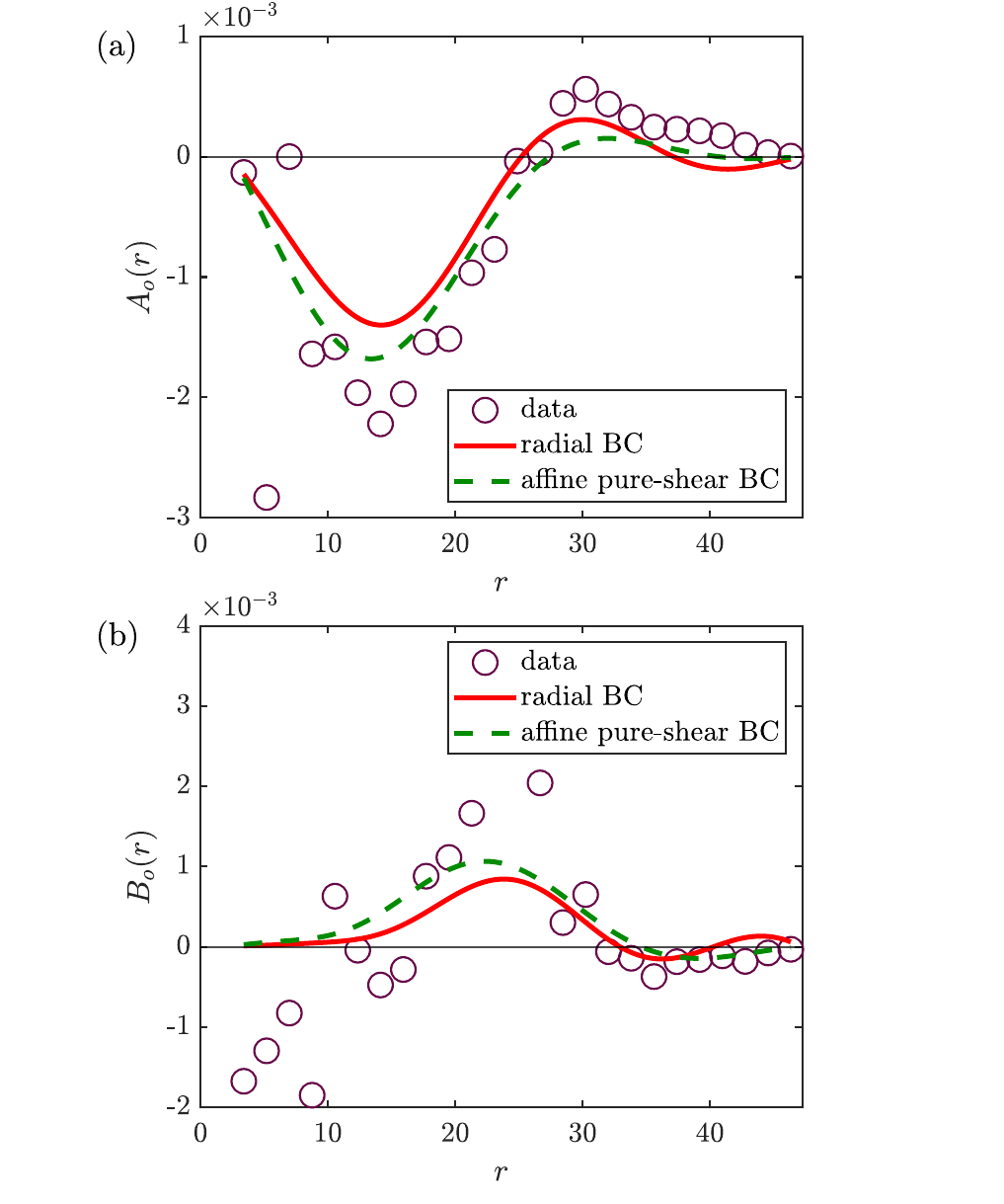}
	\caption{Top and bottom panel corresponding to the radial and tangential component of displacement field for odd coupling.}
	\label{fig:GRodd}
\end{figure}

As an additional variation to the previously described simulations, in the case of the granular amorphous solids we also changed the manner of introduction the Eshelby response. In the present case we did not change the inner circle to an inner ellipse in one step. Rather, 
the Eshelby response is probed by an athermal quasistatic deformation of the central inclusion at a selected packing fraction. An effectively circular outer boundary is imposed by immobilizing particles whose centers lay at $r>r_{\mathrm{out}}$, where $r$ is measured from the inclusion center and $r_{\mathrm{out}}=L/2-2R_1$. The boundary-particle set is selected before deformation and remains fixed throughout loading. The central inclusion was then transformed quasi-statically into an area-preserving ellipse with fixed center and orientation. Its semi-axes are $a_k=r_{\mathrm{in}}\delta_k$ and $b_k=r_{\mathrm{in}}/\delta_k$, aligned with the $x$ and $y$ axes, respectively. Here, $r_{\mathrm{in}}$ is the inclusion radius in the undeformed configuration and $\delta_k=1+k(\delta_f-1)/M$, with $k=1,\ldots,M$, $\delta_f=1.05$, and $M=250$. Disk--ellipse interactions used the same Hertzian energy, replacing the inclusion radius by the local radius of curvature $\rho$ at the closest point on the ellipse and the overlap by $R_i-s_i$, where $s_i$ is the normal distance from the disk center to the ellipse. After each deformation increment, the mobile particles are relaxed from the preceding equilibrium configuration using the same FIRE procedure and force criterion.

The Lam\'e coefficients $\mu$ and $\lambda$ are estimated from the undeformed packings under periodic boundary conditions. All particles, including the circular central inclusion, are relaxed using FIRE. The elastic tensor is then calculated using the Hessian-based method of Lema\^{i}tre and Maloney~\cite{06ML}, including non-affine relaxation. The effective Lam\'e coefficients are obtained from the isotropic part of the calculated elastic tensor.

Pronounced plastic rearrangements occur during the quasistatic Eshelby deformation of the granular packing. Figures~\ref{fig:GReven} and~\ref{fig:GRodd} show a representative response at packing fraction $\phi=0.845$, where the first plastic event occurs at $\delta=1.0046$. The inner and outer radii are $r_{\mathrm{in}}=2.5$ and $r_{\mathrm{out}}=47.25$, and the independently determined Lam\'e coefficients of the undeformed packing are $\mu=0.00104$ and $\lambda=0.0230$. The displacement field is measured relative to the undeformed configuration and averaged within concentric annuli. Figure~\ref{fig:GReven} shows the usual quadrupolar components $a(r)$ and $b(r)$, while Fig.~\ref{fig:GRodd} shows the odd components $A_o(r)$ and $B_o(r)$ within the same $m=2$ sector. These odd components have appreciable amplitudes and cannot be neglected in the comparison with theory. Keeping the measured elastic coefficients fixed, we jointly fit all four profiles using the screened solution with either radial or affine pure-shear boundary conditions at the inclusion. The fitted parameters are $(\kappa_E,\kappa_O)=(0.0113,0.0180)$ for the radial condition and $(0.0098,0.0179)$ for affine pure shear. Both prescriptions capture the main spatial trends, including non-monotonic profiles and sign changes, although noticeable deviations remain close to the inclusion. This is expected since there the use of a long wavelength theory suffers from the important granular structure. 

The important conclusion from the present simulations is that the basic agreement between the simulation results and the theory persists independently whether the deformation of the inner circle to an inner ellipse is done in one step or quasi-statically, as long as the quasi-static deformation is stopped with the first plastic avalanche event that creates the final displacement field. We stress again the analyzed displacement field is defined as the difference between the final and the initial configurations before any deformation is applied.  

\section{Summary and Concluding Remarks}\label{summary}

In summary, this paper provided a detailed analytical treatment of the outer Eshelby problem in amorphous solids in two dimensions. The main point is that whenever the Eshelby transformation of an inner circle to an inner ellipse results in an avalanche of system spanning displacement field, the classical solution of Eshelby loses relevance in favor of an anomalous displacement field characterized by screening and symmetry breaking. 

On the positive side, we have shown that the equations that we propose to describe the anomalous response, i.e Eqs.~(\ref{L1}) and (\ref{defgam}), can be solved analytically given boundary conditions at the inner and outer boundaries. One can imagine a number of boundary conditions, and we offered two of them, the first named ``radial boundary conditions" and the second ``affine pure shear" boundary condition. It turned out that the first was better suited for the radial component of the displacement field, and the second for the tangential component. It is not surprising that the long wavelength field theory can run into issues at the inner boundary, where the continuum approximation is very sensitive to the actual existence of an amorphous configuration of interacting particles. Nevertheless the analytic solutions presented above fitted quite well the measured ones, confirming without any doubt the very strong deviations from the classical Eshelby predictions. The effect of screening cannot be ignored once the avalanche of plastic events creates a complex and rich displacement field, and it does not resemble at all the analytic theory of Eshelby.

By examining two types of Eshelby deformations, a sudden one in the case of point particles and a quasi-static one in the case of granular matter, we confirmed that the basic equations Eqs.~(\ref{L1}) and (\ref{defgam}) provide very reasonable agreement between the analytic solutions and the measured data. Deviations exist, and we explained that they can be explained by higher order ($m\ne 2$) components that can be easily computed in the theory, but also from the random nature of the particle configurations, especially in the vicinity of the boundaries where the continuum theory is always having issues. These last issues are expected to reduce with increasing the system size, where the continuum theory is expected to become better and better. 

For the immediate future our plans include an experimental verification of the above conclusions. Data already exist, indicating very favorable agreements with the theory. We hope and trust that these experimental results will be offered for the scrutiny of the community quite soon.

\acknowledgments
IP acknowledges that this work was carried out in part at the Sino-Europe Complex Science Center, North University of China, Taiyuan, Shanxi 030051, China. Thanks to Profs. Guiquan Sun, Xiaofeng Luo, for helpful discussions.

\appendix

\section{Harmonic functions}
\label{harmonic}

Let $A=A(x,y)$, and $B=B(x,y)$. Then
\begin{eqnarray}
	&&\B \nabla A =   \begin{pmatrix}
		A_x (x,y)\\ 
		A_y(x,y) 
	\end{pmatrix}; ~~  \B  \nabla B =   \begin{pmatrix}
		B_x (x,y)\\ 
		B_y(x,y) 
	\end{pmatrix}; ~~\nonumber \\ &&\text{and}~~ J\B \nabla B = \begin{pmatrix}
		-B_y (x,y)\\ 
		B_x(x,y) 
	\end{pmatrix}.
\end{eqnarray}

Thus,
\begin{equation}
\B	\nabla A -J\B \nabla B =0, ~~\text{hence}~~ \B \nabla A= J\B \nabla B,
\end{equation}
and 
\begin{equation}
	A_x = - B_y, ~~ A_y = B_x, \implies A_{xx} = - B_{xy},~~ A_{yy} = B_{yx},
\end{equation}
Thus
\begin{equation}
	A_{xx}+A_{yy} =0; \implies \nabla^2A = 0; ~~\text{Similarly},~~~ \nabla^2B = 0.
\end{equation}
This shows that $A$ and $B$ are harmonic.

\section{Boundary conditions and the matrix}
\label{Boundary}

The four boundary equations are obtained by equating the coefficients multiplying $(e^{2i\theta})$. At $(r=r_{in})$,

\begin{align}
	&C_A
	\left(
	\kappa_A J_2'(\kappa_A r_{in})
	+
	\frac{2il_A}{r_{in}}J_2(\kappa_A r_{in})
	\right) \nonumber	\\
	+{}&
	D_A
	\left(
	\kappa_A Y_2'(\kappa_A r_{in})
	+
	\frac{2il_A}{r_{in}}Y_2(\kappa_A r_{in})
	\right) \nonumber	\\
	+{}&
	C_B
	\left(
	\kappa_B J_2'(\kappa_B r_{in})
	+
	\frac{2il_B}{r_{in}}J_2(\kappa_B r_{in})
	\right) \nonumber \\
	+{}&
	D_B
	\left(
	\kappa_B Y_2'(\kappa_B r_{in})
	+
	\frac{2il_B}{r_{in}}Y_2(\kappa_B r_{in})
	\right) = \epsilon r_{in},
\end{align}
and
\begin{align}
	&C_A
	\left(
	\frac{2i}{r_{in}}J_2(\kappa_A r_{in})
	-
	l_A  \kappa_A J_2'(\kappa_A r_{in})
	\right) \nonumber \\
	+{}&
	D_A
	\left(
	\frac{2i}{r_{in}}Y_2(\kappa_A r_{in})
	-
	l_A \kappa_A Y_2'(\kappa_A r_{in})
	\right) \nonumber \\
	+{}&
	C_B
	\left(
	\frac{2i}{r_{in}}J_2(\kappa_B r_{in})
	-
	l_B \kappa_B J_2'(\kappa_B r_{in})
	\right) \nonumber \\
	+{}&
	D_B
	\left(
	\frac{2i}{r_{in}}Y_2(\kappa_B r_{in})
	-
	l_B \kappa_B Y_2'(\kappa_B r_{in})
	\right)=0
\end{align}

At $(r=r_{out})$:
\begin{align}
	&C_A
	\left(
	\kappa_A J_2'(\kappa_A r_{out})
	+
	\frac{2il_A}{r_{out}}J_2(\kappa_A r_{out})
	\right) \nonumber \\
	+{}&
	D_A
	\left(
	\kappa_A Y_2'(\kappa_A r_{out})
	+
	\frac{2il_A}{r_{out}}Y_2(\kappa_A r_{out})
	\right)	\nonumber \\
	+{}&
	C_B
	\left(
	\kappa_B J_2'(\kappa_B r_{out})
	+
	\frac{2il_B}{r_{out}}J_2(\kappa_B r_{out})
	\right) \nonumber \\
	+{}&
	D_B
	\left(
	\kappa_B Y_2'(\kappa_B r_{out})
	+
	\frac{2il_B}{r_{out}}Y_2(\kappa_B r_{out})
	\right)= 0.
\end{align}
and
\begin{align}
	&C_A
	\left(
	\frac{2i}{r_{out}}J_2(\kappa_A r_{out})	-
	l_A  \kappa_A J_2'(\kappa_A r_{out})
	\right) \nonumber \\
	+{}&
	D_A
	\left(
	\frac{2i}{r_{out}}Y_2(\kappa_A r_{out})
	-
	l_A \kappa_A Y_2'(\kappa_A r_{out})
	\right)  \nonumber	\\
	+{}&
	C_B
	\left(
	\frac{2i}{r_{out}}J_2(\kappa_B r_{out})
	-
	l_B \kappa_B J_2'(\kappa_B r_{out})
	\right) \nonumber \\
	+{}&
	D_B
	\left(
	\frac{2i}{r_{out}}Y_2(\kappa_B r_{out})
	-
	l_B \kappa_B Y_2'(\kappa_B r_{out})
	\right)=0
\end{align}

Therefore, the boundary matrix is
\begin{equation} \label{BM}
	\begin{pmatrix}
		G_{11}&G_{12}&G_{13}&G_{14}\\
		G_{21}&G_{22}&G_{23}&G_{24}\\
		G_{31}&G_{32}&G_{33}&G_{34}\\
		G_{41}&G_{42}&G_{43}&G_{44}
	\end{pmatrix}
	\begin{pmatrix}
		C_A\\
		D_A\\
		C_B\\
		D_B
	\end{pmatrix}
	=
	\begin{pmatrix}
		\epsilon r_{in} \\
		0\\
		0\\
		0
	\end{pmatrix},
\end{equation}

We write the entries of the $G$ matrix for clarity:
\begin{align}
	G_{11}&=
	\kappa_A J_2'(\kappa_A r_{in})
	+
	\frac{2il_A}{r_{in}}J_2(\kappa_A r_{in}), \nonumber \\
	G_{12}&=
	\kappa_A Y_2'(\kappa_A r_{in})
	+
	\frac{2il_A}{r_{in}}Y_2(\kappa_A r_{in}), \nonumber \\
	G_{13}&=
	\kappa_B J_2'(\kappa_B r_{in})
	+
	\frac{2il_B}{r_{in}}J_2(\kappa_B r_{in}), \nonumber \\
	G_{14}&=
	\kappa_B Y_2'(\kappa_B r_{in})
	+
	\frac{2il_B}{r_{in}}Y_2(\kappa_B r_{in}),
\end{align}
\begin{align}
	G_{21}&=
	\frac{2i}{r_{in}}J_2(\kappa_A r_{in})
	-
	l_A  \kappa_A J_2'(\kappa_A r_{in}), \nonumber \\
	G_{22}&=
	\frac{2i}{r_{in}}Y_2(\kappa_A r_{in})
	-
	l_A \kappa_A Y_2'(\kappa_A r_{in}), \nonumber \\
	G_{23}&=
	\frac{2i}{r_{in}}J_2(\kappa_B r_{in})
	-
	l_B \kappa_B J_2'(\kappa_B r_{in}), \nonumber \\
	G_{24}&=
	\frac{2i}{r_{in}}Y_2(\kappa_B r_{in})
	-
	l_B \kappa_B Y_2'(\kappa_B r_{in}),
\end{align}

\begin{align}
	G_{31}&=
	\kappa_A J_2'(\kappa_A r_{out})
	+
	\frac{2il_A}{r_{out}}J_2(\kappa_A r_{out}), \nonumber \\
	G_{32}&=
	\kappa_A Y_2'(\kappa_A r_{out})
	+
	\frac{2il_A}{r_{out}}Y_2(\kappa_A r_{out}), \nonumber \\
	G_{33}&=
	\kappa_B J_2'(\kappa_B r_{out})
	+
	\frac{2il_B}{r_{out}}J_2(\kappa_B r_{out}), \nonumber \\
	G_{34}&=
	\kappa_B Y_2'(\kappa_B r_{out})
	+
	\frac{2il_B}{r_{out}}Y_2(\kappa_B r_{out}),
\end{align}
\begin{align}
	G_{41}&=
	\frac{2i}{r_{out}}J_2(\kappa_A r_{out})
	-
	l_A  \kappa_A J_2'(\kappa_A r_{out}), \nonumber \\
	G_{42}&=
	\frac{2i}{r_{out}}Y_2(\kappa_A r_{out})
	-
	l_A \kappa_A Y_2'(\kappa_A r_{out}), \nonumber \\
	G_{43}&=
	\frac{2i}{r_{out}}J_2(\kappa_B r_{out})
	-
	l_B \kappa_B J_2'(\kappa_B r_{out}), \nonumber \\
	G_{44}&=
	\frac{2i}{r_{out}}Y_2(\kappa_B r_{out})
	-
	l_B \kappa_B Y_2'(\kappa_B r_{out}),
\end{align}
\section{Limiting Cases and Completeness of the Analytical Solution}\label{limiting_cases}

The analytical solution derived in the main text assumes nonzero odd screening,
$\kappa_o\neq0$, and two distinct screening roots,
$\kappa_A^2\neq\kappa_B^2$. Here we examine the exceptional cases in which these
assumptions are not satisfied. We show that the apparent singularities associated
with vanishing odd screening and coincident screening roots can be resolved by
choosing an appropriate representation of the solutions. We begin with the coupled equations for the radial potentials,
\begin{equation}\label{LC1}
	\begin{aligned}
		\alpha L_2\Phi+\kappa_e^2\Phi+\kappa_o^2\Psi&=0,\\
		L_2\Psi+\kappa_e^2\Psi-\kappa_o^2\Phi&=0,
	\end{aligned}	
\end{equation}
where
\begin{equation}
	\alpha=2+\frac{\lambda}{\mu},\qquad
	L_2=\frac{d^2}{dr^2}+\frac1r\frac{d}{dr}-\frac4{r^2}.	
\end{equation}
Writing $\Phi=Af(r)$, $\Psi=Bf(r)$, and choosing $L_2f=-sf$, with $s=\kappa^2$, transforms Eq.~(\ref{LC1}) into
\begin{equation}\label{LC2}
	\begin{pmatrix} \kappa_e^2-\alpha s & \kappa_o^2\\ -\kappa_o^2 & \kappa_e^2-s \end{pmatrix} \begin{pmatrix} A\\B \end{pmatrix} =0.	
\end{equation}
A nontrivial solution requires the determinant to vanish, yielding
\begin{equation}\label{LC3}
	P(s)=\alpha s^2-\kappa_e^2(\alpha+1)s+\kappa_e^4+\kappa_o^4=0.
\end{equation}
The two roots, $s_A=\kappa_A^2$ and $s_B=\kappa_B^2$, determine the Bessel
branches employed in the main text. When these roots are distinct and
$\kappa_o\neq0$, the four corresponding Bessel solutions provide the complete
radial solution.

\subsection*{A. odd screening term $\kappa_{o} =0$}

Consider first the limit $\kappa_o=0$, with $\kappa_e>0$. The coupled
equations (\ref{LC1}) become
\begin{equation}\label{LC4}
\left(L_2+\frac{\kappa_e^2}{\alpha}\right)\Phi=0, \qquad (L_2+\kappa_e^2)\Psi=0.	
\end{equation}
The longitudinal and transverse potentials therefore decouple, and their general
solutions are
\begin{equation}
	\begin{aligned}\label{LC5}
		\Phi(r)&=C_LJ_2(\kappa_Lr)+D_LY_2(\kappa_Lr),\\
		\Psi(r)&=C_TJ_2(\kappa_Tr)+D_TY_2(\kappa_Tr),
	\end{aligned}	
\end{equation}
where
\begin{equation}
\kappa_L=\frac{\kappa_e}{\sqrt{\alpha}}, \qquad \kappa_T=\kappa_e.	
\end{equation}
An apparent difficulty arises from the relation introduced in the coupled
solution,
\begin{equation}\label{LC6}
	\Psi_\nu=\ell_\nu\Phi_\nu, \quad \ell_\nu= \frac{\alpha s_\nu-\kappa_e^2}{\kappa_o^2} = \frac{\kappa_o^2}{\kappa_e^2-s_\nu};~~ \nu=A,B.	
\end{equation}
Note that $\ell_\nu=B/A$ represents the ratio of the transverse to longitudinal potential
amplitudes in an eigenvector of Eq.~(\ref{LC2}). The equivalence of the two expressions follows from the characteristic equation $P(s_\nu)=0$. When $\kappa_o=0$, the matrix in Eq.~(\ref{LC2}) becomes diagonal. Its two eigenvectors are proportional to
\begin{equation}\label{LC7}
\mathbf v_L= \begin{pmatrix} 1\\0 \end{pmatrix},
\quad
\mathbf v_T= \begin{pmatrix} 0\\1 \end{pmatrix}.	
\end{equation}

To understand the limit $\kappa_o\to0$, we consider the two characteristic branches separately, assuming $\alpha>1$ and $\kappa_e>0$. For the longitudinal branch, $s_L\to\kappa_e^2/\alpha$. The second expression gives
\begin{equation}
\frac BA= \frac{\kappa_o^2}{\kappa_e^2-s_L} \longrightarrow0,	
\end{equation}
because the denominator approaches the nonzero value $\kappa_e^2(1-1/\alpha)$. Thus the transverse potential vanishes relative to the longitudinal potential.
For the transverse branch, $s_T\to\kappa_e^2$. It is more convenient to consider the inverse ratio, obtained from the first row of Eq.~(\ref{LC2}):
\begin{equation}
\frac AB=\frac{\kappa_o^2}{\alpha s_T-\kappa_e^2} \longrightarrow0,	
\end{equation}
because the denominator approaches $(\alpha-1)\kappa_e^2\neq0$. Thus the longitudinal potential vanishes relative to the transverse potential.
	
Consequently, the two eigenvectors approach the pure longitudinal and transverse directions, respectively,
\begin{equation}
\mathbf v_L\propto(1,0)^T, \qquad \mathbf v_T\propto(0,1)^T.	
\end{equation}
\textit{The divergence of $B/A$ for the transverse branch therefore reflects the vanishing of its longitudinal amplitude relative to its transverse amplitude, rather than a physical divergence. The correct even-screening limit is obtained from the two independent decoupled Helmholtz equations. The completely unscreened case, $\kappa_e=\kappa_o=0$, must instead be treated through the classical Navier equation and its biharmonic potential formulation}.

\subsection*{B. Coincident screening roots leading to vanishing of the discriminant}
The second difficulty arises when the discriminant of the characteristic
polynomial vanishes:
\begin{equation}\label{LC8}
	\Delta=\kappa_e^4(\alpha-1)^2-4\alpha\kappa_o^4=0.	
\end{equation}
For $\kappa_o\neq0$, the two roots coincide at
\begin{equation}\label{LC9}
s_A=s_B=s_{*}=	\frac{\kappa_e^2(\alpha+1)}{2\alpha}.	
\end{equation}
The fourth-order radial equation then reduces to
\begin{equation}\label{LC10}
(L_2+s_{*})^2\Phi=0.
\end{equation}
At this point, the ordinary Bessel solutions associated with the two branches
become identical. We apparently lose two independent solutions, although
Eq.~(\ref{LC10}) remains fourth-order. To recover the missing solutions, define
\begin{equation}
	f_Z(s,r)=Z_2(\sqrt{s}\,r),
	\qquad Z=J,Y,	
\end{equation}
which satisfies
\begin{equation}\label{LC11}
(L_2+s)f_Z=0.	
\end{equation}
Differentiating with respect to $s$, keeping $L_2$ fixed, gives
\begin{equation}\label{LC12}
(L_2+s)\frac{\partial f_Z}{\partial s}=-f_Z.
\end{equation}
Applying $(L_2+s)$ again shows that
\begin{equation}
	(L_2+s)^2\frac{\partial f_Z}{\partial s}=0.	
\end{equation}
Thus the derivatives of the Bessel functions provide the additional solutions
required by the repeated-root equation. Explicitly,
\begin{equation}\label{LC13}
	\frac{\partial f_Z}{\partial s}	=\frac{r}{2\sqrt{s}}Z_2'(\sqrt{s}\,r).
\end{equation}
However, the two potentials $\Phi$ and $\Psi$ remain coupled. We must
therefore differentiate the entire array of potentials rather than only the
Bessel function. We define
\begin{equation}\label{LC14}
	\mathbf U_Z(s,r)=
	\begin{pmatrix} f_Z(s,r)\\\ell(s)f_Z(s,r) \end{pmatrix},
	\qquad	\ell(s)=\frac{\alpha s-\kappa_e^2}{\kappa_o^2}.
\end{equation}
The coupled differential operator is
\begin{equation}
	\mathcal D=
	\begin{pmatrix} \alpha L_2+\kappa_e^2 & \kappa_o^2\\
		-\kappa_o^2 & L_2+\kappa_e^2 \end{pmatrix}. 	
\end{equation}
Applying on Eq.~(\ref{LC14}) gives
\begin{equation}\label{D15}
	\mathcal D \,\mathbf U_Z(s,r)=
	\begin{pmatrix} 0\\[2mm] -\dfrac{P(s)}{\kappa_o^2}f_Z(s,r) \end{pmatrix}.
\end{equation}
At the repeated root, both $P(s_{*})=0$ and $P'(s_{*})=0$. Since $\mathcal D$
does not depend on $s$, differentiating Eq.~(\ref{D15}) immediately yields
\begin{equation}\label{LC16}
	\mathcal D \left.\frac{\partial\mathbf U_Z}{\partial s}\right|_{s=s_{*}}=0.
\end{equation}
This proves that the differentiated vector satisfies both original coupled
equations. By defining
\begin{equation}
	F_Z(r)=Z_2(\sqrt{s_{*}}\,r),\qquad
	\dot F_Z(r)=
	\frac{r}{2\sqrt{s_{*}}}Z_2'(\sqrt{s_{*}}\,r),	
\end{equation}
and $\ell_{*}=\ell(s_{*})$, the complete solution at coincident roots becomes
\begin{equation}\label{LC17}
	\begin{aligned}
		\Phi(r)={}&C_JF_J+C_YF_Y
		+D_J\dot F_J+D_Y\dot F_Y,\\
		\Psi(r)={}&\ell_{*}\Phi(r)
		+\frac{\alpha}{\kappa_o^2}
		(D_JF_J+D_YF_Y).
	\end{aligned}
\end{equation}
The additional term in $\Psi$ arises from differentiating the eigenvector
ratio, since $\ell'(s)=\alpha/\kappa_o^2$, and is necessary to satisfy the
original coupled equations.

The four radial functions are linearly independent: applying $L_2+s_{*}$ to
any vanishing linear combination eliminates the ordinary Bessel terms and leaves
a combination of $F_J$ and $F_Y$. Their independence first forces the
derivative coefficients to vanish and subsequently the remaining coefficients. Therefore, Eq.~(\ref{LC17}) provides the complete four-dimensional radial solution space at the repeated root. The coincidence of screening roots represents a degeneracy of the original Bessel basis and does not, by itself, imply a singular
physical displacement or a mechanical instability.

\section{Equation for rotated ellipse, and transformation from circle to ellipse} \label{rotated}
The Equation of ellipse in a polar form is given by 
\begin{align}
	r(\theta) = \frac{a\times b}{\sqrt{b^2\cos^2(\theta) + a^2\sin^2(\theta)}},
\end{align}
where, $a$ and $b$ denote the semi-major and semi-minor axis respectively. In the above equation, $r$ represents the distance of a point on the periphery of the ellipse from the center, and $\theta$ is the angle made by $r$ from the horizontal axis. In this case the semi-major axis is aligned with x-axis, and it makes $\theta =0$ from the horizontal or x-axis.\\ \\
On the other hand if the semi-major axis makes an angle $\phi$ from the x-axis, taken in the counterclockwise, then the ellipse is rotated in the counter-clockwise by an angle $\phi$. The equation of the rotated ellipse, in general, is given by
\begin{align}
	r(\theta,\phi) = \frac{a\times b}{\sqrt{A(\alpha)\cos^2(\theta) + B(\alpha)\sin^2(\theta) +C(\alpha) \sin(2\theta)}},
\end{align}
where
\begin{align}
	A(\alpha) &=  [b^2\cos^2(\alpha) + a^2\sin^2(\alpha)], \nonumber \\
	B(\alpha) &=  [a^2\cos^2(\alpha) + b^2\sin^2(\alpha)], \nonumber \\
	C(\alpha) &=  \frac{[a^2-b^2]}{2}\sin(2\alpha) .
\end{align}
\\ 
If we consider a circle of radius $r_{in}$ to be deformed into an ellipse with the same area as that of the circle, then the semi-major and semi-minor axes are given by
\begin{align}
	a &=  r_{in}\times \delta \nonumber \\
	b &=  r_{in}\times\frac{1.0}{\delta},
\end{align}
such that $\pi a b = \pi r_{in}^2$.
In that case, the $A, B, C$ take the following form,

\begin{align}
	A(\alpha) &=  \frac{r_{in}^2}{\delta^2}[\cos^2(\alpha) + \delta^4\sin^2(\alpha)], \nonumber \\
	B(\alpha) &=  \frac{r_{in}^2}{\delta^2}[\sin^2(\alpha) + \delta^4\cos^2(\alpha)],\nonumber \\
	C(\alpha) &=  \frac{r_{in}^2}{\delta^2} \frac{1}{2}[\delta^4-1]\sin(2\alpha).
\end{align}
Now we substitute the expressions of $A,B,C$ in the equation of the rotated ellipse, we have
\begin{widetext}
	\begin{align}\label{Ellpise2}
		r(\theta, \phi) =  \frac{r_{in} \times \delta}{\sqrt{[\cos^2(\alpha) + \delta^4\sin^2(\alpha)]\cos^2(\theta) + [\sin^2(\alpha) + \delta^4\cos^2(\alpha)]\sin^2(\theta) + \frac{1}{2}[\delta^4-1]\sin(2\alpha)\sin(2\theta)}}.
	\end{align}
\end{widetext}
Now we use the following trigonometric identities in the above equation
\begin{align}
	\cos^2(\theta) = \frac{1 + \cos(2\theta)}{2}, \nonumber \\
	\sin^2(\theta) = \frac{1 - \cos(2\theta)}{2}.
\end{align}
Therefore, Eq. \ref{Ellpise2} becomes, after simplification
	\begin{align}\label{Ellpise3}
		r(\theta, \phi) = 
		&=  \frac{\sqrt{2} \times r_{in} \times \delta} {\sqrt{(\delta^4 +1) - (\delta^4-1)\cos(2\theta + 2\alpha) }}
	\end{align}
Now break $\delta$ into an approximated form such as
\begin{align}
	\delta = 1+\epsilon.
\end{align}
Therefore we have, 
\begin{align}
	\delta^4 = (1+\epsilon)^4 = 1+ 4\epsilon + 6\epsilon^2 + 4\epsilon^3 + \epsilon^4,
\end{align}
and
\begin{align}
	(\delta^4 +1) &=  2+ 4\epsilon + 6\epsilon^2 + 4\epsilon^3 + \epsilon^4, \nonumber \\
	(\delta^4 -1) &=  4\epsilon + 6\epsilon^2 + 4\epsilon^3 + \epsilon^4.
\end{align}
Now in the limmit $\bf \epsilon \rightarrow 0$, we obtain
\begin{align}
	(\delta^4 +1) & \simeq  2+ 4\epsilon, \nonumber \\
	(\delta^4 -1) & \simeq  4\epsilon.
\end{align}
Now, substituting these expressions in the last line of Eqn.(\ref{Ellpise4}), we obtain

\begin{align}\label{Ellpise4}
	r(\theta, \phi)  &=  \frac{\sqrt{2} \times r_{in} \times (1+\epsilon)} {\sqrt{(2+4\epsilon) - 4\epsilon \cos(2\theta + 2\alpha) }},  \nonumber \\ \nonumber \\
	&= \left[ r_{in} \times (1+\epsilon)\right] \times    \left[ (1 -\epsilon) + \epsilon\cos(2\theta + 2\alpha)  \right]  ,  
\end{align}
which on taking the limit $\bf \epsilon \rightarrow 0$, we obtain the following form of the rotated ellipse \\
\begin{align}
	r(\theta, \phi) =  r_{in} \left[ 1 + \epsilon  \cos(2\theta + 2\alpha)  \right]. 
\end{align}

Now if we subtract the polar equation of circle, $i.e.,r(\theta)_{circ} =r_{in} $ from the above equation, we get the change $dr$ that takes place going from circle to ellipse. That is
\begin{align}
	dr(\theta, \phi) &= r_{in} \left[ 1 + \epsilon  \cos(2\theta + 2\alpha)  \right] -r_{in} \nonumber \\ \nonumber \\
	dr(\theta, \phi)    &= r_{in} \epsilon  \cos(2\theta + 2\alpha).
\end{align}
For the semi-major axis aligned with $x$-axis $\alpha =0$. 
\begin{equation}
	dr(\theta, \phi)  = r_{in} \epsilon  \cos(2\theta).
\end{equation}
\section{Defects: Vortices, Anti-vortices from Displacement Vector Field}\label{defects}
To further characterize the spatial organization of the non-affine response, we examine the topological structure of the coarse-grained displacement field. Vortices and antivortices are identified from the winding number of the displacement orientation around elementary plaques. Their number increases as the coarse-graining length is reduced, revealing a hierarchy of rotational structures that is progressively suppressed by spatial averaging. Moreover, the antivortices exhibit a nonuniform spatial organization, with several defects appearing along regions where the displacement field changes its direction and amplitude. This observation suggests that the topological organization of the non-affine displacement field may provide a microscopic signature of the collective polarization underlying the phenomenon of screening.

The defect computation, that is finding the number of vortices and antivortcises in a given displacement vector field, is fundamentally a statement about the geometry and topology of the coarse-grained displacement vector field $\mathbf{u}$
\begin{equation}
	\mathbf u_h(\mathbf r) = \big(u_{hx}(\mathbf r),u_{hy}(\mathbf r)\big),
\end{equation}
where $h$ represent the coarse-grained length of a cell.
The defects come from the following geometrical construction, i.e., \textbf{how the direction of $\mathbf u_h$ rotates as one travels around a closed loop}. Suppose that at every point of a two-dimensional system we have a vector
\begin{equation}
	\mathbf u(\mathbf r)
	=
	\begin{pmatrix}
		u_x(\mathbf r)\\
		u_y(\mathbf r)
	\end{pmatrix}.
\end{equation}

Whenever $|\mathbf u|\neq0$, we can write it as
\begin{equation}
	\mathbf u =
	|\mathbf u|
	\begin{pmatrix}
		\cos\phi\\
		\sin\phi
	\end{pmatrix},
\end{equation}
where
\begin{equation}
	\phi(\mathbf r) = \operatorname{atan2}(u_y,u_x)
\end{equation}
is the orientation of the vector.

Now we imagine drawing a small closed curve $C$ in some region of the material and follow the direction of $\phi$ continuously while we make one round along the curve. If nothing special lies inside the curve, the orientation comes back to itself without accumulating a net rotation, 
\begin{equation}
	\Delta\phi_C=0.
\end{equation}
But suppose that during one counterclockwise movement the vector direction itself makes one complete counter-clockwise revolution. Then
$
\Delta\phi_{C}=2\pi.
$
Therefore, we define
\begin{equation}
	q=\frac{1}{2\pi} \oint_C d\phi
\end{equation}
and therefore
$
q=+1.
$
On the other hand if the vector orientation makes one complete clockwise revolution,
$
\Delta\phi_C=-2\pi,
$
and
$
q=-1.
$
This integer $q$ is the \textbf{topological charge}, or winding number. That is the basic origin of our calculation.

To understand why we consider traversing a loop and what special happens inside such a loop, let us take the case of 
$
q=+1.
$
In this case the vector orientation has rotated through $2\pi$ around the loop. Now we ask, could the vector field remain nonzero and perfectly regular everywhere inside?

For an ordinary continuous two-component vector field on a disk, it cannot do so while carrying a nonzero winding on the boundary. The vector field must become undefined in orientation somewhere inside the loop. For our displacement field, this happens where
\begin{equation}
	\mathbf u_h(\mathbf r_0)=0
\end{equation}
or, numerically,
\begin{equation}
	|\mathbf u_h(\mathbf r_0)|\approx0.
\end{equation}
At this point,
\begin{equation}
	u_{hx}=0,\qquad u_{hy}=0,
\end{equation}
so
\begin{equation}
	\phi= \text{atan2}(0,0)
\end{equation}
has no physical direction. 

Thus a topological defect is fundamentally associated with a \textit{zero of the vector field}. This immediately explains why low-magnitude regions in our coarse-grained color map are physically very important. They are not used to \emph{define} the defect, but topological defects should be associated with zeros or near-zeros of the coarse-grained displacement field. Note that the quantity $\log_{10}|\mathbf u_h|$ in the color map is only a way of displaying this field. For detail description of defects such as vortices and anti-vortices, please look at the N.D. Mermin's Review in Ref.~\cite{79Mer}. 

\subsection{Computation of vortices and anti-vortices}
Starting from the discrete particle positions $\mathbf{r}_i = (x_i, y_i)$ and displacement vectors $\mathbf{d}_i = (d_{xi}, d_{yi})$, we constructed a continuous displacement field on a regular Cartesian grid. The grid covered $-80 \le x, y \le 80$ with $\Delta x = \Delta y = 2.5$ (64 nodes along each direction). At every grid node $\mathbf{R}_{jk}$ we evaluated the normalized Gaussian average
\begin{equation}
	\mathbf{u}_h(\mathbf{R}_{jk}) = \frac{\sum_i w_{ijk} \mathbf{d}_i}{\sum_i w_{ijk}},	
\end{equation}
where
\begin{equation}
	w_{ijk} = \exp\left[-\frac{|\mathbf{R}_{jk} - \mathbf{r}_i|^2}{2h^2}\right].	
\end{equation}
\begin{figure}
	\centering
	\includegraphics[width=0.42\textwidth]{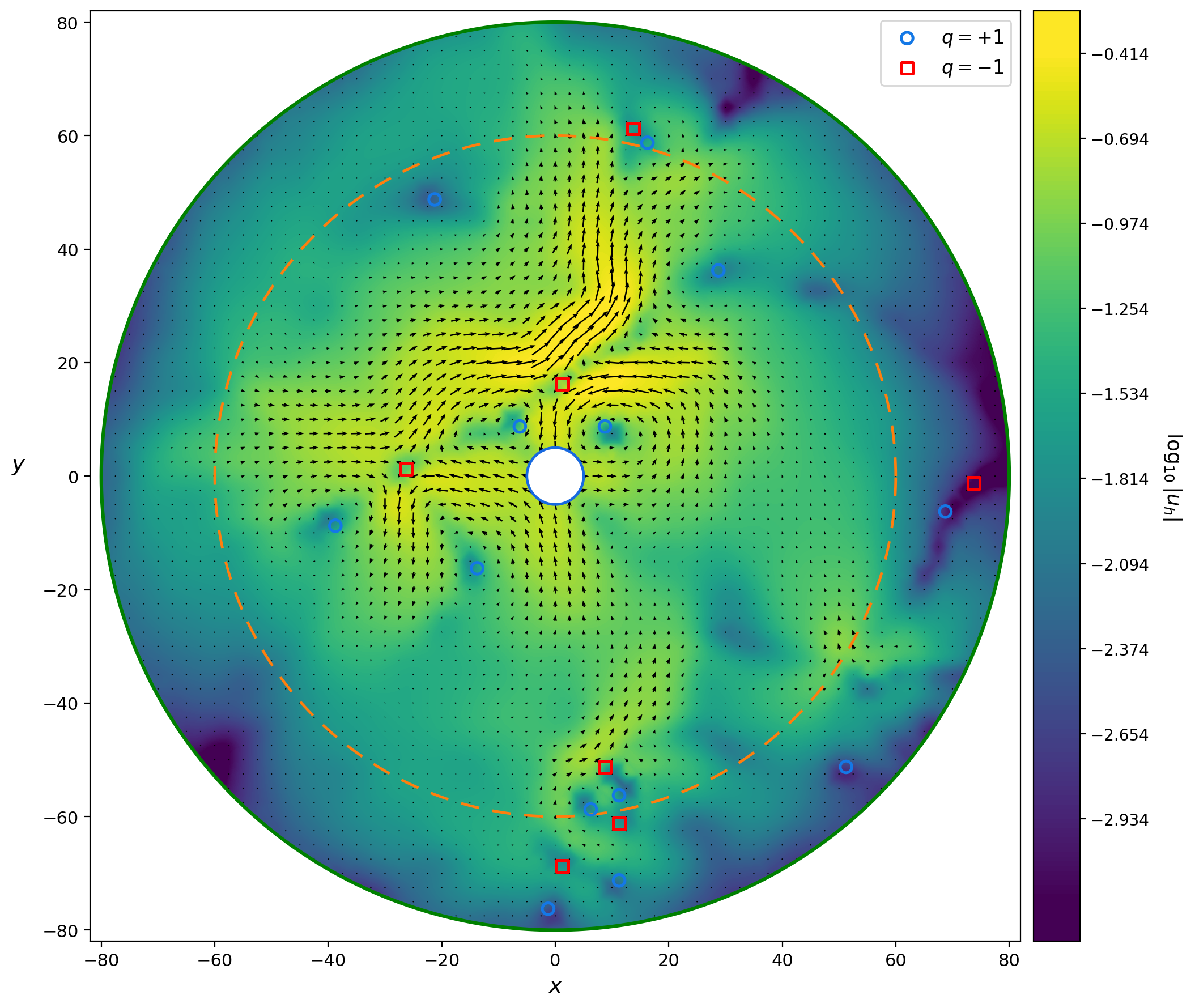}
	\caption{Computation of vortices and anti-vortices for the total displacement vector field in Fig.~\ref{fig:anomalousMap}. The vortices correspond to $q=+1$ representing the spiral like or circulating vector fields around a core of vanishing field. The anti-vortices correspond to $q=-1$ which representing saddle like displacement vector fields or the Eshelby quadrupolar fields, \cite{79Mer}.}
	\label{fig:defectsMap}
\end{figure}
The parameter $h$ is the coarse-graining length and sets the spatial scale over which neighboring particle displacements are averaged; for the displayed map we used $h = 2.5$ and truncated the Gaussian sum at $r_c = 3h = 7.5$. We then calculated
\begin{equation}
	|\mathbf{u}_h| = \left(u_{hx}^2 + u_{hy}^2\right)^{1/2}
\end{equation}
and
\begin{equation}
	\phi = \text{atan2}(u_{hy}, u_{hx})
\end{equation}
at every grid node. 
\begin{figure}
	\centering
	\includegraphics[width=0.42\textwidth]{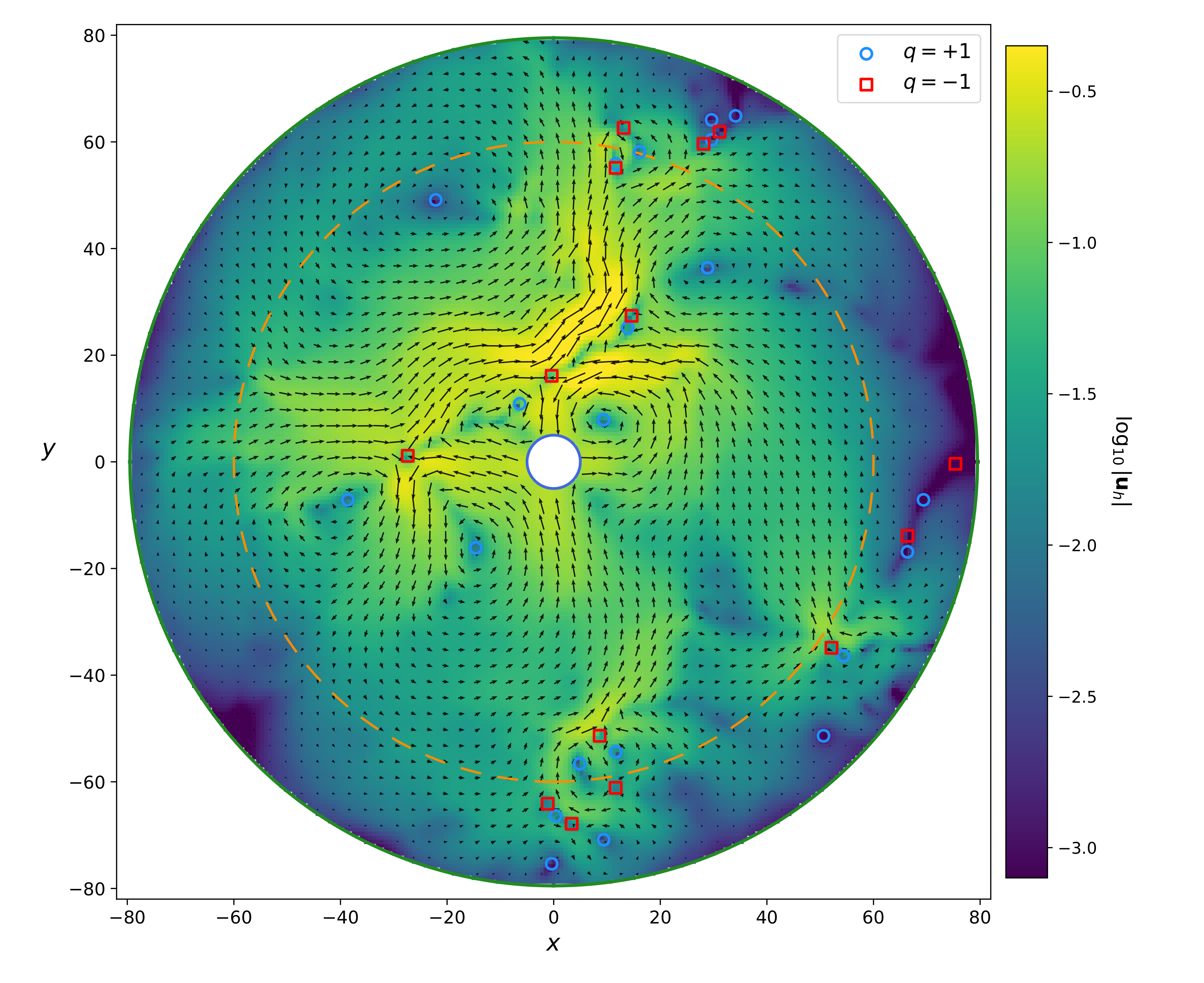}
	\caption{The coarse-graining length is $h=1.25$ in this case. By reducing the Gaussian coarse-graining length the number of defects increases. However, their amplitude is low. }
	\label{fig:VMap}
\end{figure}
Topological defects were determined from the winding of $\phi$ around each elementary grid cell, traversed counterclockwise. For every edge, the phase increment was wrapped into $(-\pi, \pi)$ to remove the \texttt{atan2} branch-cut discontinuity, and the cell winding was
\begin{equation}
	W = \frac{1}{2\pi} \sum_{\text{edges}} \operatorname{wrap}(\Delta \phi).
\end{equation}
Cells with $W > 0.5$ and $W < -0.5$ were classified as $q = +1$ and $q = -1$ defects, respectively, with the initial defect coordinate assigned to the cell center. Candidate cores were retained within the fixed physical detection domain $7 \le r \le 76$. The separate radius $r = 60$ was used only as an interior analysis boundary and did not enter the construction of the coarse-grained field.

The defects computation has been shown in Fig.~\ref{fig:defectsMap}. The scalar background in the figure represents $\log_{10} |\mathbf{u}_h|$, while the arrows indicate the local direction of $\mathbf{u}_h$.
Robustness was assessed independently by varying the coarse-graining length $h$ and, separately, by refining the sampling grid at fixed $h$. This distinction is important: changing $h$ changes the observed coarse-grained field, whereas changing $\Delta x$ and $\Delta y$ changes only its numerical sampling. In Fig.~\ref{fig:VMap} we have shown the defect map corresponding to Gaussian coarse-graining length $h=1.25$. We observe few more defects appearing which is due to fine resolution. 

\bibliography{ALL,All.Anomalous}
\end{document}